\documentclass[a4paper,review]{cas-sc}
\usepackage[section]{placeins} % For '\input{file}'
\usepackage{amssymb}
\usepackage{amsthm}
\usepackage{amsmath}
\usepackage{upgreek}
\usepackage{pdflscape}
\usepackage{listings}
\usepackage{multirow}
\usepackage[percent]{overpic}
\usepackage{multicol}
\usepackage{color}
\usepackage{stmaryrd}
\usepackage{xfrac}
\usepackage[capitalise]{cleveref}
\usepackage{booktabs}
\usepackage[section]{placeins} % For '\input{file}'
\usepackage[section]{algorithm}
\usepackage{calc}
\usepackage[titletoc]{appendix}
\usepackage{siunitx}
\usepackage{mathtools}
\usepackage[sort&compress,square,numbers]{natbib}

\usepackage{tikz-dimline}
\usepackage{graphicx}
\usepackage{graphics}
\usepackage{wrapfig}
\usepackage{float}
\usepackage{subfig}
\usepackage[percent]{overpic}
\usepackage{varwidth}
\usepackage{tikz}
\usepackage{pgfplots}
\usepackage{adjustbox}
\usetikzlibrary{arrows,matrix,positioning,fit}
\usetikzlibrary{shapes,positioning}
\usetikzlibrary{backgrounds}
\usepackage{tikz-layers}
\usepgfplotslibrary{fillbetween}
\usetikzlibrary{intersections}
\pgfplotsset{compat=1.14}
\usepgfplotslibrary{colorbrewer}
\usepgfplotslibrary{patchplots}
\usepgfplotslibrary[colorbrewer]
\usetikzlibrary{pgfplots.colorbrewer}
\usetikzlibrary[pgfplots.colorbrewer]
\usepgfplotslibrary{units}
\usetikzlibrary{spy}
\usepackage{pgfplotstable}
\usepackage{arrayjobx}
\graphicspath{ {./figs/} }
\usetikzlibrary{external}

\newlength\myheight
\newlength\mydepth
\settototalheight\myheight{Xygp}
\newcommand*\inlinegraphics[1]{%
  \settototalheight\myheight{Xygp}%
  \settodepth\mydepth{Xygp}%
  \raisebox{-\mydepth}{\includegraphics[height=\myheight]{#1}}%
}
\newcommand\orcid[1]{\href{https://orcid.org/#1}{\inlinegraphics{orcid_16x16.png}}}

\makeatletter
\def\BState{\State\hskip-\ALG@thistlm}
\makeatother

\newdefinition{definition}{Definition}[section]

\begin{document}

\title[mode=title]{Validation of wall boundary conditions for simulating complex fluid flows via the Boltzmann equation: Momentum transport and skin friction}
\shorttitle{Wall boundary conditions for the Boltzmann--BGK equation}
\shortauthors{T. Dzanic \textit{et al.}}

\author[1]{T. Dzanic}[orcid=0000-0003-3791-1134]
\cormark[1]
\cortext[cor1]{Corresponding author}
\ead{tdzanic@tamu.edu}
\author[2]{F. D. Witherden}[orcid=0000-0003-2343-412X]
\author[1]{L. Martinelli}[orcid=0000-0001-6319-7646]

\address[1]{Department of Mechanical and Aerospace Engineering, Princeton University, Princeton, NJ 08544, USA}
\address[2]{Department of Ocean Engineering, Texas A\&M University, College Station, TX 77843, USA}

\begin{abstract}
The influence and validity of wall boundary conditions for non-equilibrium fluid flows described by the Boltzmann equation remains an open problem. The substantial computational cost of directly solving the Boltzmann equation has limited the extent of numerical validation studies to simple, often two-dimensional, flow problems. Recent algorithmic advancements for the Boltzmann--BGK (Bhatnagar--Gross--Krook) equation introduced by the authors (Dzanic \textit{et al}., \textit{J. Comput. Phys.}, 486, 2023), consisting of a highly-efficient high-order spatial discretization augmented with a discretely-conservative velocity model, have made it feasible to accurately simulate unsteady three-dimensional flow problems across both the rarefied and continuum regimes. This work presents a comprehensive evaluation and validation of wall boundary conditions across a variety of flow regimes, primarily for the purpose of exploring their effects on momentum transfer in the low Mach limit. Results are presented for a range of steady and unsteady wall-bounded flow problems across both the rarefied and continuum regimes, from canonical two-dimensional laminar flows to unsteady three-dimensional transitional and turbulent flows, the latter of which are the first instances of wall-bounded turbulent flows computed by directly solving the Boltzmann equation. We show that approximations of the molecular gas dynamics equations can accurately predict both non-equilibrium phenomena and complex hydrodynamic flow instabilities and show how spatial and velocity domain resolution affect the accuracy. The results indicate that an accurate approximation of particle transport (i.e., high spatial resolution) is significantly more important than particle collision (i.e., high velocity domain resolution) for predicting flow instabilities and momentum transfer consistent with that predicted by the hydrodynamic equations and that these effects can be computed accurately even with very few degrees of freedom in the velocity domain. These findings suggest that highly-accurate spatial schemes (e.g., high-order schemes) are a promising approach for solving molecular gas dynamics for complex flows and that the direct solution of the Boltzmann equation can be performed at a reasonable cost when compared to hydrodynamic simulations at the same level of resolution. 
\end{abstract}


%\linenumbers

% ================================================================================

\maketitle

 %!TEX root = ./main.tex
\section{Introduction}
\label{sec:intro}

For the majority of application areas in computational fluid dynamics (CFD), the assumption that the fluid can be treated as continuum is generally well-posed, and the fluid flow can be reliably described solely via its macroscopic state. This assumption justifies the choice of models based on the macroscopic conservation laws for mass, momentum and energy: the governing equations of continuum fluid mechanics (e.g., Navier--Stokes equations) for which computational fluid dynamics methods have primarily focused on over the years. However, in certain applications such as rarefied gas dynamics, microflows, and hypersonic aeronautics, the flow can deviate significantly from thermodynamic equilibrium, such that the approximation of these flows using governing equations based on the continuum assumption can give exceedingly erroneous predictions. For these non-equilibrium flows, it is necessary to instead rely on methods derived from the kinetic theory of gases such as the governing equations of molecular gas dynamics which underpin the macroscopic behavior of the fluid.

\begin{figure}[tbhp]
        \centering
        \adjustbox{width=0.4\linewidth, valign=b}{\begin{tikzpicture}[scale=1]
\draw[->, black, thick] (0,0) -- (5,0);
\draw[->, black, thick] (0,0) -- (0,5);
\node[text=black] at (4.4,-0.5) { \begin{tabular}{c} Reynolds \\  number \end{tabular}};
\node[text=black, rotate=90] at (-0.5, 4.5) {\begin{tabular}{c} Mach \\ number \end{tabular}};

\draw[-, black, dotted, thick] (0,2.5) -- (5, 2.5);
\draw[-, black, dotted, thick] (2.5, 0) -- (2.5, 5);
\node[text=black] at (3.75, 1.25) { \footnotesize \begin{tabular}{c} \textit{Low-speed} \\  \textit{equilibrium flows} \end{tabular}};
\node[text=black] at (3.75, 3.75) { \footnotesize  \begin{tabular}{c} \textit{High-speed} \\  \textit{non-equilibrium} \\ \textit{flows} \end{tabular}};
\node[text=black] at (1.25, 3.75) { \footnotesize  \begin{tabular}{c} \textit{High-speed} \\  \textit{rarefied flows} \end{tabular}};
\node[text=black] at (1.25, 1.25) { \footnotesize  \begin{tabular}{c} \textit{Low-speed} \\  \textit{rarefied flows} \end{tabular}};

\draw[red, dashed, thick, dash pattern=on 4pt off 4pt,dash phase=4pt] (0.1, 0.1) -- (0.1, 4.9) -- (2.4, 4.9) -- (2.4, .1) -- (0.1, 0.1) ;
\draw[blue, dashed, thick, dash pattern=on 4pt off 4pt,dash phase=0pt] (0.1, 0.1) -- (0.1, 2.4) -- (4.9, 2.4) -- (4.9, 0.1) -- (0.1, 0.1) ;

\end{tikzpicture}}
        \caption{\label{fig:regimes} 
        Diagram of the various flow regimes for which the Boltzmann equation is valid across. Red line encloses the regimes the Boltzmann equation is typically applied to. Blue line encloses the regimes which are explored in this work. 
        }
    \end{figure}
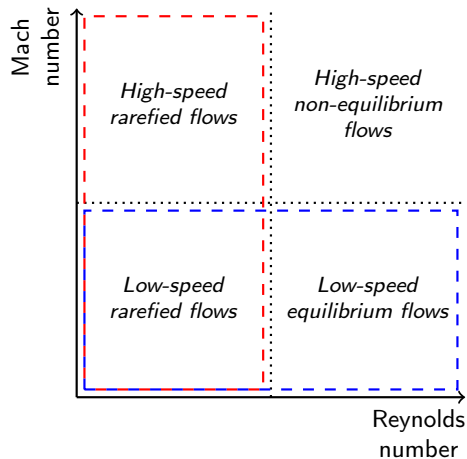

One such kinetic approach can be given by approximating the Boltzmann equation which characterizes molecular gas dynamics through particle transport and collision. In this approach, the underlying flow physics are straightforwardly described via the evolution of a scalar \textit{particle distribution function}. From this description at the mesoscale level, one can recover the evolution of the macroscopic flow state that is equally valid for rarefied and continuum flow regimes by simply making use of the collision invariants of the governing equations \citep{Cercignani1988}. The broad-range validity of this model makes the Boltzmann equation an attractive approach for simulating complex multi-scale flows, particularly ones that span multiple flow regimes. In fact, the Boltzmann equation provides a single framework for simulating flows across a variety of regimes, shown in \cref{fig:regimes}, ranging from low-speed rarefied flows to high-speed non-equilibrium flows. However, the direct simulation of the Boltzmann equation has typically been focused strictly on the low-Reynolds number rarefied regime, with applications ranging from microflows (i.e., low-speed rarefied flows) to atmospheric re-entry (i.e., high-speed rarefied flows), as these are typically the most common conditions where continuum fluid mechanics descriptions break down. Nevertheless, solutions of the Boltzmann equation in other flow regimes may potentially offer new insights and provide more robust and accurate numerical approaches for complex problems in  fluid dynamics~\citep{McMullen2022}. For example, the approximation of low-speed equilibrium (i.e., continuum) flows via the Boltzmann equation would provide a description of fluid flow through the evolution of a probability density function which encodes the dynamics of the flow in a radically different manner than the macroscopic conservation laws. As such, analysis of  solutions to the Boltzmann equation may lead to enhancing our fundamental understanding  of  flow phenomena such as transition to turbulence since the complex or even chaotic flow behavior may actually stem from much simpler dynamics in higher-dimensional representations such as the one provided by the Boltzmann equation~\citep{Chen2003}. Furthermore, for high-speed non-equilibrium flows (e.g., hypersonic aeronautics), the strong multi-scale nature of the flow poses a challenge for current numerical approaches which rely on the Navier--Stokes equations~\citep{Longo2007}, and the approximation of these flows through the molecular gas dynamics perspective offered by the Boltzmann equation may provide a better framework for simulating the complex high-temperature aerothermodynamic effects which are encountered in these applications~\citep{Titarev2018,Titarev2020}.

While these potential benefits certainly motivate the use of the Boltzmann equation as a numerical approach for simulating complex multi-scale flows, they come at the expense of a significantly larger computational cost than the hydrodynamic equations. This cost stems from two main sources: high dimensionality and modeling of the particle interactions. For the former, the distribution function, which represents a measure of the probability density of a particle existing at some location with some velocity, is necessarily defined over a phase space which can be up to 7-dimensional. As such, the cost of approximating this phase space grows extremely quickly with respect to the resolution. For the latter, modeling particle collision can be an extremely computationally intensive task, requiring integration over even higher-dimensional spaces, such that the process can become the majority of the computational effort of solving the governing equations. As a result, directly solving the Boltzmann equation has typically been restricted to simpler applications (e.g., two-dimensional steady flows), and its use for complex engineering flows has widely been considered intractable.

To broaden the applicability of the Boltzmann equation, more efficient approaches must be used to mitigate the cost of directly computing the collision term and the high-dimensionality of the equations. A common approach for the former is to approximate this process with a suitable model, the most ubiquitous being the Bhatnagar–Gross–Krook (BGK) \citep{Bhatnagar1954} model. This approach, which approximates collisions as a relaxation process toward thermodynamic equilibrium, can drastically reduce the cost of the modeling particle interactions while still retaining many of the desirable properties of the full collision integral, such as its entropy-satisfying properties embedded in Boltzmann’s H-theorem and convergence to the continuum approximation in the asymptotic limit. For the latter, the curse of dimensionality can be stymied by utilizing more efficient numerical methods which can offer equivalent accuracy with lower resolution, reducing the number of degrees of freedom necessary to approximate the high-dimensional phase space. These efficiency improvements can be realized through the use of spatially high-order schemes \citep{Xiao2021, Jaiswal2022}, which offer better spatial resolution at a lower computational cost, in conjunction with discretely-conservative velocity models \citep{Mieussens2000, Mieussens2000JCP}, which can reduce the unnecessary degrees of freedom in the velocity domain typically needed to mitigate integration errors to ensure conservation. The combination of these approaches for the Boltzmann--BGK equation was introduced by the authors in \citet{Dzanic2023}, consisting of a positivity-preserving high-order flux reconstruction scheme augmented with a discretely-conservative nodal velocity model efficiently implemented on large-scale GPU computing architectures. These advancements have drastically expanded the complexity of the flows that can be simulated by directly solving the Boltzmann equation, allowing for its application to even three-dimensional turbulent flows, which is, to the authors' knowledge, the first instance of such flows computed in this manner.

 The purpose of this work is to verify and validate our approach to molecular gas dynamics for complex wall-bounded fluid flows by leveraging the ability to directly solve the Boltzmann-BGK partial differential equation for problems that were previously intractable. As the Boltzmann equation has not yet been applied to these more complex flows (e.g., transitional and turbulent flows, unsteady non-equilibrium flows, etc.), it is not evident whether this approach can even predict these fundamental flow phenomena and properties such as: 1) hydrodynamic instabilities leading to transition to turbulence; 2) momentum transfer effects and wall shear stresses across the continuum and rarefied regimes; 3) energy transfer effects and surface heat fluxes in the high Mach regime; and 4) shock-driven flow instabilities and interactions. Uncertainties in the ability of predicting these fundamental flow features are further compounded by the fact that the correct choice and validity of wall boundary conditions for the Boltzmann equation remain an open problem \citep{Evans2011, Williams2001}, particularly so for unsteady flows and flows in the rarefied regime. Thus,  an extensive validation of the model is necessary prior to its application to complex engineering problems (e.g., hypersonic transitional flows). This paper focuses on the influence of wall boundary conditions on momentum transfer, with the goal of presenting a comprehensive evaluation and validation of approximations of the Boltzmann equation for wall-bounded flows across a variety of operating conditions, including two- and three-dimensional laminar and transitional/turbulent flows across the rarefied and continuum regimes. To isolate the effects of momentum transfer from energy transfer, calculations are performed in the low Mach limit, and particular attention is dedicated to observing the effects of spatial and velocity domain resolution on predicting non-equilibrium and unsteady flows. 

The remainder of this manuscript is organized as follows. In \cref{sec:methodology}, the governing equations and numerical approach are presented. Implementation details are then shown in \cref{sec:implementation}, and the results of numerical experiments are presented in \cref{sec:results}. A discussion on the results in then given in \cref{sec:discussion}, and conclusions are finally drawn in \cref{sec:conclusion}.

%!TEX root = ./main.tex
\section{Methodology}\label{sec:methodology}
The Boltzmann equation can be given in terms of a linear conservation law with a nonlinear source term as
\begin{equation}\label{eq:boltzmann}
    \partial_t f (\mathbf{x}, \mathbf{u}, t) + \mathbf{u} {\cdot} \nabla f = \mathcal C(f, f'),
\end{equation}
where $\mathbf{x} \in \Omega^{\mathbf{x}}$ is the physical space for some physical domain $\Omega^{\mathbf{x}} \subseteq \mathbb R^d$ and spatial dimension $d$, $\mathbf{u} \in \Omega^{\mathbf{u}}$ is the associated velocity space for some velocity domain $\Omega^{\mathbf{u}} \in \mathbb R^m$ and velocity dimension $m \geq d$, $f (\mathbf{x}, \mathbf{u}, t)\in \mathbb R$ is a scalar particle distribution function, and $\mathcal C(f, f')$ is the collision operator which models particle interactions \citep{Cercignani1988}. The distribution function, which represents a probability density of a particle existing at some location $\mathbf{x}$ with some velocity $\mathbf{u}$, can be related to a unique macroscopic state of the system. By taking the moments of the distribution function, the conserved flow variables $\mathbf{Q}(\mathbf{x}, t)$ can be recovered as 
\begin{equation}\label{eq:moments}
    \mathbf{Q}(\mathbf{x}, t) = \left[\rho, \rho \mathbf{U}, E \right]^T = 
    \int_{\mathbb R^m} f (\mathbf{x}, \mathbf{u}, t)\ \boldsymbol{\psi} (\mathbf{u}) \ \mathrm{d}\mathbf{u},
\end{equation}
where $\rho$ is the density, $\rho \mathbf{U}$ is the momentum vector, $E$ is the total energy, and $\boldsymbol{\psi} (\mathbf{u}) \coloneqq [1, \mathbf{u}, (\mathbf{u}\cdot\mathbf{u})/2]^T$ is the vector of collision invariants. Furthermore, it is sometimes convenient to represent the solution in terms of the primitive variables $\mathbf{q} = [\rho, \mathbf{U}, P]^T$, where $\mathbf{U} = \rho \mathbf{U} / \rho$ is the macroscopic velocity and $P = (\gamma - 1)(E - \frac{1}{2}\rho \mathbf{U}\cdot\mathbf{U})$ is the pressure. To differentiate between the microscopic and macroscopic velocity, we utilize the notation that the lowercase symbol $\mathbf{u}$ refers to the microscopic velocity whereas the uppercase symbol $\mathbf{U}$ refers to the macroscopic velocity.

The primary characteristic quantity for the Boltzmann equation is the Knudsen number, defined as
\begin{equation}
    Kn = \frac{\lambda}{L},
\end{equation}
where $\lambda$ is the particle mean free path and $L$ is some characteristic length scale of the problem. The Knudsen number can be considered as a measure of the non-equilibrium of the flow, with approximate ranges in Knudsen number corresponding to varying flow regimes (e.g., continuum flow for $Kn < 10^{-3}$, slip flow for $10^{-3} < Kn < 10^{-1}$, transitional flow for $10^{-1} < Kn < 10$, and free molecular flow for $Kn > 10$ \citep{Darabi2012}). This quantity can be related to the Mach number $M$ and Reynolds number $Re$ in the flow as
\begin{equation}
    Kn = \sqrt{\frac{\gamma \pi}{2}} \frac{M}{Re},
\end{equation}
where $\gamma$ is the specific heat ratio. 

Without a model for internal degrees of freedom for a molecule (e.g., rotation, vibration), the Boltzmann equation as given by \cref{eq:boltzmann} approximates a monatomic molecule for which only translational degrees of freedom exist. Therefore, for an $m$-dimensional velocity space, there are $m$ degrees of freedom, such that the specific heat ratio can be computed as $\gamma = 1 + 2/m$. In this work, we neglect internal degrees of freedom and use a velocity domain of equal dimensionality as the spatial domain (i.e., $m = d$), which fixes the specific heat ratio to $\gamma = 2$ and $\gamma = 5/3$ for two- and three-dimensional problems, respectively. While proper representation of the specific heat ratio as well as suitable modeling of internal degrees of freedom play an important role for high-temperature flows \citep{Baranger2020}, these effects are negligible at low Mach numbers which is the focus of this work.

\subsection{Collision operator}
To avoid the computational complexity of directly computing the collision operator $C(f, f')$, the effect of particle interactions can be approximated using a suitable collision model. In the BGK model \citep{Bhatnagar1954}, particle collision is approximated as a multi-dimensional relaxation process of a gas tending towards thermodynamic equilibrium, given as 
\begin{equation}
    C(f, f') \approx \frac{g(\mathbf{x}, \mathbf{u}, t) - f(\mathbf{x}, \mathbf{u}, t)}{\tau}.
\end{equation}
Here, $g(\mathbf{x}, \mathbf{u}, t)$ represents the distribution of a state in thermodynamic equilibrium and $\tau$ represents some collision time scale. 

The distribution of equilibrium state stems closely from Boltzmann's H-theorem as it is the state that minimizes the entropy $H(\mathbf{z})$, i.e.,
\begin{equation}
    g(\mathbf{u}) = \underset{\mathbf{z}}{\arg \min}\ H(\mathbf{z}), \quad \quad H(\mathbf{z}) = \int_{\mathbb R^m} \mathbf{z} \log (\mathbf{z}) \ \mathrm{d} \mathbf{z},
\end{equation}
subject to the compatibility condition
\begin{equation}\label{eq:compatibility}
    \int_{\mathbb R^m} g (\mathbf{x}, \mathbf{u}, t)\ \boldsymbol{\psi} (\mathbf{u}) \ \mathrm{d}\mathbf{u} = \int_{\mathbb R^m} f (\mathbf{x}, \mathbf{u}, t)\ \boldsymbol{\psi} (\mathbf{u}) \ \mathrm{d}\mathbf{u}.
\end{equation}
This compatibility condition ensures that the macroscopic states of $g (\mathbf{x}, \mathbf{u}, t)$ and $f (\mathbf{x}, \mathbf{u}, t)$ are identical, such that conservation of the macroscopic variables is enforced. In the absence of internal degrees of freedom, the state that minimizes the entropy and (analytically) satisfies the compatibility condition is a Maxwellian distribution of the form
\begin{equation}
    g(\mathbf{x}, \mathbf{u}, t) = \frac{\rho (\mathbf{x}, t)}{\left[2 \pi \theta(\mathbf{x}, t) \right]^{d/2}}\exp \left [-\frac{ \|\mathbf{u} - \mathbf{U}(\mathbf x, t) \|_2^2}{2 \theta (\mathbf x, t)} \right],
\end{equation}
where $\theta = P/\rho$ is a scaled temperature. We additionally use the notation $g(\mathbf{Q})$ to denote the Maxwellian distribution corresponding to the macroscopic state variables $\mathbf{Q}$. 

It can be shown \citep{Cercignani1988} that the collision time $\tau$ is related to the dynamic viscosity $\mu$ of the flow as 
\begin{equation}
    \mu = \tau P. 
\end{equation}
By setting the collision time as a constant value, unphysical behavior of the transport coefficient is recovered where the viscosity varies proportionally with pressure. Instead, it is more appropriate to adapt the collision time to recover a more physically consistent temperature-viscosity power law, i.e.,
\begin{equation}
    \mu \propto \theta^\omega,
\end{equation}
where $\omega$ is the viscosity law exponent that depends on the gas being simulated. This is achieved by setting the collision time as
\begin{equation}
    \tau = \frac{\mu_{\mathrm{ref}}}{P_{\mathrm{ref}}}\left( \frac{\theta}{\theta_{\mathrm{ref}}} \right)^\omega,
\end{equation}
where the subscript $\mathrm{ref}$ denotes the reference quantities.

In its standard form, the BGK model assumes a single relaxation time for both momentum transfer and heat transfer, such that it recovers a unit Prandtl number. While relatively simple modifications can be applied to the model to recover proper thermal behavior (e.g., Shakhov model \citep{Shakhov1972}, ellipsoidal BGK \citep{Holway1966}), the application and validation of these heat transfer models are left as a topic of future research as this work focuses on the low Mach regime where temperature variations due to fluid compressibility are minimal. 

\subsection{Spatial discretization}\label{ssec:spatial}
\begin{figure}[tbhp]
    \begin{centering}
    \adjustbox{width=0.6\linewidth, valign=b}{\input{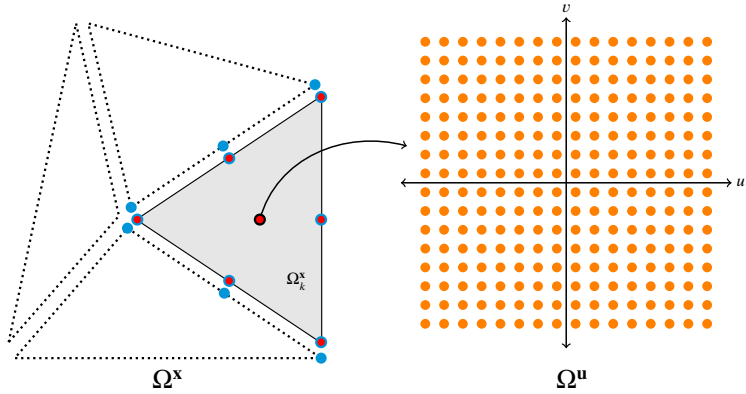}}
    \caption{\label{fig:scheme} Schematic of a two-dimensional phase space discretization using an unstructured spatial domain $\Omega^{\mathbf{x}}$ with $\mathbb P_2$ elements and a velocity domain $\Omega^{\mathbf{u}}$ with $N_v = 16^2$. Circles denote the spatial solution nodes (red), interface flux nodes (blue), and velocity space nodes (orange), respectively.}
    \end{centering}
\end{figure}

The phase-space in the Boltzmann equation possesses a useful property in that for a fixed location $\mathbf{u}_0$ in velocity space, the governing equations reduce to an advection equation with a linear flux $\mathbf{F}(f)$ and nonlinear source term $S$, given as 
\begin{equation}\label{eq:transport}
    \partial_t f (\mathbf{x}, t) +  \boldsymbol{\nabla} {\cdot} \mathbf{F}(f) = S \quad \mathrm{and} \quad \mathbf{F}(f) = \mathbf{u}_0 f.
\end{equation}
For each location in velocity space, the left-hand side of \cref{eq:transport} is discretized using the flux reconstruction scheme of \citet{Huynh2007}, a high-order discontinuous spectral element method that can be considered to be a generalization of the nodal discontinuous Galerkin method \citep{Hesthaven2008DG}. A brief overview of this scheme as applied to the Boltzmann equation is presented here, but a more detailed description can be found in \citet{Dzanic2023}, Section 3.1. 

In this approach, the spatial domain $\Omega^{\mathbf{x}}$ is partitioned into $N_e$ elements $\Omega^{\mathbf{x}}_k $ such that $\Omega^{\mathbf{x}} = \bigcup_{N_e}\Omega^{\mathbf{x}}_k$ and $\Omega^{\mathbf{x}}_i\cap\Omega^{\mathbf{x}}_j=\emptyset$ for $i\neq j$. An example of this tessellation is shown on the left-hand side of \cref{fig:scheme}. For each element $\Omega^{\mathbf{x}}_k$, a nodal interpolating polynomial approximation of the solution is given in the form of
\begin{equation}
    f (\mathbf{x}) = \sum_{i = 1}^{N_s} f (\mathbf{x}^s_i) {\phi}_i (\mathbf{x}),
\end{equation}
where $\mathbf{x}^s_i \in \Omega^{\mathbf{x}}_k \ \forall \ i \in \{1,..., N_s\}$ is a set of $N_s$ solution nodes and ${\phi}_i (\mathbf{x})$ is a set of nodal basis functions with the property ${\phi}_i (\mathbf{x}^s_j) = \delta_{ij}$. We utilize the notation that $\mathbb P_p$ denotes the order of the approximation, taken as the maximal order of $f (\mathbf{x})$. 

Per the flux reconstruction methodology, the flux is calculated through a collocation projection onto the solution nodes augmented with a correction term to allow for communication between elements. 
\begin{equation}
    \mathbf{F}(\mathbf{x}) = \mathbf{u}_0 f (\mathbf{x}) + \sum_{i = 1}^{N_f} \left[F^I_i - \mathbf{u}_0\cdot \mathbf{n}_i f (\mathbf{x}^f_i) \right] \mathbf{g}_i (\mathbf{x}).
\end{equation}
Here, $\mathbf{x}^f_i \in \partial \Omega^{\mathbf{x}}_k \ \forall \ i \in \{1,..., N_f\}$ is a set of $N_f$ interface flux nodes and $\mathbf{n}_i$ is their associated outward-facing normal vector. Furthermore, $F^I_i$ is a common interface flux, simply taken as the upwind-biased value at the interfaces, i.e., 
\begin{equation}
    F_i^I = \begin{cases}
    u_n f_i^-, \quad \quad \mathrm{if} \ u_n > 0,\\
    u_n f_i^+, \quad \quad \mathrm{else},
    \end{cases}
\end{equation}
where $u_n = \mathbf{u}_0 \cdot \mathbf{n}_i$ and the superscripts $-$ and $+$ denote the interior value (from the element of interest) and the exterior value (from the face-adjacent element) of the solution at the interface, respectively. Finally, $\mathbf{g}_i$ are the set of correction functions associated with the given flux nodes $\mathbf{x}_i^f$, which have the properties that
\begin{equation}
    \mathbf{n}_i \cdot\mathbf{g}_j (\mathbf{x}^f_i) = \delta_{ij} \quad \mathrm{and} \quad \sum_{i = 1}^{N_f} \mathbf{g}_i (\mathbf{x}) \in RT_p,
\end{equation}
where $RT_p$ is the Raviart--Thomas space of order $p$. These correction functions are typically chosen such as to recover the nodal discontinuous Galerkin approach \citep{Huynh2007, Hesthaven2008DG, Trojak2021}.

Due to the use of high-order interpolating polynomials for the distribution function, an initially strictly-positive solution may not remain strictly-positive as these high-order schemes typically do not preserve a maximum principle. Since the distribution function is a probability measure, the presence of negative values would result in unphysical predictions. Furthermore, this may also result in numerical instabilities as the equilibrium distribution function is only well-defined for a strictly-positive macroscopic density and temperature, which is ensured if the distribution function is strictly-positive. As such, the spatial discretization is augmented with the high-order, positivity-preserving limiter of \citet{Zhang2010}. For each spatial element and each location in velocity space, a contraction to the element-wise mean $\bar{f}$ is performed if the distribution function attains a negative value at any spatial node, i.e.,
\begin{equation}
    \hat{f}(\mathbf{x}_i) = \bar{f} + \beta\left[ f(\mathbf{x}_i) - \bar{f} \right],
\end{equation}
where
\begin{equation}
    \beta = \min \left [ \left | \frac{\bar{f}}{\bar{f} - f^{\min}}\right |, 1\right] \quad \mathrm{and} \quad  f^{\min} = \min f(\mathbf{x}_i)\ \forall \ i \in \{1, ..., N_s\}.
\end{equation}
This limiting approach retains the high-order accuracy of the underlying discontinuous spectral element method while ensuring positivity of the distribution function and the resulting macroscopic density and temperature. 

\subsection{Velocity discretization}
The velocity space is represented by truncating the infinite domain onto a finite domain $\Omega^{\mathbf{u}} \subset \mathbb R^m$ and discretizing this space by $N_v$ nodal points, shown on the right-hand side of \cref{fig:scheme}. A Cartesian grid with a uniform distribution in velocity space was chosen --- this differs from the original scheme presented in \citet{Dzanic2023} which utilizes a non-uniform polar/spherical velocity space distribution as it was found to be less generalizable and robust than the uniform Cartesian approach. The velocity space is uniquely defined by its extent and its resolution. For the numerical experiments in this work, we denote the number of nodal points in the velocity domain as $N_v$ and utilize a uniform resolution and extent for each velocity dimension centered at the origin, characterized by some maximum velocity $u_{\max}$, i.e., 
\begin{equation}
    \Omega^{\mathbf{u}} = [-u_{\max}, u_{\max}]^m.
\end{equation}
A standard approach for computing this velocity bound is to set it as some factor of the maximum thermal and macroscopic velocities in the domain \citep{Evans2011}, i.e.,
\begin{equation}
    u_{\max} = \underset{\mathbf{x}}{\max}\ \left ( \| \mathbf{U}(\mathbf{x})\|_2 + k\sqrt{\theta(\mathbf{x})}\right),
\end{equation}
where $k=4$ in this work which, for a Maxwellian distribution, results in $99\%$ of the distribution function being contained within $\Omega^{\mathbf{u}}$. This value is computed from the initial conditions and set as constant throughout the simulation as these bounds are not expected to vary much for low Mach number flows. 

With this formulation, the velocity space and the associated distribution function at a given spatial location are represented by discrete vectors $\mathbf{u}$ and $\mathbf{f}$, respectively, each with $N_v$ entries. To compute the moments of the distribution function, a discrete integration operator $\mathbf{M}$ with strictly-positive entries is introduced, such that
\begin{equation}
    \mathbf{M} \cdot \mathbf{f} \approx \int_{\mathbb R^m} f (\mathbf{u}) \ \mathrm{d}\mathbf{u}.
\end{equation}
Due to the use of a uniform Cartesian distribution in the velocity domain, the integration operation is computed by the trapezoidal rule which offers spectral convergence for smooth, compactly-supported, periodic functions on uniform grids \citep{Trefethen2014}. However, as this integration operator is not exact, the compatibility condition in \cref{eq:compatibility}, which holds under exact integration, is not necessarily satisfied discretely, i.e.
\begin{equation}\label{eq:discrete_compatibility}
    \mathbf{M} \cdot \left[\mathbf{f} \otimes \boldsymbol{\psi} \right] = \mathbf{Q} \neq \mathbf{M} \cdot \left[\mathbf{g}(\mathbf{Q}) \otimes \boldsymbol{\psi} \right].
\end{equation}
As a consequence of this inexactness, a standard discrete nodal implementation of the BGK collision operator introduces conservation errors on the macroscopic state proportional to the integration error. While this error can be mitigated with increased velocity space resolution, it was shown in \citet{Dzanic2023} that the primary source of approximation error in the scheme stemmed from the conservation error, and if the scheme can be formulated such as to ensure discrete conservation by satisfying the \textit{discrete} compatibility condition, accurate approximation of the collision operator could be performed with much fewer degrees of freedom in the velocity domain. 

This discrete compatibility condition is enforced using the discrete velocity model (DVM) approach of \citet{Mieussens2000}. It was shown that there exists a unique strictly-positive discrete distribution function which satisfies the discrete compatibility condition and minimizes the discrete log-entropy. This equilibrium distribution function is a Maxwellian, referred to as a modified Maxwellian, which is formed around a slightly modified macroscopic state $\mathbf{Q}'$. The modified macroscopic state is sought to satisfy the discrete compatibility condition as
\begin{equation}
    \mathbf{M} \cdot \left[\mathbf{f} \otimes \boldsymbol{\psi} \right] = \mathbf{M} \cdot \left[\mathbf{g}(\mathbf{Q}') \otimes \boldsymbol{\psi} \right],
\end{equation}
and it can easily be shown that $\mathbf{Q}' \to \mathbf{Q}$ in the limit of infinite velocity space resolution. There does not exist a closed-form expression for computing $\mathbf{Q}'$, and therefore it must be computed numerically. However, this nonlinear optimization process can be easily and efficiently computed due to its low dimensionality ($d+2$) and the presence of a closed-form expression of the Jacobian. As such, discrete conservation can be ensured to near machine-zero levels with as few as two iterations of Newton's method. For an in-depth overview of this numerical approach, the reader is referred to \citet{Dzanic2023}, Section 3.4. 

\subsection{Boundary conditions}
For the Boltzmann equation, enforcing boundary conditions can be highly nontrivial, particularly so for more complex boundary effects such as wall-fluid interactions. In this work, the boundary conditions are enforced weakly via the standard approach in discontinuous spectral element methods where an exterior boundary state is prescribed and the interface flux is computed from the interior and exterior solution states. Due to the upwinding nature of the interface flux computation, outgoing particles are not affected by the boundary state. Similarly to \cref{ssec:spatial}, we utilize the notation $f^+(\mathbf{u})$ to refer to the exterior (boundary) state and $f^-(\mathbf{u})$ to refer to the interior state from the boundary-adjacent element. 

The most straightforward boundary condition to enforce via the weak formulation is a Neumann (free) boundary condition, where the exterior state is set to mimic the interior state as
\begin{equation}
    f^+(\mathbf{u}) = f^-(\mathbf{u}).
\end{equation}
For this boundary condition, the solution is effectively unaffected by the boundary state. Similarly, one can impose a Dirichlet-type (fixed) boundary condition by prescribing an exterior state for the distribution function. In most applications, this is typically taken to be a Maxwellian distribution centered around a particular macroscopic state $\mathbf{Q}$, i.e.,
\begin{equation}
    f^+(\mathbf{u}) = g(\mathbf{Q}').
\end{equation}
Note that the modified Maxwellian is used for the boundary state as this is critical to ensuring conservation. For a constant boundary state, this modified Maxwellian can either be precomputed or computed on the fly. We remark here that due to the natural enforcement of the characteristic direction of information propagation at the boundary interfaces as a result of the upwinding performed at the particle level, the use of fixed boundary states for farfield-type boundaries was found to be much more robust and suitable than mixed states commonly encountered in Navier--Stokes approximations (e.g., fixed density/momentum and extrapolated pressure at inlets, fixed pressure and extrapolated density/momentum at outlets). 

The enforcement of wall boundary conditions for the Boltzmann equation can be a highly nontrivial task, and in many applications, it is an open problem as to how to suitably model particle-wall interactions \citep{Williams2001}. Two models originally suggested by \citet{Maxwell1867} assume that molecules are either reflected directly back into the domain, referred to as specular reflection, or absorbed by the wall and re-emitted back into the domain in equilibrium with the wall, referred to as diffuse reflection \citep{Evans2011}. One may further consider a convex combination of these two boundary conditions based on some absorption coefficient $\alpha \in [0,1]$, sometimes referred to as Maxwellian reflection, although this formulation is not considered in this work \citep{Evans2011}. In the continuum limit, the specular and diffuse wall boundary conditions mimic adiabatic slip-wall and isothermal no slip-wall boundary conditions for the Navier--Stokes equations, respectively.

Specular wall boundary conditions can simply be represented as a distribution function with a reflection about the wall normal direction $\mathbf{n}$ as 
\begin{equation}
    f^+(\mathbf{u}) = f^-(\mathbf{u} - 2 \mathbf{u}_\perp), \quad \mathrm{where} \quad \mathbf{u}_\perp = (\mathbf{u}\cdot\mathbf{n}) \mathbf{n}.
\end{equation}
For arbitrary normal vectors, this reflection operation may require an interpolation in velocity space. However, if the normal direction aligns with the Cartesian axes, which will be the case for the numerical experiments to be presented, the boundary condition can be implemented as a simple transformation of the velocity space. It can easily be shown that this boundary condition preserves the macroscopic density, wall-tangential momentum, and total energy of the interior state. 

Diffuse wall boundary conditions can be represented as a modified Maxwellian distribution, given as
\begin{equation}
    f^+(\mathbf{u}) = \eta g(\mathbf{Q}_w'),
\end{equation}
where $\mathbf{Q}_w$ is the macroscopic state of the wall and $\eta$ is a scaling parameter to be defined momentarily which is used to enforce the no-penetration condition. The macroscopic wall state corresponds to the primitive state 
\begin{equation}
    \mathbf{q}_w = [\rho^-, \mathbf{U}_w, \rho^- \theta_w]^T,
\end{equation}
where $\rho^-$ is the interior macroscopic density, $\mathbf{U}_w$ is wall velocity vector, and $\theta_w$ is the wall temperature. Since the interface flux is computed using an upwind approach (i.e., outgoing particles cannot affect the interior state), it is necessary to scale the wall distribution function to ensure zero mass flux across the boundary. The scaling parameter is computed as
\begin{equation}
    \eta = \frac{\sum_{i=1}^{N_v} M_i h^-_i}{\sum_{i=1}^{N_v} M_i h^+},
\end{equation}
where $M_i$ is the $i$-th component of the discrete integration operator,
\begin{equation}
    h^-_i = \begin{cases}
        f^-_i, \quad \quad \mathrm{if}\ \mathbf{u}_i\cdot\mathbf{n} > 0, \\
        0, \quad \quad  \ \ \ \mathrm{else},
    \end{cases}
\end{equation}
and
\begin{equation}
    h^+_i = \begin{cases}
        0, \quad \quad \ \ \ \mathrm{if}\ \mathbf{u}_i\cdot\mathbf{n} > 0, \\
        g_i(\mathbf{Q}_w'), \ \,  \mathrm{else}.
    \end{cases}
\end{equation}

%!TEX root = ./main.tex
\section{Implementation}\label{sec:implementation}
The numerical scheme in this work was implemented in PyFR \citep{Witherden2014}, a massively-parallel high-order flux reconstruction solver that can be efficiently deployed on CPU and GPU computing architectures. Computation was performed on parallel GPU computing architectures on up to 80 NVIDIA V100 GPUs. Closed solution node distribution were used, corresponding to Gauss--Lobatto and $\alpha$-optimized points \citep{Hesthaven2008DG} for tensor-product and simplex elements, respectively. An explicit, fourth-order, four-stage Runge--Kutta scheme was used for temporal integration with a fixed time step based on the minima of the collision time and a CFL-based time step bound with $\mathrm{CFL} = 0.5$ (see \citet{Dzanic2023}, Section 4). As the explicit time step limits imposed by highly-resolved unsteady numerical simulations do not drastically differ from the limits imposed by the stiffness of the source term in the continuum limit, the use of an explicit time stepping scheme was deemed preferable due to the significant advantages in computational efficiency for GPU computing architectures.  

Initial conditions were set as the modified Maxwellian based on the initial macroscopic state. Two iterations of the DVM were performed for both the initial conditions and every subsequent evaluation of the equilibrium distribution function as this was found to be sufficient to ensure macroscopic conservation \cite{Dzanic2023}. Unless otherwise stated, the viscosity law exponent was set to $\omega = 0.81$. In certain numerical experiments, comparisons were performed between the Boltzmann--BGK approximation and a standard Navier--Stokes approximation. For these comparisons, the Navier--Stokes results were computed via the same codebase using identical values of the viscosity, specific heat ratio, and Prandtl number and the corresponding wall boundary conditions (e.g., no-slip isothermal wall boundary conditions for diffuse wall boundary conditions). Furthermore, the inviscid interface fluxes were computed using a Godunov-type exact Riemann solver \citep{Godunov1959} and the viscous interface fluxes were computed using the BR2 approach of \citet{Bassi2000}. In many scenarios, the nonlinearity of the Navier--Stokes equations caused numerical instabilities for under-resolved flows. These instabilities were mitigated through anti-aliasing via overintegration with a sufficiently strong quadrature rule. 
%!TEX root = ./main.tex
\section{Results}\label{sec:results} 
The results of numerical experiments 
spanning a wide range of parameters of interests, shown by the blue box in \cref{fig:regimes}, are presented in this section. First, numerical solutions obtained for flows in the rarefied regime, where the Boltzmann equation is more commonly applied, are validated against established data in the literature. Novel results for a three-dimensional rarefied flow are also presented to highlight the capabilities of our model and to document a well-posed problem which could be used as a validation test for future studies. Afterwards, numerical solutions for problems in the continuum regime are presented together with an analysis of the results obtained with the molecular gas dynamics model for wall-bounded transitional and turbulent flows. 

\subsection{Couette flow}
An initial validation of the numerical approach was performed for planar Couette flow across a range of Knudsen numbers. For this problem, the fluid rests between two walls moving tangentially relative to each other, inducing fluid flow by imparting a wall shear stress. At higher Knudsen numbers, strong non-equilibrium effects can result in significant slip velocity at the wall, introducing complexity into the flow and affecting the wall shear stress. The problem was defined on the domain $y \in [-H, H]$, where $H$ is the channel half-height. The top and bottom walls, located at $y = \pm H$, were given a fixed velocity $\pm U_w$, respectively, with a fixed wall temperature $\theta_w$. Wall boundary conditions were enforced as diffuse wall boundaries.
The problem was set up with a two-dimensional spatial domain (and corresponding velocity domain) using a one element-wide mesh with periodicity along the $x$ direction, such that the problem was effectively one-dimensional. The initial flow field was set as constant with zero velocity and unit density, and the pressure was set to yield a Mach number of $M = 0.2$ based on the wall velocity. The wall temperature was set in equilibrium with initial flow field. The Reynolds number and, by extent, the Knudsen number, was then varied by modifying the collision time. The problem was integrated to a steady-state using a $\mathbb P_3$ approximation with $N_e = 16$ elements and $N_v = 16^2$ velocity nodes. 

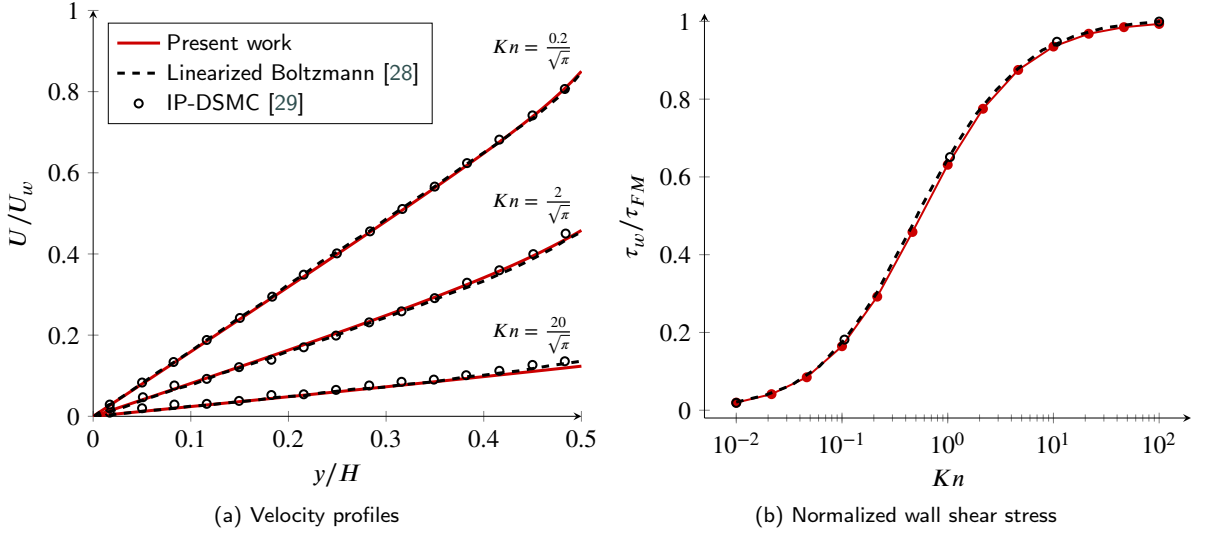
\begin{figure}[tbhp]
        \centering
        \subfloat[Velocity profiles]{\adjustbox{width=0.49\linewidth, valign=b}{\label{fig:couette_vel}\begin{tikzpicture}[spy using outlines={rectangle, height=3cm,width=2.3cm, magnification=3, connect spies}]
	\begin{axis}[name=plot1,
		axis line style={latex-latex},
	    axis x line=left,
        axis y line=left,
		xlabel={$y/H$},
    	xmin=0, xmax=0.5,
    	xtick={0, 0.1, 0.2, 0.3, 0.4, 0.5},
    	ylabel={$U/U_w$},
    	ymin=0,ymax=1.0,
    	ytick={0, 0.2, 0.4, 0.6, 0.8, 1.0},
        clip mode=individual,
    	legend style={at={(0.03, 0.97)},anchor=north west,font=\small},
    	legend cell align={left},
    	style={font=\normalsize}]
    	
        \addplot[color=red!80!black, style={very thick}]
        table[x=y, y=u, col sep=comma]{./figs/data/couette_bgk_velocity_high.csv};
        \addlegendentry{Present work }; 
        
        \addplot[color=black, style={very thick, dashed}]
        table[x=y, y=u, col sep=comma]{./figs/data/couette_sone_velocity_high.csv};
        \addlegendentry{Linearized Boltzmann \citep{Sone1990}}; 
        
        \addplot[color=black, style={thick}, only marks, mark=o, mark options={scale=0.8}]
        table[x=y, y=u, col sep=comma]{./figs/data/couette_fan_velocity_high.csv};
        \addlegendentry{IP-DSMC \citep{Fan2001}};
        \node at (0.45, 0.9) {\footnotesize$Kn = \frac{0.2}{\sqrt{\pi}}$};

        \addplot[color=red!80!black, style={very thick}]
        table[x=y, y=u, col sep=comma]{./figs/data/couette_bgk_velocity_medium.csv};
        
        \addplot[color=black, style={very thick, dashed}]
        table[x=y, y=u, col sep=comma]{./figs/data/couette_sone_velocity_medium.csv}; 

        \addplot[color=black, style={thick}, only marks, mark=o, mark options={scale=0.8}]
        table[x=y, y=u, col sep=comma]{./figs/data/couette_fan_velocity_medium.csv};
        \node at (0.45, 0.52) {\footnotesize$Kn = \frac{2}{\sqrt{\pi}}$};

        \addplot[color=red!80!black, style={very thick}]
        table[x=y, y=u, col sep=comma]{./figs/data/couette_bgk_velocity_low.csv};
        
        \addplot[color=black, style={very thick, dashed}]
        table[x=y, y=u, col sep=comma]{./figs/data/couette_sone_velocity_low.csv};
        
        \addplot[color=black, style={thick}, only marks, mark=o, mark options={scale=0.8}]
        table[x=y, y=u, col sep=comma]{./figs/data/couette_fan_velocity_low.csv};
        \node at (0.45, 0.2) {\footnotesize$Kn = \frac{20}{\sqrt{\pi}}$};

	\end{axis}
\end{tikzpicture}}}
        \subfloat[Normalized wall shear stress]{\label{fig:couette_stress}\adjustbox{width=0.47\linewidth, valign=b}{\begin{tikzpicture}[spy using outlines={rectangle, height=3cm,width=2.3cm, magnification=3, connect spies}]
	\begin{semilogxaxis}[name=plot1,
		axis line style={latex-latex},
	    axis x line=left,
        axis y line=left,
		xlabel={$Kn$},
    	xmin=5e-3, xmax=2e2,
    	xtick={1e-2, 1e-1, 1e0, 1e1, 1e2},
    	ylabel={$\tau_w/\tau_{FM}$},
    	ymin=-0.02,ymax=1.02,
    	ytick={0, 0.2, 0.4, 0.6, 0.8, 1.0},
        clip mode=individual,
    	legend style={at={(0.97, 0.03)},anchor=south east,font=\small},
    	legend cell align={left},
    	style={font=\normalsize}]
    	
        \addplot[color=red!80!black, style={thick}, mark=*, mark options={scale=0.8}]
        table[x=kn, y=tw, col sep=comma]{./figs/data/couette_bgk_tw.csv};
        % \addlegendentry{Present work}; 
        
        \addplot[color=black, style={very thick, dashed}]
        table[x=kn, y=tw, col sep=comma]{./figs/data/couette_sone_tw.csv};
        % \addlegendentry{\citet{Sone1990}}; 
        
        \addplot[color=black, style={thick}, only marks, mark=o, mark options={scale=0.8}]
        table[x=kn, y=tw, col sep=comma]{./figs/data/couette_fan_tw.csv};
        % \addlegendentry{\citet{Fan2001}};
	\end{semilogxaxis}
\end{tikzpicture}}}
        \caption{\label{fig:couette} Normalized velocity profiles (left) and wall shear stress (right) for the Couette flow problem computed using a $\mathbb P_3$ approximation with $N_v = 16^2$ at varying Knudsen numbers. }
    \end{figure}

The resulting normalized velocity profiles at Knudsen numbers of $0.2/\sqrt{\pi}$, $2/\sqrt{\pi}$, and $20/\sqrt{\pi}$ are shown in \cref{fig:couette_vel} in comparison to the linearized Boltzmann results of \citet{Sone1990} and information-preserving direct simulation Monte Carlo (IP-DSMC) results of \citet{Fan2001}. Very good agreement was observed between the present work and the reference results for all values of the Knudsen number. Furthermore, the effect of the Knudsen number of the wall shear stress was analyzed, where the shear stress was computed directly from the distribution function as
\begin{equation}\label{eq:wallshear}
    \tau_{ij}(\mathbf{x}, t) = \int_{\mathbb R^m} \mathbf{u}_i \mathbf{u}_j f(\mathbf{x}, \mathbf{u}, t)\ \mathrm{d} \mathbf{u}.
\end{equation}
The behavior of the wall shear stress, normalized by the free molecular wall shear stress $\tau_{FM} = \rho U_w^2/(M\sqrt{\pi})$, is shown for varying Knudsen numbers in \cref{fig:couette_stress}. It can be seen that the results show excellent agreement with the reference data across the entire range of Knudsen number, from the slip flow regime to the free molecular regime. 

\tikzexternaldisable
\tikzexternalenable

\subsection{Rarefied flat plate}
The extension to two-dimensional non-equilibrium flows was performed through the simulation of a flat plate at zero incidence, a fundamental flow problem in fluid dynamics. At low Reynolds numbers, the strong interaction between the boundary layer and the freestream result in non-negligible rarefaction in the flow, such that the Navier--Stokes equations start to give erroneous results. The problem consists of a flat plate of finite length $L$ and zero thickness held stationary within a fixed freestream flow. For a validation against established results such as the works of \citet{Sun2004} and \citet{Zhu2017}, we consider the cases of a freestream Mach number of $M = 0.2$ and Reynolds numbers of 10 and 50, yielding flow conditions which sit within the slip flow and continuum regimes, respectively. 

The domain was set identically to the work of \citet{Zhu2017} as  $\Omega^{\mathbf{x}} = [-20.5L, 20.5L] \times [0, 20L]$, with the plate defined over the boundary $-L/2 \leq x \leq L/2, y = 0$. Due to the symmetry of the problem at zero incidence, only a half-domain was considered, with symmetry boundary conditions (i.e., specular wall boundary conditions) applied for the remaining portion of the $x$-axis. Dirichlet boundary conditions corresponding to the freestream flow conditions were applied at the farfield boundaries, and a diffuse wall boundary condition was applied for the plate with the temperature set identical to the freestream. The viscosity temperature exponent was set as $\omega = 0.77$. 

A quadrilateral mesh was generated with a streamwise and normal mesh spacing of $\Delta x = \Delta y = L/32$ at the wall. The mesh spacing was proportionally increased away from the wall to yield 32 elements in each direction, such that the final mesh consisted of $N_e = 96 \times 32$ elements. A $\mathbb P_3$ approximation with $N_v = 16^2$ velocity nodes was used, resulting in a total of approximately 12.5 million degrees of freedom. The solution was advanced to a steady state and the resulting wall skin friction coefficient as derived from \cref{eq:wallshear} was computed.

\tikzexternaldisable
   \begin{figure}[tbhp]
        \centering
        \adjustbox{width=0.6\linewidth, valign=b}{\begin{tikzpicture}[spy using outlines={rectangle, height=3cm,width=2.3cm, magnification=3, connect spies}]
	\begin{axis}[name=plot1,
		axis line style={latex-latex},
	    axis x line=left,
        axis y line=left,
		xlabel={$x/L$},
    	xmin=-0.5, xmax=0.5,
    	xtick={-0.5, -0.25,  0, 0.25,  0.5},
    	ylabel={$C_f$},
    	ymin=0,ymax=1.6,
        clip mode=individual,
    	legend style={at={(0.97, 0.97)},anchor=north east,font=\small},
    	legend cell align={left},
    	style={font=\normalsize},
        reverse legend,
        width=1.5*\axisdefaultheight,
        height=\axisdefaultheight]

        \addplot[color=black, style={very thin}, only marks, mark=o, mark options={scale=0.9}]
        table[x expr = \thisrow{x} - 0.5, y=cf, col sep=comma]{./figs/data/flatplate_zhu_cf_Re50.csv};
        
        \addplot[color=black, style={very thin}, only marks, mark=triangle*, mark options={scale=0.9}]
        table[x=x, y=cf, col sep=comma]{./figs/data/flatplate_sun_ip_cf_Re50.csv};   
        
        \addplot[color=black, style={}]
        table[x=x, y=cf, col sep=comma]{./figs/data/flatplate_sun_dsmc_cf_Re50.csv};    
        
        \addplot[color=black, style={thick, dashed}]
        table[x=x, y=cf, col sep=comma]{./figs/data/flatplate_sun_ns_cf_Re50.csv};         
    	
    	\addplot[color=red!80!black, style={very thick}]
        table[x=x, y=cf, col sep=comma]{./figs/data/flatplate_bgk_cf_Re50.csv};   

        \addplot[color=black, style={very thin}, only marks, mark=o, mark options={scale=0.9}]
        table[x expr = \thisrow{x} - 0.5, y=cf, col sep=comma]{./figs/data/flatplate_zhu_cf_Re10.csv};
        \addlegendentry{UGKS \citep{Zhu2017}}; 
        
        \addplot[color=black, style={very thin}, only marks, mark=triangle*, mark options={scale=0.9}]
        table[x=x, y=cf, col sep=comma]{./figs/data/flatplate_sun_ip_cf_Re10.csv};        
        \addlegendentry{IP-DSMC \citep{Sun2004}}; 
        
        \addplot[color=black, style={}]
        table[x=x, y=cf, col sep=comma]{./figs/data/flatplate_sun_dsmc_cf_Re10.csv};        
        \addlegendentry{DSMC \citep{Sun2004}}; 
        
        \addplot[color=black, style={thick, dashed}]
        table[x=x, y=cf, col sep=comma]{./figs/data/flatplate_sun_ns_cf_Re10.csv};        
        \addlegendentry{Navier--Stokes \citep{Sun2004}}; 

    	\addplot[color=red!80!black, style={very thick}]
        table[x=x, y=cf, col sep=comma]{./figs/data/flatplate_bgk_cf_Re10.csv};
        \addlegendentry{Present work}; 
        
        \node at (0.35, 0.7) {\footnotesize$Re = 10$};
        \node at (0.35, 0.3) {\footnotesize$Re = 50$};

	\end{axis}
\end{tikzpicture}}
        \caption{\label{fig:flatplate_cf} Skin friction coefficient for the rarefied flat plate problem at varying Reynolds numbers computed using a $\mathbb P_3$ approximation with $N_v = 16^2$ in comparison to the Navier--Stokes, DSMC, and IP-DSMC results of \citet{Sun2004} and UGKS results of \citet{Zhu2017}.}
    \end{figure}
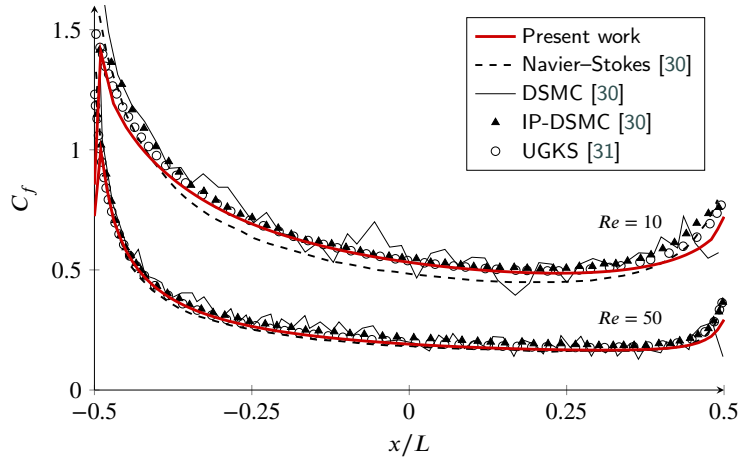
\tikzexternalenable

The predicted skin friction coefficient for the cases of $Re = 10$ and $Re = 50$ is shown in \cref{fig:flatplate_cf} in comparison to the Navier--Stokes, DSMC, and IP-DSMC results of \citet{Sun2004} and the unified gas kinetic scheme (UGKS) results of \citet{Zhu2017}. It can be seen that in the continuum case of $Re = 50$, the varying approaches show similar predictions, generally within the statistical scatter of the DSMC results. For the higher Knudsen number case at $Re = 10$, the effects of rarefaction become more obvious as the present results show good agreement with the varying kinetic approaches, which all notably differ from the Navier--Stokes results. At the trailing edge where non-equilibrium effects can be quite strong, the present results for both Reynolds numbers show less wall shear stress than the IP-DSMC and UGKS results while showing better agreement with the DSMC results, and it is expected that IP-DSMC approaches cannot capture all the necessary physics in these strong non-equilibrium regions \citep{Sun2004}. These differences highlight the potential benefits in accuracy obtained by directly solving the Boltzmann equation. Furthermore, it was observed for the given mesh, which contains high aspect ratio cells that can be problematic for high-order discontinuous spectral element methods, the simulation of the Boltzmann equation was stable and yielded consistent results whereas the simulation of the Navier--Stokes equations diverged at approximation orders above $\mathbb P_1$ irrespective of time step and degree of anti-aliasing. This suggests that the approximations of the Boltzmann equation may be less prone to numerical stability issues than the Navier--Stokes equations, likely due to the linear nature of the transport term and the lack of second-order viscous terms. 

\subsection{Bent microchannel}
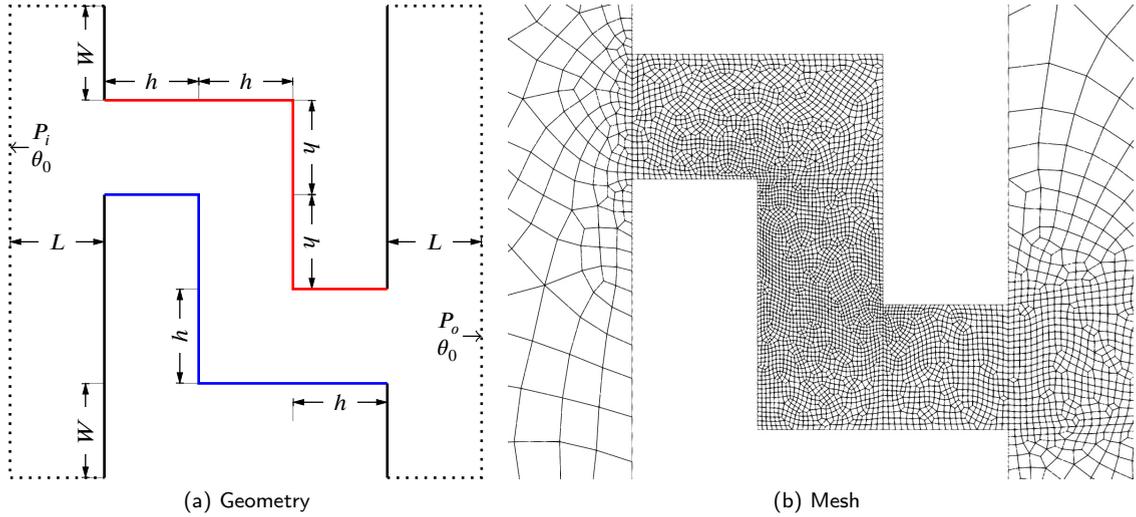
\begin{figure}[tbhp]
        \centering
        \subfloat[Geometry]{
        \adjustbox{width=0.4\linewidth, valign=b}{\begin{tikzpicture}[scale=1]

    \dimline[extension start length=0.2 cm, extension end length=0.2 cm] {(0, 0.2 )}{(1, 0.2)}{\scriptsize $h$};
    \dimline[extension start length=0.2 cm, extension end length=0.2 cm] {(1, 0.2 )}{(2, 0.2)}{\scriptsize $h$};
    \dimline[extension start length=0.2 cm, extension end length=0.2 cm] {(2.2, 0 )}{(2.2, -1)}{\scriptsize $h$};
    \dimline[extension start length=0.2 cm, extension end length=0.2 cm] {(2.2, -1)}{(2.2, -2)}{\scriptsize $h$};
    \dimline[extension start length=0.2 cm, extension end length=0.2 cm] {(0.8, -3)}{(0.8, -2)}{\scriptsize $h$};
    \dimline[extension start length=0.2 cm, extension end length=0.2 cm] {(2, -3.2 )}{(3, -3.2)}{\scriptsize $h$};
    \dimline[extension start length=0 cm, extension end length=0 cm] {(4, -1.5)}{(3, -1.5)}{\scriptsize $L$};
    \dimline[extension start length=0 cm, extension end length=0 cm] {(0, -1.5)}{(-1, -1.5)}{\scriptsize $L$};
    \dimline[extension start length=0.2 cm, extension end length=0.2 cm] {(-0.2, 0)}{(-0.2, 1)}{\scriptsize $W$};
    \dimline[extension start length=0.2 cm, extension end length=0.2 cm] {(-0.2, -4)}{(-0.2, -3)}{\scriptsize $W$};
    
    \draw[black, thick] (0, 1) -- (0, 0);
    \draw[black, thick] (3, -2) -- (3, 1);
    \draw[black, thick] (0, -4) -- (0, -1) ;
    \draw[black, thick] (3, -3) -- (3, -4);
    \draw[red, thick] (0, 0) -- (2, 0) -- (2, -2) -- (3, -2);
    \draw[blue, thick] (0, -1) -- (1, -1) -- (1, -3) -- (3, -3);
    
    \draw[black, thick, dotted] (0, 1) -- (-1, 1) -- (-1, -4) -- (0, -4);
    \draw[black, thick, dotted] (3, -4) -- (4, -4) -- (4, 1) -- (3, 1);

    \draw [black, <-] (-1, -0.5) -- (-0.8, -0.5) node[right,xshift=-0.3cm] {\scriptsize \begin{tabular}{l} $P_i$  \\ $\theta_0$ \end{tabular}};
    \draw [black, ->] (3.8, -2.5) -- (4, -2.5) node[left,xshift=0.1cm] {\scriptsize \begin{tabular}{l} $P_o$  \\ $\theta_0$ \end{tabular}};

\end{tikzpicture}}}
        \subfloat[Mesh]{
        \adjustbox{width=0.5\linewidth, valign=b}{\includegraphics[width=\textwidth]{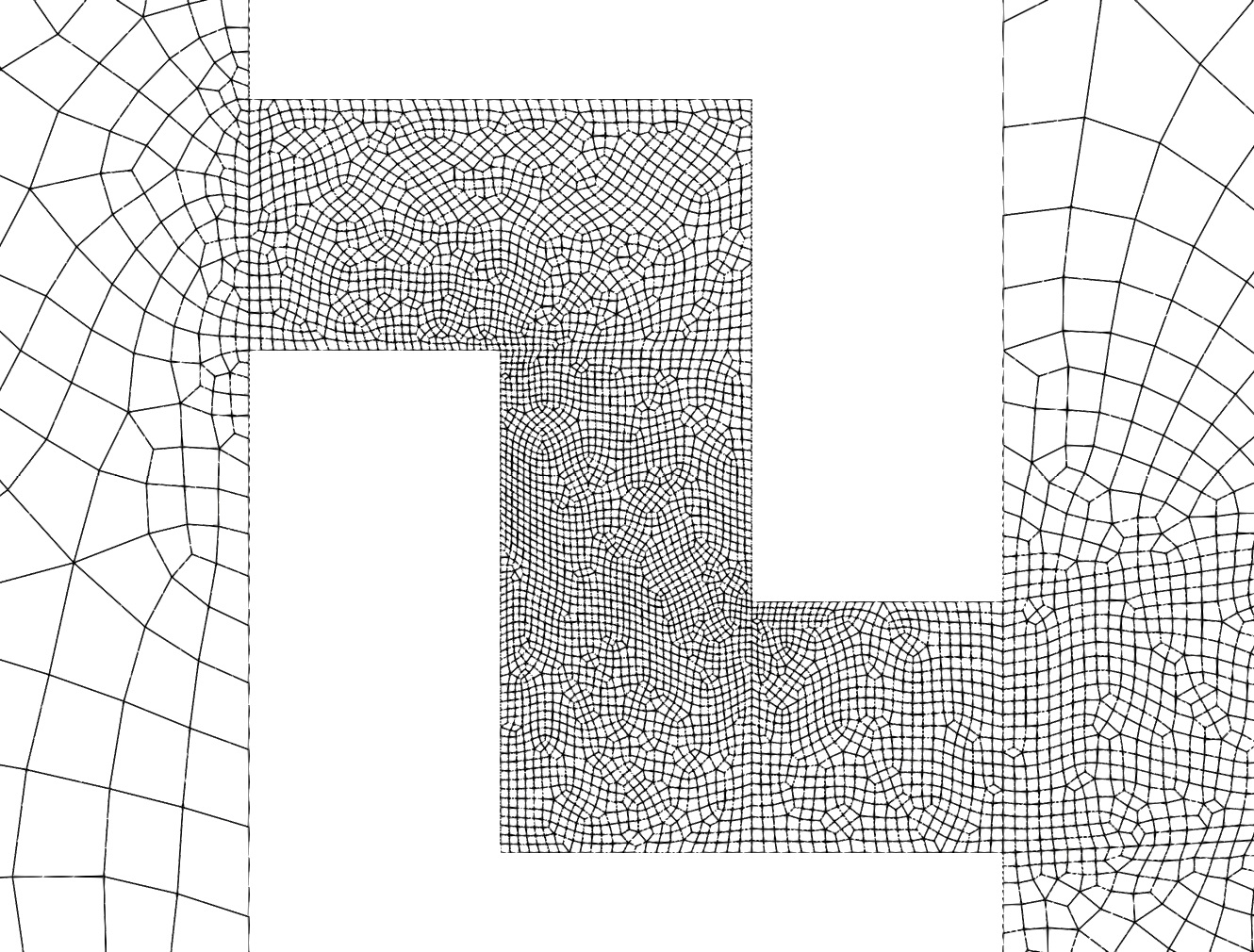}}}
        \caption{\label{fig:microchannel_geo} Schematic of the bent microchannel geometry (left) and simulation mesh (right). Reservoirs not drawn to scale. Wall boundaries represented by solid lines, inlet/outlet boundaries represented by dotted lines. Top and bottom walls represented by red and blue lines, respectively. }
    \end{figure}

For a more complex validation case with practical applications, we consider the rarefied flow in a microchannel with bends which has been investigated in works such as that of \citet{Ho2020}, \citet{Agrawal2009}, \citet{Liu2018}, and \citet{Varade2015}. In these microsystems, the interaction of flow separation and reattachment around corners with rarefaction effects such as wall slip presents challenging flow physics that are difficult to accurately predict. For this problem, we attempt to replicate the numerical experiments of \citet{Ho2020} in which the flow through a doubly-bent channel is simulated for flow conditions in the slip flow regime. The geometry consists of a rectangular channel of width $h$ and centerline length $5h$ feeding from an inlet reservoir to an outlet reservoir, shown in \cref{fig:microchannel_geo}. The reservoir dimensions are taken as $L = 10$ and $W = 4$.

The flow was driven by a pressure gradient across the inlet and outlet reservoirs, represented by a pressure ratio $\beta$. At the inlet, fixed pressure $P_i$ and temperature $\theta_0$ were imposed, with the corresponding outlet pressure and temperature set as $P_o = \beta P_i$ and $\theta_0$, respectively. For consistency with the work of \citet{Ho2020}, the pressure ratio was set as $\beta = 0.5$. At the walls, diffuse wall boundary conditions were used with the wall temperature set as $\theta_w = \theta_0$. For the given setup, the Mach and Reynolds numbers were computed with respect to the bulk velocity and inlet density. Three operating flow conditions from \citet{Ho2020} were considered: 1) $Kn \sim 0.01$, with $Re = 40.65$ and $M = 0.251$; 2) $Kn \sim 0.02$, with $Re = 17.80$ and $M = 0.22$; and 3) $Kn \sim 0.05$, with $Re = 4.72$ and $M = 0.146$. We remark here that since the specific heat ratio used in \citet{Ho2020} ($\gamma = 5/3$) differs from the present work ($\gamma = 2$), we attempt to replicate the flow Mach and Reynolds numbers and let the Knudsen number vary slightly, less than $10\%$. As such, minor differences in the predicted flow fields are to be expected. 

\tikzexternaldisable
   \begin{figure}[htbp!]
        \centering
        \subfloat[Slip velocity]{\adjustbox{width=0.48\linewidth, valign=b}{\begin{tikzpicture}[spy using outlines={rectangle, height=3cm,width=2.3cm, magnification=3, connect spies}]
	\begin{axis}[name=plot1,
		axis line style={latex-latex},
	    axis x line=left,
        axis y line=left,
		xlabel={$d/h$},
    	xmin=0, xmax=5,
    	xtick={0, 1, 2, 3, 4, 5},
    	ylabel={$U_s/u_m$},
    	ymin=-0.1,ymax=0.45,
    	ytick={-0.1, 0, 0.1, 0.2, 0.3, 0.4},
        clip mode=individual,
    	legend style={at={(0.03, 1.0)},anchor=north west,font=\small},
    	legend cell align={left},
    	style={font=\normalsize},
        reverse legend]
    	
        \addplot[color=black, style={very thin}, only marks, mark=triangle*, mark options={scale=0.6}]
        table[x=x, y=u, col sep=comma]{./figs/data/microchannel_ho_slip_lower.csv};
        \addlegendentry{\citet{Ho2020} (Lower)}
        
        \addplot[color=black, style={very thin}, only marks, mark=o, mark options={scale=0.6}]
        table[x=x, y=u, col sep=comma]{./figs/data/microchannel_ho_slip_upper.csv};
        \addlegendentry{\citet{Ho2020} (Upper)}
        
        \addplot[color=red!80!black, style={very thick, dashed}]
        table[x=d, y=u_l, col sep=comma ]{./figs/data/microchannel_bgk_Kn0p01.csv};
        \addlegendentry{Present work (Lower)}
        
        \addplot[color=red!80!black, style={thick}]
        table[x=d, y=u_u, col sep=comma]{./figs/data/microchannel_bgk_Kn0p01.csv};
        \addlegendentry{Present work (Upper)}
        
	\end{axis}
\end{tikzpicture}}}
        \subfloat[Pressure]{\adjustbox{width=0.48\linewidth, valign=b}{\begin{tikzpicture}[spy using outlines={rectangle, height=3cm,width=2.3cm, magnification=3, connect spies}]
	\begin{axis}[name=plot1,
		axis line style={latex-latex},
	    axis x line=left,
        axis y line=left,
		xlabel={$d/h$},
    	xmin=0, xmax=5,
    	xtick={0, 1, 2, 3, 4, 5},
    	ylabel={$p/p_i$},
    	ymin=0.2,ymax=1,
    	ytick={0.2, 0.4, 0.6, 0.8, 1.0},
        clip mode=individual,
    	legend style={at={(0.03, 0.97)},anchor=north west,font=\small},
    	legend cell align={left},
    	style={font=\normalsize}]
    	
        \addplot[color=black, style={very thin}, only marks, mark=triangle*, mark options={scale=0.6}]
        table[x=x, y=p, col sep=comma]{./figs/data/microchannel_ho_pressure_lower.csv};
        
        \addplot[color=black, style={very thin}, only marks, mark=o, mark options={scale=0.6}]
        table[x=x, y=p, col sep=comma]{./figs/data/microchannel_ho_pressure_upper.csv};
        
        % \addplot[color=black, style={thick, dotted}]
        % table[x=x, y=p, col sep=comma]{./figs/data/microchannel_ho_pressure_center.csv};
        
        \addplot[color=red!80!black, style={thick}]
        table[x=d, y=p_u, col sep=comma]{./figs/data/microchannel_bgk_Kn0p01.csv};
        
        \addplot[color=red!80!black, style={very thick, dashed}]
        table[x=d, y=p_l, col sep=comma ]{./figs/data/microchannel_bgk_Kn0p01.csv};
        
        % \addplot[color=red!80!black, style={thick, dotted}]
        % table[x=d, y=p_m, col sep=comma]{./figs/data/microchannel_bgk_Kn0p01.csv};

	\end{axis}
\end{tikzpicture}}}
        \caption{\label{fig:microchannel_profiles} Profiles of normalized wall slip velocity (left) and normalized pressure (right) along the upper and lower channel walls for the bent microchannel problem at $Kn \sim 0.01$ ($Re = 40.65$, $M = 0.251$) computed using a $\mathbb P_3$ approximation with $N_v = 32^2$ in comparison to the results of \citet{Ho2020}.}
    \end{figure}
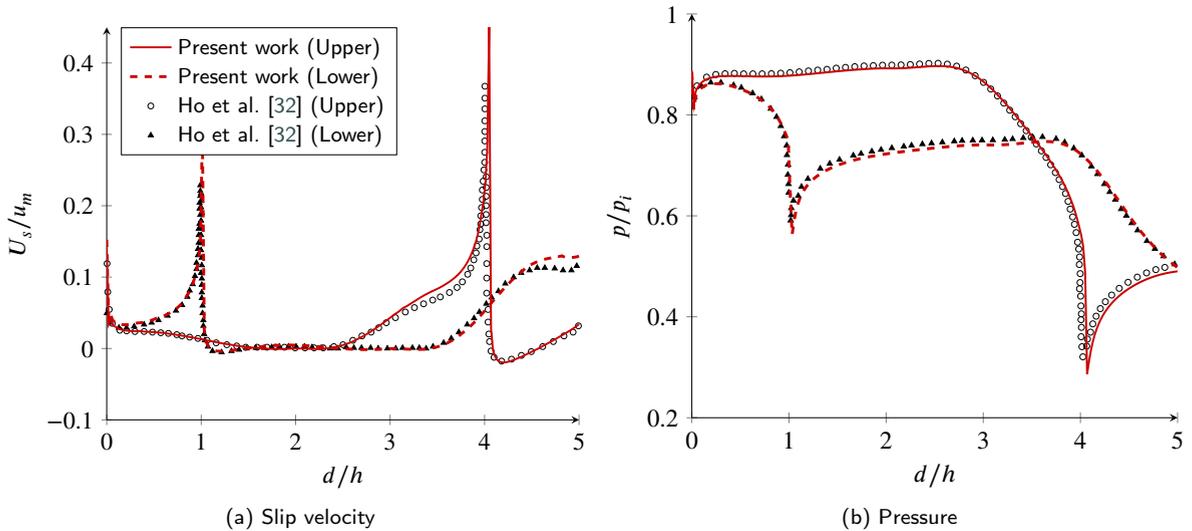
\tikzexternalenable

The problem was solved using a $\mathbb P_3$ approximation with $N_v = 32^2$ on an unstructured quadrilateral mesh with an approximate edge length of $h/40$ within the channel, shown in \cref{fig:microchannel_geo}. The mesh consisted of $N_e = 6,510$ elements, yielding a total of $106.7$ million degrees of freedom. The flow was advanced to a steady state, after which the flow along the top and bottom channel walls was analyzed. The profiles of the wall slip velocity and pressure, normalized by the most probable molecular speed $u_m = \sqrt{2 \theta_0}$ and inlet pressure, respectively, is shown in \cref{fig:microchannel_profiles} for the case of $Kn \sim 0.01$. Very good agreement could be observed between the present work and the results of \citet{Ho2020}, with nearly identical predictions of the wall slip velocity along both the upper and lower walls throughout the channel, even around the convex and concave corners. Some differences could be seen in the wall slip on the lower wall near the channel exit, but these differences were quite minor. Additionally, very similar profiles in the wall pressure distributions were observed, with the peaks around the convex corners well-predicted. 

\begin{figure}[htbp!]
    \centering
    \subfloat[Present work ($Kn \sim 0.01$)] {
    \adjustbox{width=0.3\linewidth,valign=b}{\includegraphics[width=\textwidth]{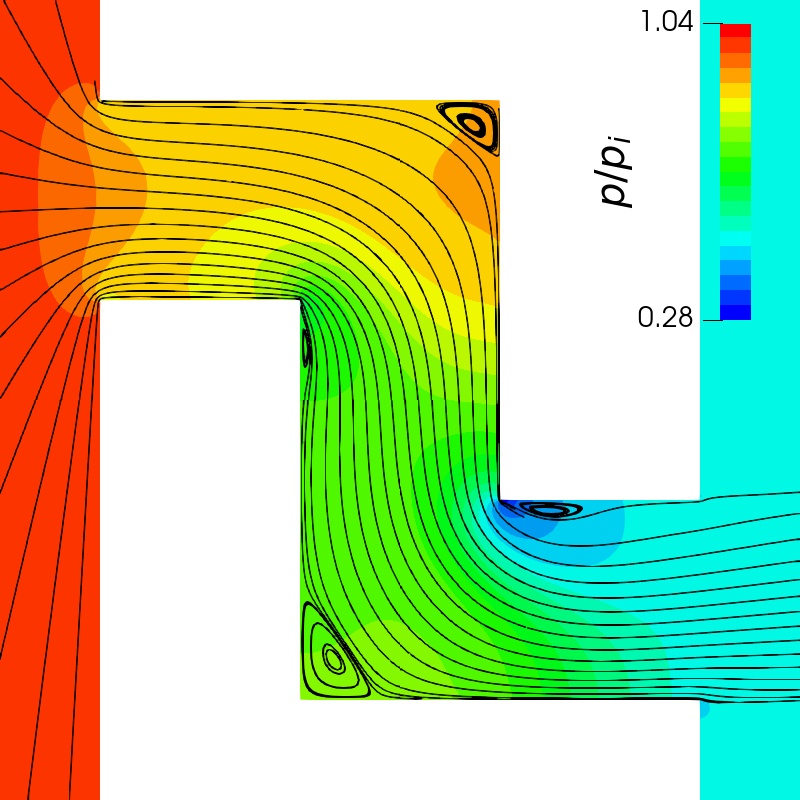}}}
    \hfill
    \subfloat[Present work ($Kn \sim 0.02$)] {
    \adjustbox{width=0.3\linewidth,valign=b}{\includegraphics[width=\textwidth]{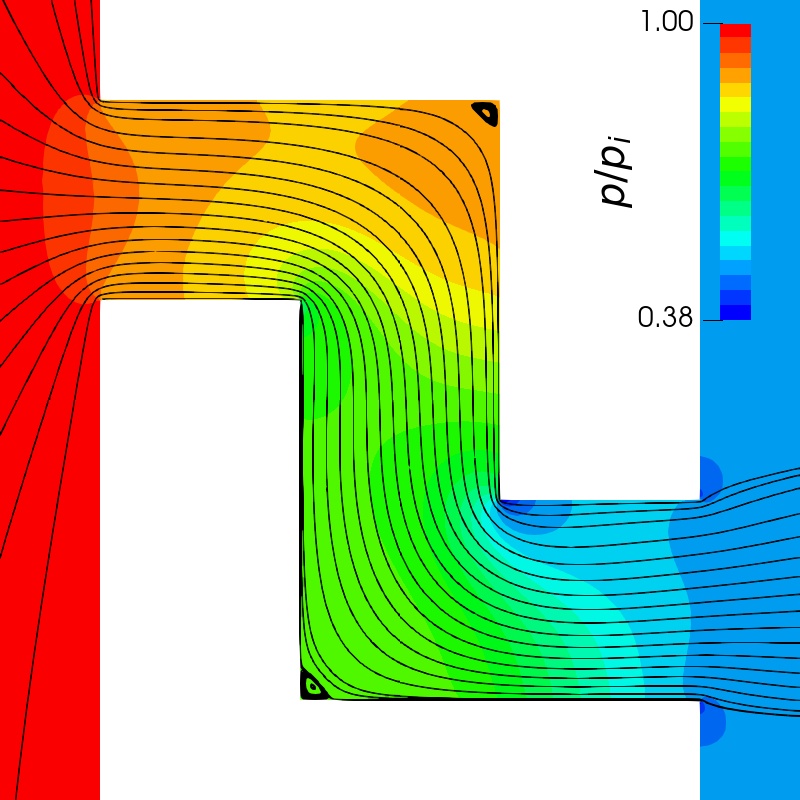}}}
    \hfill
    \subfloat[Present work ($Kn \sim 0.05$)] {
    \adjustbox{width=0.3\linewidth,valign=b}{\includegraphics[width=\textwidth]{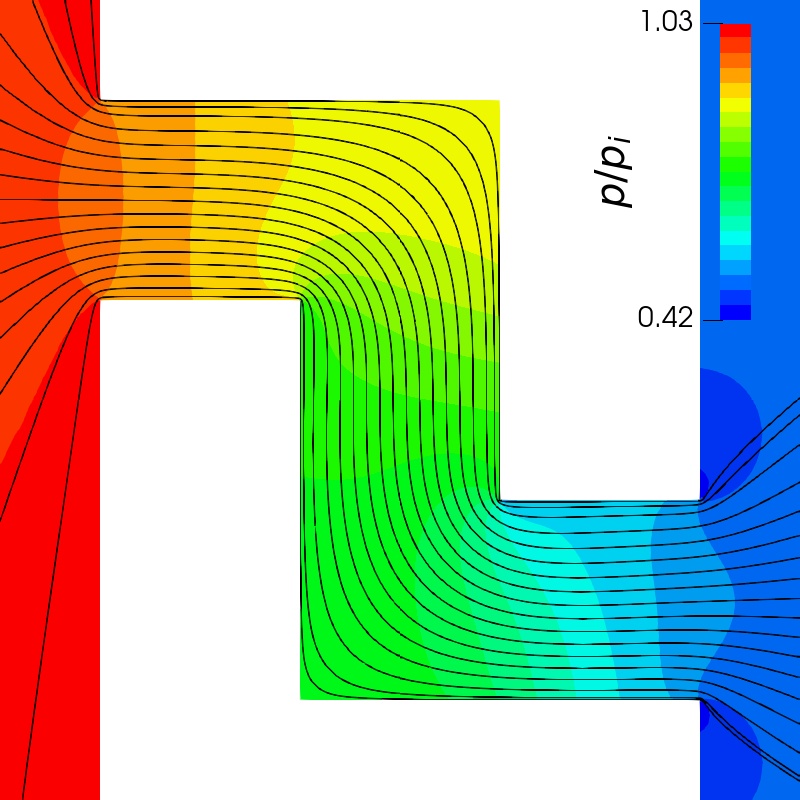}}}
    \newline
    \subfloat[\citet{Ho2020} ($Kn \sim 0.01$)]{
    \adjustbox{width=0.3\linewidth,valign=b}{\includegraphics[width=\textwidth]{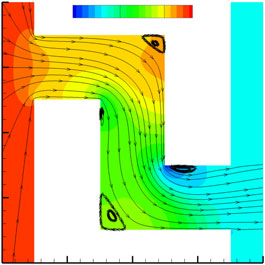}}}
    \hfill
    \subfloat[\citet{Ho2020} ($Kn \sim 0.02$)]{
    \adjustbox{width=0.3\linewidth,valign=b}{\includegraphics[width=\textwidth]{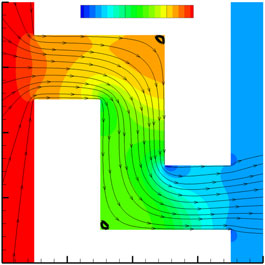}}}
    \hfill
    \subfloat[\citet{Ho2020} ($Kn \sim 0.05$)]{
    \adjustbox{width=0.3\linewidth,valign=b}{\includegraphics[width=\textwidth]{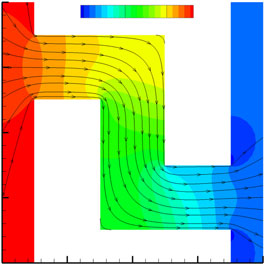}}}
    \caption{\label{fig:microchannel_contours} Contours of normalized pressure overlaid with velocity streamlines for the bent microchannel problem at $Kn \sim 0.01$ (left), $Kn \sim 0.02$ (middle), and $Kn \sim 0.05$ (right) computed using a $\mathbb P_3$ approximation with $N_v = 32^2$ in comparison to the results of \citet{Ho2020}.}
\end{figure}

A further comparison was performed against the results of \citet{Ho2020} through the visualization of the resulting flow fields. The contours of pressure overlaid with velocity streamlines are shown in \cref{fig:microchannel_contours} for the three operating conditions. Excellent agreement between the predicted pressure and velocity fields was observed across the range of Knudsen numbers, with nearly identical pressure contours between the present work and reference results. Furthermore, the velocity streamlines show that the complex flow behavior in the slip flow regime is well predicted, both in terms of the separation bubbles at the convex and concave corners at lower Knudsen number to their subsequent reattachment with increasing Knudsen number.

  \begin{figure}[htbp!]
        \centering
        \subfloat[$N_v = 8^2$] {
        \adjustbox{width=0.3\linewidth,valign=b}{\includegraphics[width=\textwidth]{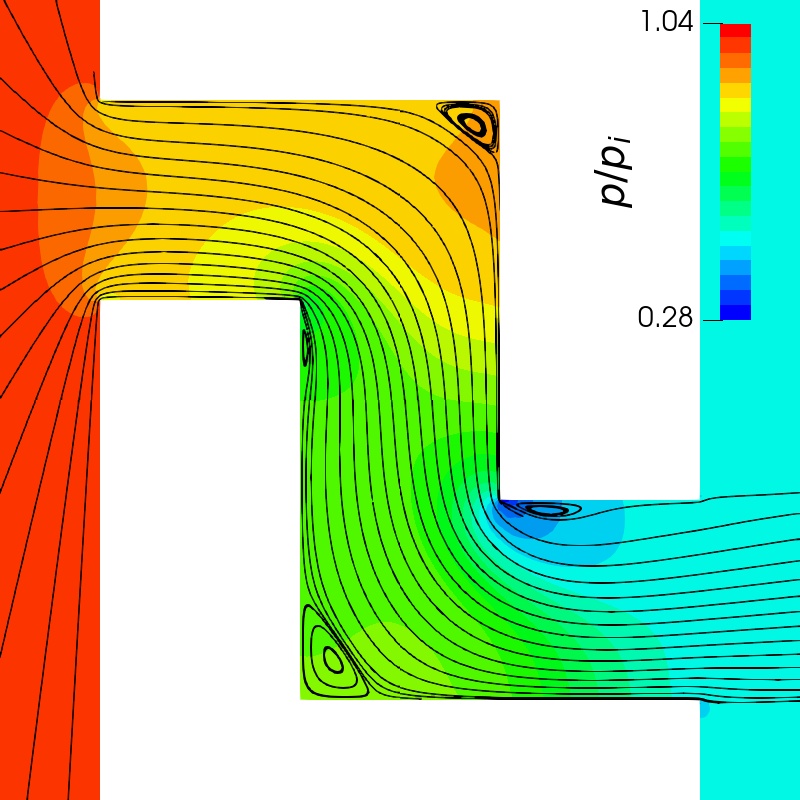}}}
        \hfill
        \subfloat[$N_v = 16^2$] {
        \adjustbox{width=0.3\linewidth,valign=b}{\includegraphics[width=\textwidth]{figs/microchannel_bgk_contours_Kn0p01_N32.jpg}}}
        \hfill
        \subfloat[$N_v = 32^2$] {
        \adjustbox{width=0.3\linewidth,valign=b}{\includegraphics[width=\textwidth]{figs/microchannel_bgk_contours_Kn0p01_N32.jpg}}}
        \caption{\label{fig:microchannel_comp_Kn0p01} Comparison of contours of normalized pressure overlaid with velocity streamlines for the bent microchannel problem at $Kn \sim 0.01$ computed using a $\mathbb P_3$ approximation with varying velocity space resolution.}
    \end{figure}

    \begin{figure}[htbp!]
        \centering
        \subfloat[$N_v = 8^2$] {
        \adjustbox{width=0.3\linewidth,valign=b}{\includegraphics[width=\textwidth]{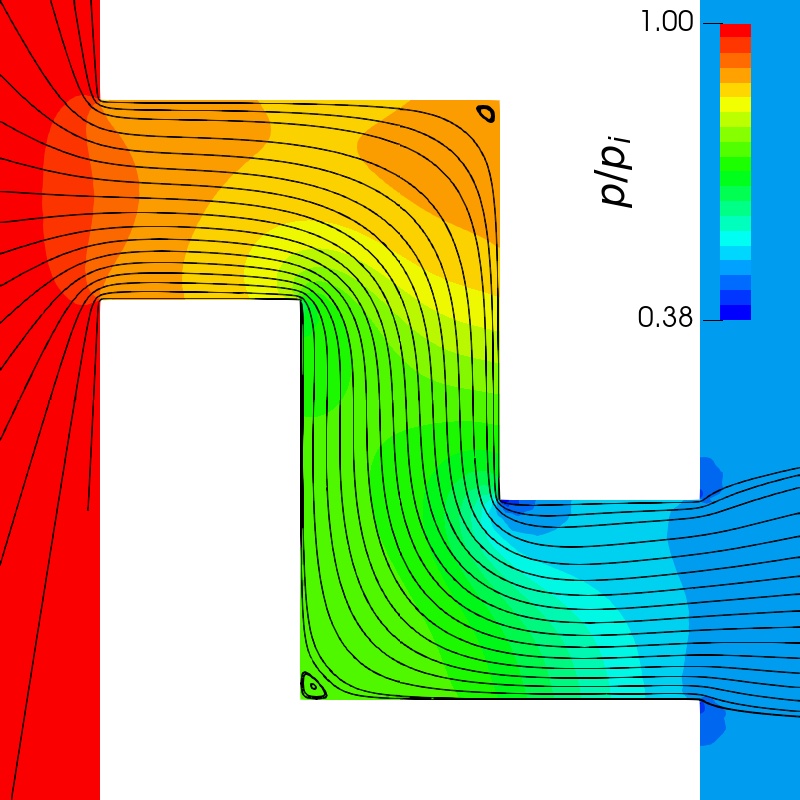}}}
        \hfill
        \subfloat[$N_v = 16^2$] {
        \adjustbox{width=0.3\linewidth,valign=b}{\includegraphics[width=\textwidth]{figs/microchannel_bgk_contours_Kn0p02_N32.jpg}}}
        \hfill
        \subfloat[$N_v = 32^2$] {
        \adjustbox{width=0.3\linewidth,valign=b}{\includegraphics[width=\textwidth]{figs/microchannel_bgk_contours_Kn0p02_N32.jpg}}}
        \caption{\label{fig:microchannel_comp_Kn0p02} Comparison of contours of normalized pressure overlaid with velocity streamlines for the bent microchannel problem at $Kn \sim 0.02$ computed using a $\mathbb P_3$ approximation with varying velocity space resolution.}
    \end{figure}
      \begin{figure}[htbp!]
        \centering
        \subfloat[$N_v = 8^2$] {
        \adjustbox{width=0.3\linewidth,valign=b}{\includegraphics[width=\textwidth]{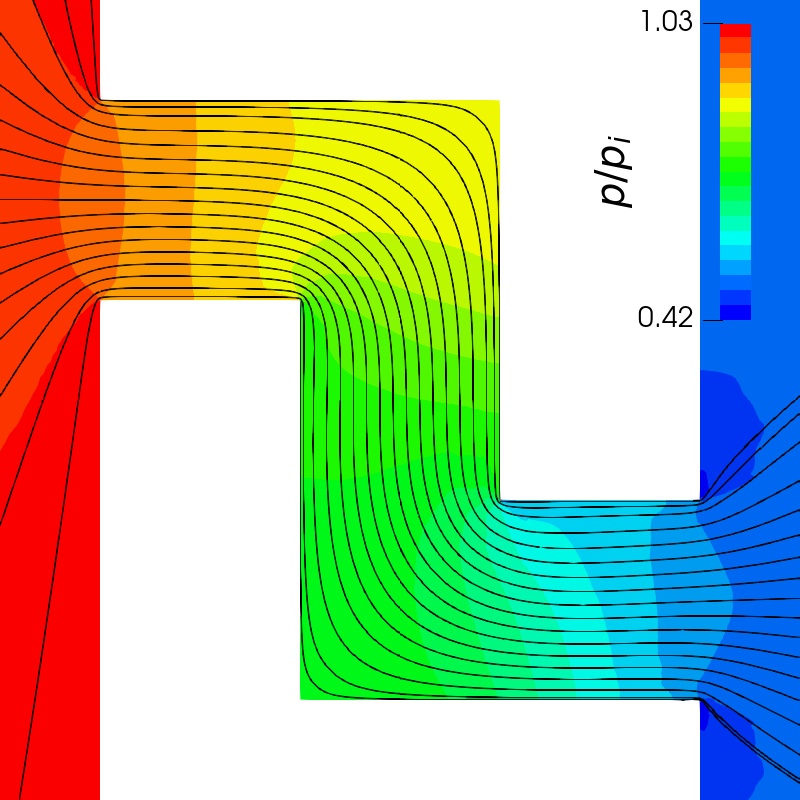}}}
        \hfill
        \subfloat[$N_v = 16^2$] {
        \adjustbox{width=0.3\linewidth,valign=b}{\includegraphics[width=\textwidth]{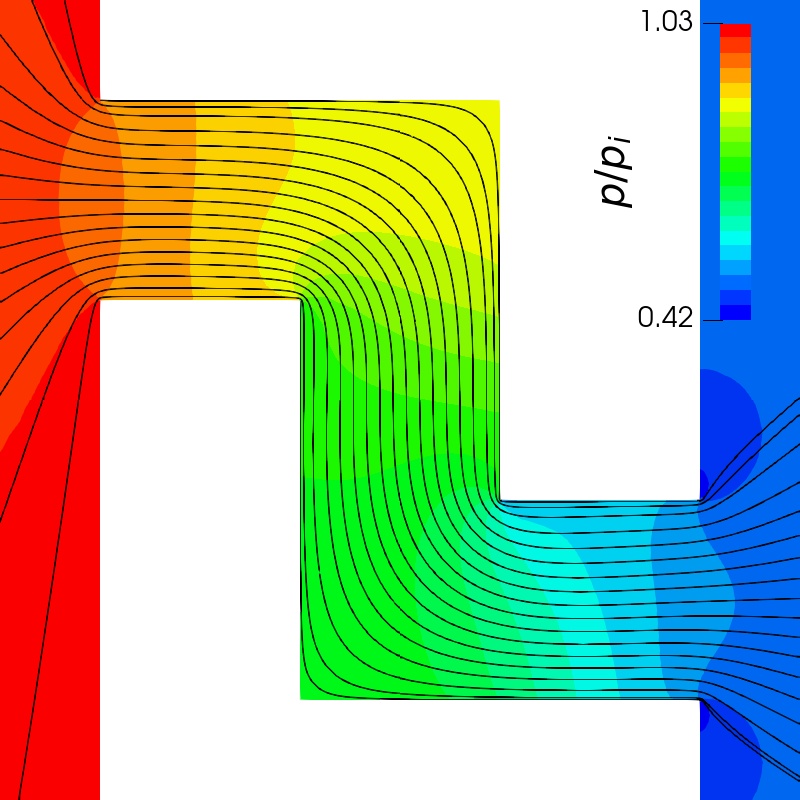}}}
        \hfill
        \subfloat[$N_v = 32^2$] {
        \adjustbox{width=0.3\linewidth,valign=b}{\includegraphics[width=\textwidth]{figs/microchannel_bgk_contours_Kn0p05_N32.jpg}}}
        \caption{\label{fig:microchannel_comp_Kn0p05} Comparison of contours of normalized pressure overlaid with velocity streamlines for the bent microchannel problem at $Kn \sim 0.05$ computed using a $\mathbb P_3$ approximation with varying velocity space resolution.}
    \end{figure}

To observe the effects of velocity space resolution on the accuracy of the method for predicting flows in the rarefied regime, a qualitative convergence study was performed for the three operating conditions. The numerical experiments were repeated using identical problem setups with three levels of velocity space resolution, a coarse space with $N_v = 8^2$, a medium space with $N_v = 16^2$, and a fine space with $N_v = 32^2$, the latter of which was the resolution level used for the previous numerical experiments. The contours of pressure overlaid with velocity streamlines computed with the varying velocity space resolution levels at $Kn \sim 0.01$, $0.02$, and $0.05$ are shown in \cref{fig:microchannel_comp_Kn0p01}, \cref{fig:microchannel_comp_Kn0p02}, and \cref{fig:microchannel_comp_Kn0p05}, respectively. It can be seen that when decreasing the velocity space resolution from $N_v = 32^2$ to $N_v = 16^2$, the results were indistinguishable regardless of the Knudsen number.  Furthermore, when the velocity space resolution was further decreased to $N_v = 8^2$, the results remained effectively identical, with only very minor differences in the velocity streamlines at lower Knudsen numbers that may even be attributed to sensitivity in the post-processing method. These observations indicate very promising results in that flows with strong non-equilibrium effects, at least in the low Mach regime, may be well-resolved with as few as eight velocity nodes per dimension, a level of resolution which makes the direct simulation of higher-dimensional problems via the Boltzmann equation completely feasible. However, it must be noted that this 
is only achievable given a discretely-conservative velocity model such as the one used in the present work as it was shown in \citet{Dzanic2023} that the conservation errors stemming from underresolved velocity spaces are significantly detrimental to the accuracy and stability of the numerical approach. 

\subsection{Rarefied three-dimensional T-junction}

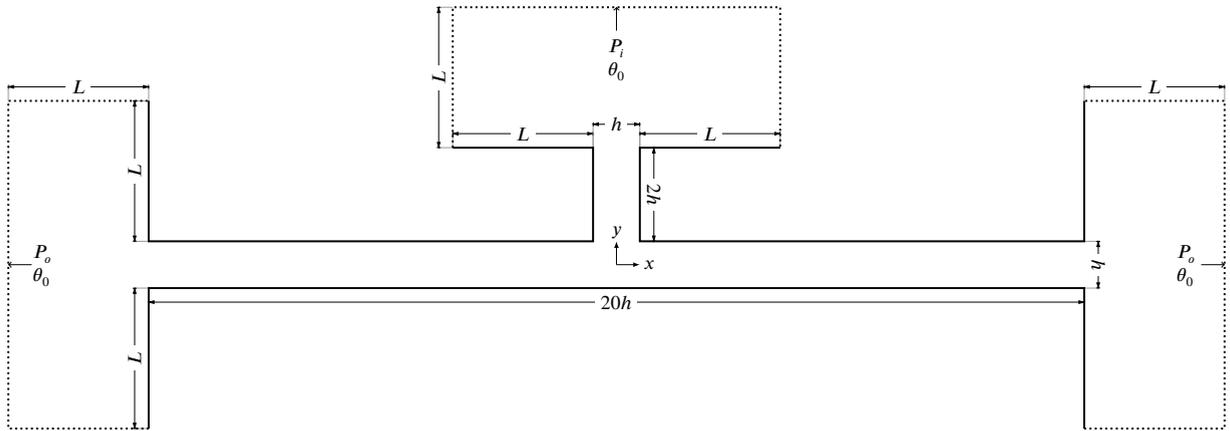
\begin{figure}[tbhp]
        \centering
        \adjustbox{width=0.98\linewidth, valign=b}{\begin{tikzpicture}[scale=1]

    \dimline[extension start length=0.3 cm, extension end length=0.3 cm] {(-3.5, 3.3 )}{(-0.5, 3.3)}{\large $L$};
    \dimline[extension start length=0.5 cm, extension end length=0.5 cm] {(-0.5, 3.5 )}{(0.5, 3.5)}{\large $h$};
    \dimline[extension start length=0.3 cm, extension end length=0.3 cm] {(0.5, 3.3 )}{(3.5, 3.3)}{\large $L$};
    \dimline[extension start length=0.3 cm, extension end length=0.3 cm] {(0.8, 3 )}{(0.8, 1)}{\large $2h$};
    \dimline[extension start length=-0.3 cm, extension end length=-0.3 cm] {(-10, -0.3 )}{(10, -0.3)}{\large $20h$};
    \dimline[extension start length=0.3 cm, extension end length=0.3 cm] {(10.3, 1)}{(10.3, 0)}{\large $h$};
    \dimline[extension start length=0.3 cm, extension end length=0.3 cm] {(-10.3, 1)}{(-10.3, 4)}{\large $L$};
    \dimline[extension start length=0.3 cm, extension end length=0.3 cm] {(-10.3, -3)}{(-10.3, 0)}{\large $L$};
    \dimline[extension start length=0.3 cm, extension end length=0.3 cm] {(-3.8, 3)}{(-3.8, 6)}{\large $L$};
    \dimline[extension start length=0.3 cm, extension end length=0.3 cm] {(-13, 4.3)}{(-10, 4.3)}{\large $L$};
    \dimline[extension start length=0.3 cm, extension end length=0.3 cm] {(10, 4.3)}{(13, 4.3)}{\large $L$};

    \draw[black, very thick] (-3.5, 3) -- (-0.5, 3) -- (-0.5, 1) -- (-10, 1)-- (-10, 4);
    \draw[black, very thick] (3.5, 3) -- (0.5, 3) -- (0.5, 1) --  (10, 1) --  (10, 4);
    \draw[black, very thick] (-10, -3) -- (-10, 0) -- (10, 0) -- (10, -3);
    \draw[black, very thick, dotted] (-10, -3) -- (-13, -3) -- (-13, 4) -- (-10, 4);
    \draw[black, very thick, dotted] (10, -3) -- (13, -3) -- (13, 4) -- (10, 4);
    \draw[black, very thick, dotted] (-3.5, 3) -- (-3.5, 6) -- (3.5, 6) -- (3.5, 3);

    \draw [black, <-] (-13, 0.5) -- (-12.5, 0.5) node[right,xshift=-0.3cm] {\large \begin{tabular}{l} $P_o$  \\ $\theta_0$ \end{tabular}};
    \draw [black, ->] (12.5, 0.5) -- (13, 0.5) node[left,xshift=-0.3cm] {\large \begin{tabular}{l} $P_o$  \\ $\theta_0$ \end{tabular}};
    \draw [black, ->] (0, 5.5) -- (0, 6) node[below,yshift=-0.5cm] {\large \begin{tabular}{l} $P_i$  \\ $\theta_0$ \end{tabular}};

    \draw[black, -latex] (0, 0.5) -- (0.5, 0.5);
    \node[text=black] at (0.7, 0.5) {\large $x$};
    \draw[black, -latex] (0, 0.5) -- (0, 1);
    \node[text=black] at (0, 1.2) {\large $y$};

\end{tikzpicture}}
        \caption{\label{fig:tjunction_geo} 
        Schematic of the rarefied three-dimensional T-junction domain on the cross-section $z = 0$. Wall boundaries represented by solid lines, inlet/outlet boundaries represented by dotted lines.
        }
    \end{figure}

To highlight the ability of the numerical approach in simulating three-dimensional problems and to present novel results for three-dimensional non-equilibrium flows, the simulation of a rarefied three-dimensional T-junction was performed via the Boltzmann--BGK approach. The flow within a three-dimensional T-junction can contain complex flow physics that are not present in its two-dimensional counterpart, and the T-shape geometry is frequently encountered in both microfluidic and standard hydraulic devices as a method for mixing fluids \citep{Hussain2019, Vigolo2014}. We consider a similar case to that of the bent microchannel, with a square channel fed from an inlet reservoir to an outlet reservoir. A cross-section of the geometry on the plane $z = 0$ is shown in \cref{fig:tjunction_geo}. Along the $z$ direction, the channel extent was set as $[-h/2, h/2]$ while the reservoir extent was set as $[-h/2 -L,  h/2 + L]$ with $L = 3h$.

The reservoir and wall boundary conditions were set identically to the bent microchannel case with flow conditions corresponding to a Mach number of $M = 0.1$ and a Reynolds number of $Re = 10$, yielding a Knudsen number of $Kn \sim 0.016$ which places the flow well within the slip regime. The problem was solved using a $\mathbb P_3$ approximation on a structured hexahedral mesh with an approximate edge length of $h/10$ within the channel, yielding approximately $N_e = 3.5{\cdot}10^4$ elements. Per the observations in the velocity space convergence study for the bent microchannel case, a velocity space resolution of $N_v = 8^3$ was deemed sufficient for the numerical experiment. The resulting total number of degrees of freedom for the numerical experiment was approximately 1.1 billion. The computational cost of the simulation per characteristic time $t_c = U_b/h$ was approximately 8.5 GPU hours for the Boltzmann--BGK approach and 2.7 GPU hours for the Navier--Stokes approach, with the relatively small increase in computational cost primarily a result of the larger maximum admissible time step of the Boltzmann--BGK approach for low Reynolds number flows. 

      \begin{figure}[htbp!]
        \centering
        \subfloat[Navier--Stokes] {
        \adjustbox{width=0.48\linewidth,valign=b}{\includegraphics[width=\textwidth]{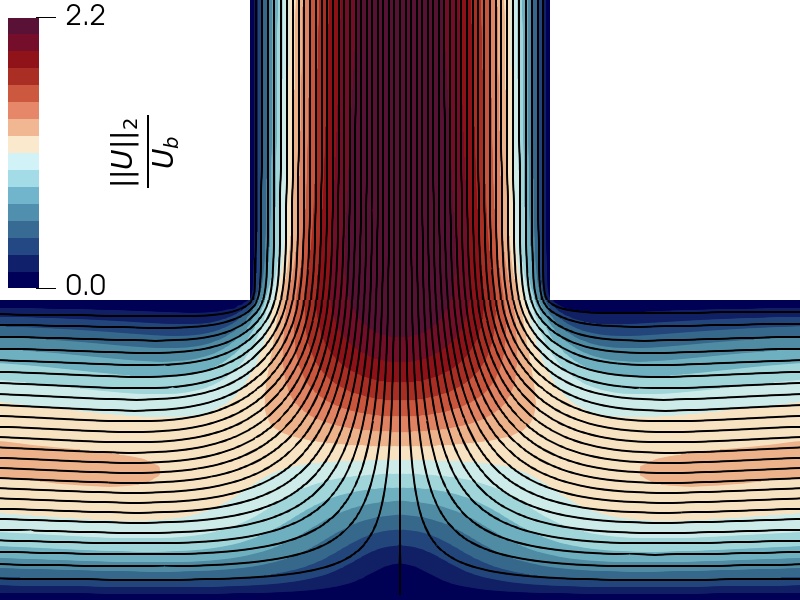}}}
        \hfill
        \subfloat[Boltzmann--BGK] {
        \adjustbox{width=0.48\linewidth,valign=b}{\includegraphics[width=\textwidth]{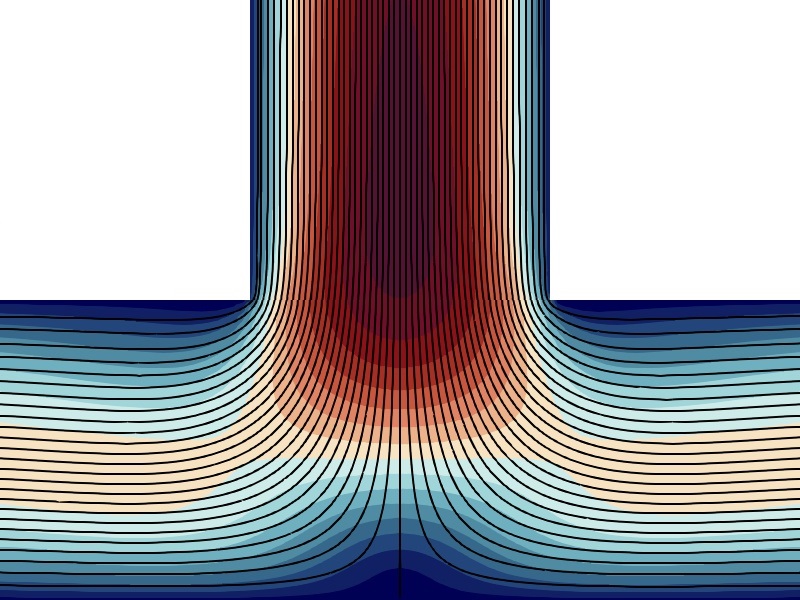}}}
        \caption{\label{fig:tjunction_rarefied_zslice} Contours of normalized velocity magnitude overlaid with velocity streamlines for the rarefied three-dimensional T-junction problem on the cross-section $z = 0$ computed using a $\mathbb P_3$ approximation and $N_v = 8^3$ with the Navier--Stokes equations (left) and the Boltzmann--BGK equation (right).}
    \end{figure}

    \begin{figure}[htbp!]
        \centering
        \subfloat[$x/h = 0$] {
        \adjustbox{width=0.16\linewidth,valign=b}{\includegraphics[width=\textwidth]{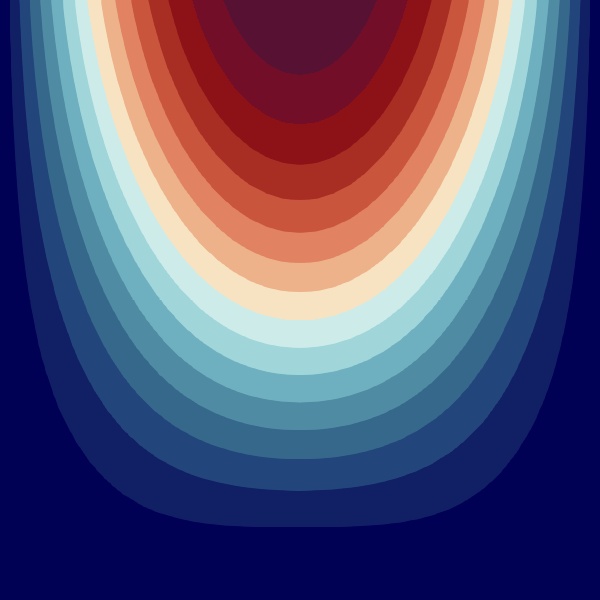}}}
        \subfloat[$x/h = 0.2$] {
        \adjustbox{width=0.16\linewidth,valign=b}{\includegraphics[width=\textwidth]{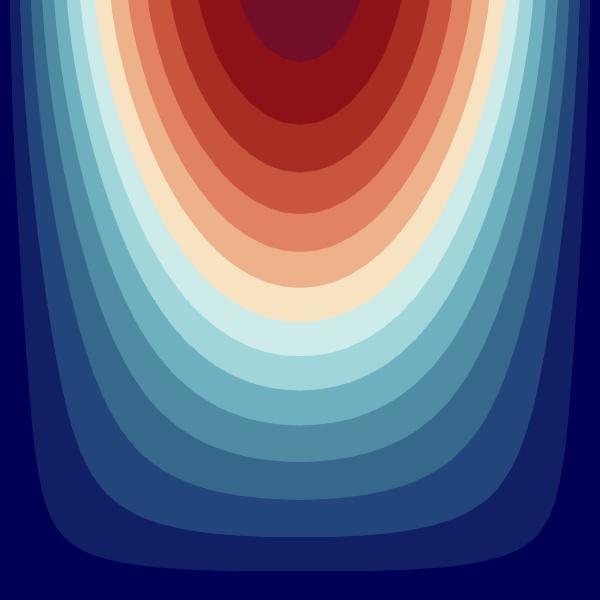}}}
        \subfloat[$x/h = 0.4$] {
        \adjustbox{width=0.16\linewidth,valign=b}{\includegraphics[width=\textwidth]{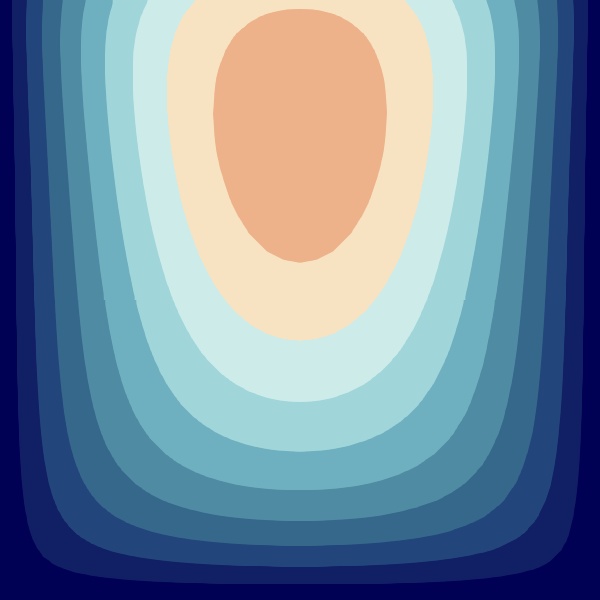}}}
        \subfloat[$x/h = 0.6$] {
        \adjustbox{width=0.16\linewidth,valign=b}{\includegraphics[width=\textwidth]{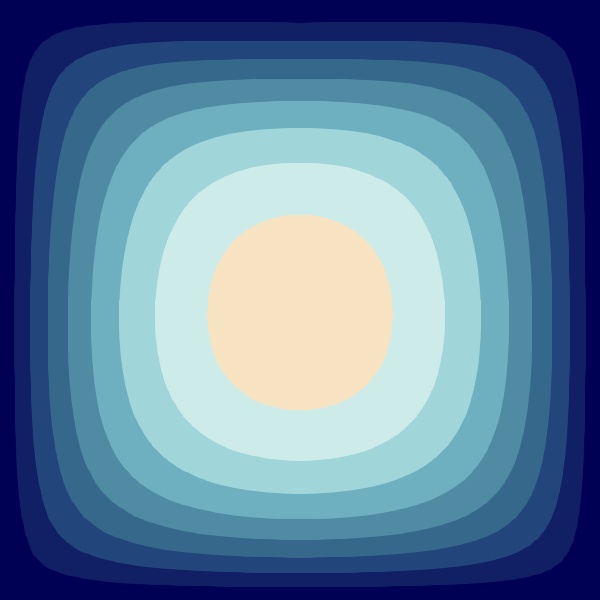}}}
        \subfloat[$x/h = 0.8$] {
        \adjustbox{width=0.16\linewidth,valign=b}{\includegraphics[width=\textwidth]{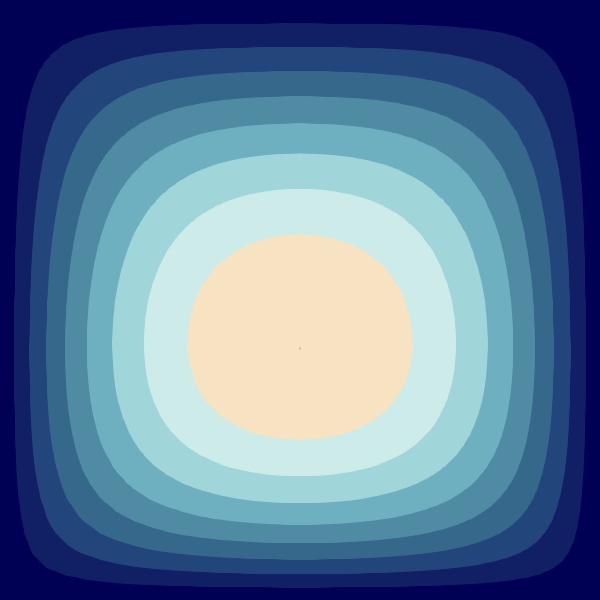}}}
        \subfloat[$x/h = 1$] {
        \adjustbox{width=0.16\linewidth,valign=b}{\includegraphics[width=\textwidth]{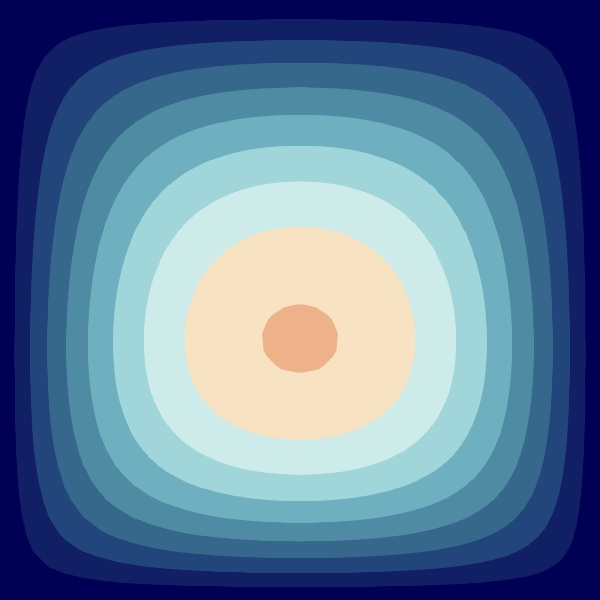}}}
        \caption{\label{fig:tjunction_rarefied_xslice_ns} 
        Contours of normalized velocity magnitude for the rarefied three-dimensional T-junction problem on cross-sections at varying $x$ locations computed using a $\mathbb P_3$ approximation with the Navier--Stokes equations. Legend identical to \cref{fig:tjunction_rarefied_zslice}.
        }
    \end{figure}
    \begin{figure}[htbp!]
        \centering
        \subfloat[$x/h = 0$] {
        \adjustbox{width=0.16\linewidth,valign=b}{\includegraphics[width=\textwidth]{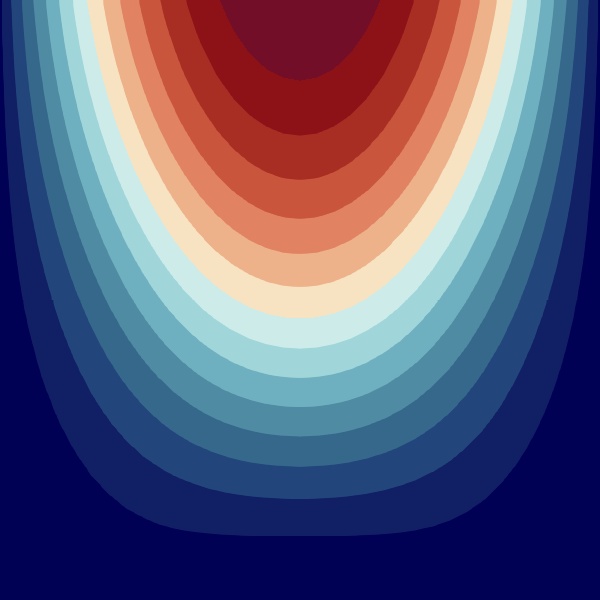}}}
        \subfloat[$x/h = 0.2$] {
        \adjustbox{width=0.16\linewidth,valign=b}{\includegraphics[width=\textwidth]{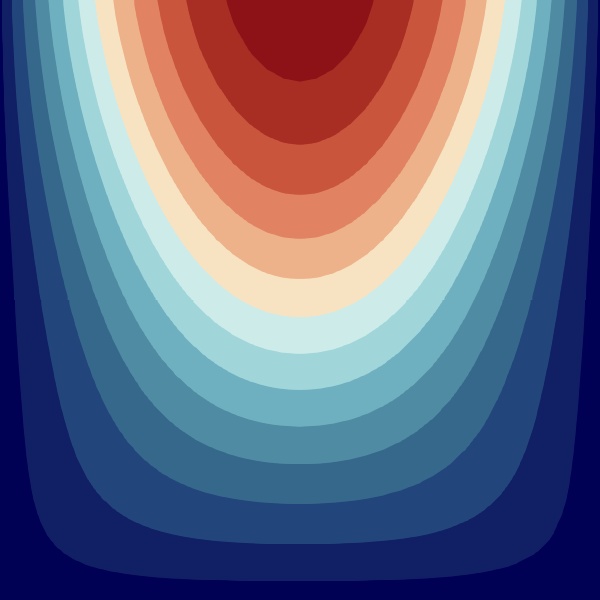}}}
        \subfloat[$x/h = 0.4$] {
        \adjustbox{width=0.16\linewidth,valign=b}{\includegraphics[width=\textwidth]{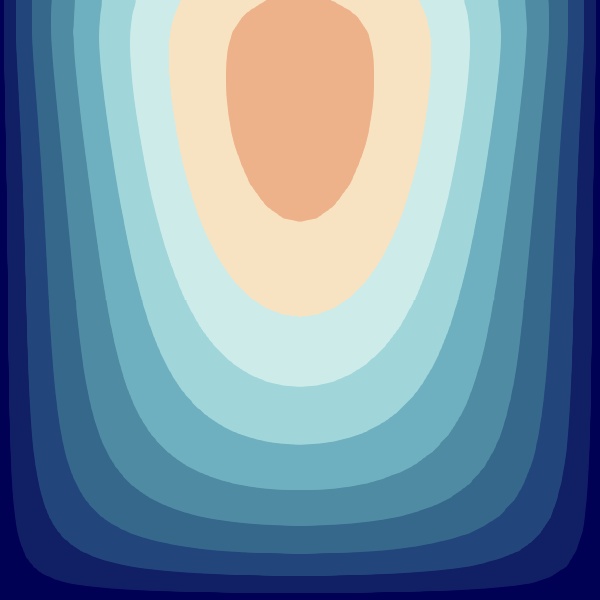}}}
        \subfloat[$x/h = 0.6$] {
        \adjustbox{width=0.16\linewidth,valign=b}{\includegraphics[width=\textwidth]{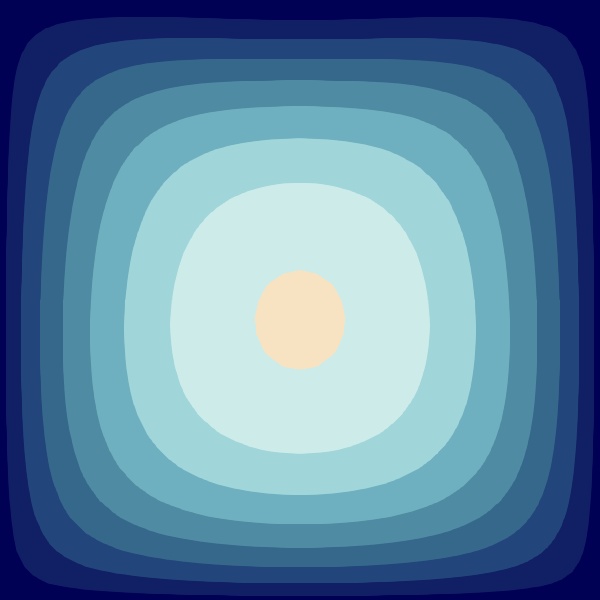}}}
        \subfloat[$x/h = 0.8$] {
        \adjustbox{width=0.16\linewidth,valign=b}{\includegraphics[width=\textwidth]{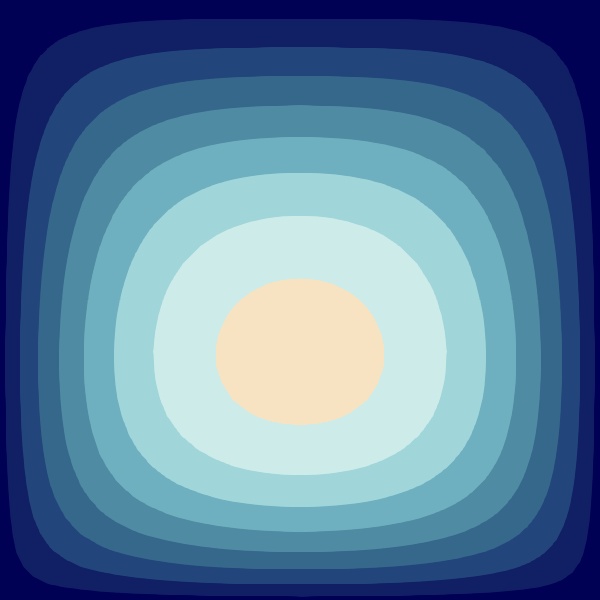}}}
        \subfloat[$x/h = 1$] {
        \adjustbox{width=0.16\linewidth,valign=b}{\includegraphics[width=\textwidth]{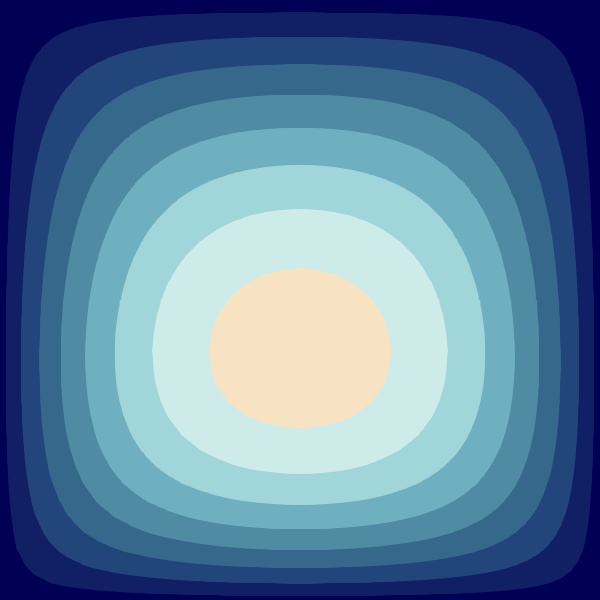}}}
        \caption{\label{fig:tjunction_rarefied_xslice_bgk} 
        Contours of normalized velocity magnitude for the rarefied three-dimensional T-junction problem on cross-sections at varying $x$ locations computed using a $\mathbb P_3$ approximation and $N_v = 8^3$ with the Boltzmann--BGK equation. Legend identical to \cref{fig:tjunction_rarefied_zslice}.
        }
    \end{figure}
\begin{figure}[htbp!]
        \centering
        \adjustbox{width=0.98\linewidth, valign=b}{     \begin{tikzpicture}[spy using outlines={rectangle, height=3cm,width=2.3cm, magnification=3, connect spies}]
		\begin{axis}[name=plot1,
		    axis line style={latex-latex},
		    axis x line=left,
            axis y line=left,        
            width=\axisdefaultwidth,
            height=0.4*\axisdefaultwidth,
            clip mode=individual,
		    xlabel={$U/U_b$},
		    xtick={0, 2, 4, 6, 8},
    		xmin=0.0,
    		xmax=9,
    		x tick label style={
        		/pgf/number format/.cd,
            	fixed,
            	precision=1,
        	    /tikz/.cd},
    		ylabel={$y/H$},
    		ytick={0, 0.2, 0.4, 0.6, 0.8, 1.0},
    		ymin=0,
    		ymax=1,
    		y tick label style={
        		/pgf/number format/.cd,
            	% fixed,
            	% precision=1,
        	    /tikz/.cd},
    		legend style={at={(.9,1.0)},anchor=north west,font=\scriptsize},
    		legend cell align={left},
    		style={font=\small},
    		scale = 2]

    		\addplot[color={black}, style={very thick, dashed}] table[x = u1, y = y, col sep=comma, unbounded coords=jump]{./figs/data/tjNS_Re10_yslices.csv};
            \addplot[color={red!80!black}, style={very thick}]
            table[x expr={\thisrow{u1}}, y = y, col sep=comma, unbounded coords=jump]{./figs/data/tjBGK_Re10_yslices.csv};

    		\addplot[color={black}, style={very thick, dashed}] table[x expr={\thisrow{u2} + 2}, y = y, col sep=comma, unbounded coords=jump]{./figs/data/tjNS_Re10_yslices.csv};
            \addplot[color={red!80!black}, style={very thick}]
            table[x expr={\thisrow{u2} + 2}, y = y, col sep=comma, unbounded coords=jump]{./figs/data/tjBGK_Re10_yslices.csv};

    		\addplot[color={black}, style={very thick, dashed}] table[x expr={\thisrow{u3} + 4}, y = y, col sep=comma, unbounded coords=jump]{./figs/data/tjNS_Re10_yslices.csv};
            \addplot[color={red!80!black}, style={very thick}]
            table[x expr={\thisrow{u3} + 4}, y = y, col sep=comma, unbounded coords=jump]{./figs/data/tjBGK_Re10_yslices.csv};

    		\addplot[color={black}, style={very thick, dashed}] table[x expr={\thisrow{u4} + 6}, y = y, col sep=comma, unbounded coords=jump]{./figs/data/tjNS_Re10_yslices.csv};
            \addplot[color={red!80!black}, style={very thick}]
            table[x expr={\thisrow{u4} + 6}, y = y, col sep=comma, unbounded coords=jump]{./figs/data/tjBGK_Re10_yslices.csv};

            \node at (1, 1) {\scriptsize$x/H = 1$};
            \node at (3, 1) {\scriptsize$x/H = 2$};
            \node at (5, 1) {\scriptsize$x/H = 3$};
            \node at (7.2, 1) {\scriptsize$x/H = 4$};
    		
    		\addlegendentry{Navier--Stokes}
    		\addlegendentry{Boltzmann--BGK}

		\end{axis}

	\end{tikzpicture}}
        \caption{\label{fig:tjunction_rarefied_velocity} 
        Velocity profiles for the rarefied three-dimensional T-junction problem on the cross-section $z = 0$ at varying $x$ locations computed using a $\mathbb P_3$ approximation and $N_v = 8^3$ with the Navier--Stokes equations (black, dashed) and the Boltzmann--BGK equation (red, solid). Profiles are shifted $+0$, $+2$, $+4$, $+6$ along the abscissa, respectively.}
    \end{figure}
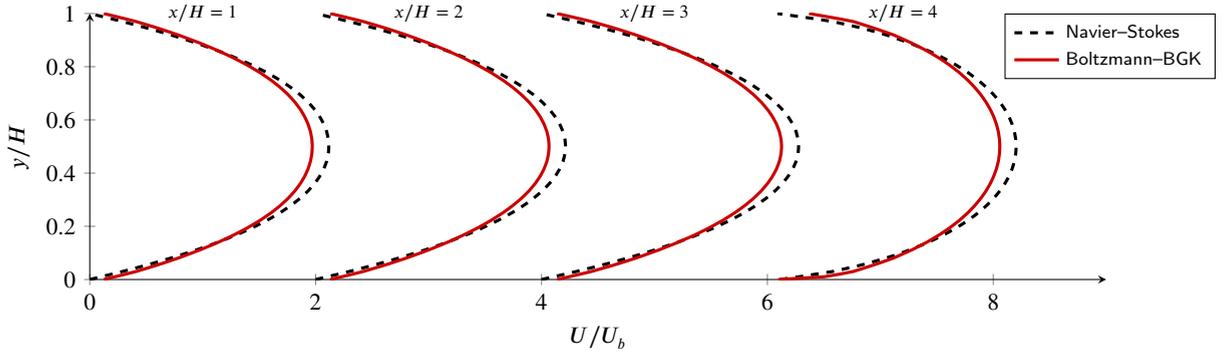

The contours of velocity magnitude, normalized by the bulk velocity $U_b$ and overlaid with velocity streamlines, are shown on the cross-section $z=0$ in \cref{fig:tjunction_rarefied_zslice} as computed by both the Boltzmann--BGK equation and the Navier--Stokes equations on the same mesh. It can be seen that in this flow regime, where slip velocity becomes non-negligible, the overarching flow structure between the two approaches was similar but noticeable differences could be observed. The Boltzmann--BGK prediction showed a lower maximum velocity along the centerline which can be attributed to the slip velocity at the wall. As a result, the velocity flow field appeared more diffused with the Boltzmann--BGK approach than the Navier--Stokes approach. These observations were consistent with the velocity magnitude contours on cross-sections at various $x$ locations, shown in \cref{fig:tjunction_rarefied_xslice_ns} and \cref{fig:tjunction_rarefied_xslice_bgk} for the Navier--Stokes and Boltzmann--BGK approach, respectively. The differences in the maximum centerline velocity were most pronounced aft of the channel corner ($x/h > 0.5$). A more detailed comparison of the velocity profiles can be seen in \cref{fig:tjunction_rarefied_velocity}, shown in terms of the normalized velocity on the cross-section $z = 0$. The slip velocity at the wall is evident in the Boltzmann--BGK results, which results in the lower centerline velocity in comparison to the Navier--Stokes results. Overall, a difference of approximately $7\%$ in the maximum velocity was observed between the two approaches. 

   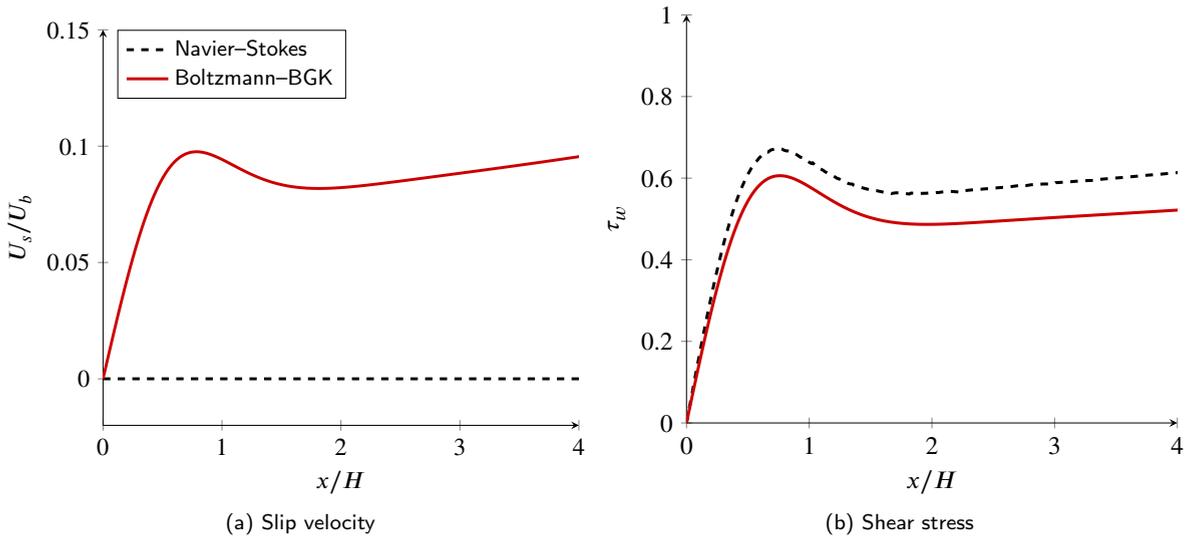
\begin{figure}[tbhp]
        \centering
        \subfloat[Slip velocity]{\adjustbox{width=0.48\linewidth, valign=b}{\begin{tikzpicture}[spy using outlines={rectangle, height=3cm,width=2.3cm, magnification=3, connect spies}]
	\begin{axis}[name=plot1,
		axis line style={latex-latex},
	    axis x line=left,
        axis y line=left,
		xlabel={$x/H$},
    	xmin=0, xmax=4,
    	xtick={0, 1, 2, 3, 4},
    	ylabel={$U_s/U_b$},
    	ymin=-0.02,ymax=0.15,
    	ytick={0, 0.05, 0.1, 0.15},
        clip mode=individual,
        y tick label style={/pgf/number format/.cd, fixed, precision=2, /tikz/.cd},
    	legend style={at={(0.03, 1.0)},anchor=north west,font=\small},
    	legend cell align={left},
    	style={font=\normalsize}
        ]
    	        
        \addplot[color={black}, style={very thick, dashed}] table[x expr = \thisrow{x}, y = us, col sep=comma, unbounded coords=jump]{./figs/data/tjNS_Re10_wall3.csv};
        \addlegendentry{Navier--Stokes};
        
        \addplot[color={red!80!black}, style={very thick}] table[x expr = \thisrow{x}, y = us, col sep=comma, unbounded coords=jump]{./figs/data/tjBGK_Re10_wall3.csv};
        \addlegendentry{Boltzmann--BGK};

	\end{axis}
\end{tikzpicture}}}
        % \subfloat[Pressure]{\adjustbox{width=0.33\linewidth, valign=b}{\input{figs/tj10_slip_bottom}}}
        \subfloat[Shear stress]{\adjustbox{width=0.48\linewidth, valign=b}{\begin{tikzpicture}[spy using outlines={rectangle, height=3cm,width=2.3cm, magnification=3, connect spies}]
	\begin{axis}[name=plot1,
		axis line style={latex-latex},
	    axis x line=left,
        axis y line=left,
		xlabel={$x/H$},
    	xmin=0, xmax=4,
    	xtick={0, 1, 2, 3, 4},
    	ylabel={$\tau_w$},
    	ymin=0,ymax=1,
        clip mode=individual,
    	legend style={at={(0.03, 1.0)},anchor=north west,font=\small},
    	legend cell align={left},
    	style={font=\normalsize}]
    	        
        \addplot[color={black}, style={very thick, dashed}] table[x expr = \thisrow{x}, y = tw, col sep=comma, unbounded coords=jump]{./figs/data/tjNS_Re10_wall3.csv} ;
        % \addlegendentry{Navier--Stokes}
        
        \addplot[color={red!80!black}, style={very thick}] table[x expr = \thisrow{x}, y = tw, col sep=comma, unbounded coords=jump]{./figs/data/tjBGK_Re10_wall3.csv};
        % \addlegendentry{Boltzmann--BGK}

	\end{axis}
\end{tikzpicture}}}
        
        \caption{\label{fig:tjunction_rarefied_botwall} Profiles of normalized wall slip velocity (left) and normalized wall shear stress (right) along the bottom wall ($y = -h/2, z = 0$) for the rarefied three-dimensional T-junction problem computed using a $\mathbb P_3$ approximation and $N_v = 8^3$ with the Navier--Stokes equations (black, dashed) and the Boltzmann--BGK equation (red, solid). }
    \end{figure}

   \begin{figure}[tbhp]
        \centering
        \subfloat[Slip velocity]{\adjustbox{width=0.48\linewidth, valign=b}{\begin{tikzpicture}[spy using outlines={rectangle, height=3cm,width=2.3cm, magnification=3, connect spies}]
	\begin{axis}[name=plot1,
		axis line style={latex-latex},
	    axis x line=left,
        axis y line=left,
		xlabel={$x/H$},
    	xmin=0.5, xmax=4,
    	xtick={0.5, 1, 2, 3, 4},
    	ylabel={$U_s/U_b$},
    	ymin=-0.02,ymax=0.15,
    	ytick={ 0, 0.05, 0.1, 0.15},
        clip mode=individual,
        y tick label style={/pgf/number format/.cd, fixed,  precision=2, /tikz/.cd},
    	legend style={at={(0.03, 1.0)},anchor=north west,font=\small},
    	legend cell align={left},
    	style={font=\normalsize}]
    	        
        \addplot[color={black}, style={very thick, dashed}] table[x expr = \thisrow{x} + 0.5, y = us, col sep=comma, unbounded coords=jump]{./figs/data/tjNS_Re10_wall2.csv};
        \addlegendentry{Navier--Stokes};
        
        \addplot[color={red!80!black}, style={very thick}] table[x expr = \thisrow{x} + 0.5, y = us, col sep=comma, unbounded coords=jump]{./figs/data/tjBGK_Re10_wall2.csv};
        \addlegendentry{Boltzmann--BGK};

	\end{axis}
\end{tikzpicture}}}
        % \subfloat[Pressure]{\adjustbox{width=0.33\linewidth, valign=b}{\input{figs/tj10_slip_top}}}
        \subfloat[Shear stress]{\adjustbox{width=0.48\linewidth, valign=b}{\begin{tikzpicture}[spy using outlines={rectangle, height=3cm,width=2.3cm, magnification=3, connect spies}]
	\begin{axis}[name=plot1,
		axis line style={latex-latex},
	    axis x line=left,
        axis y line=left,
		xlabel={$x/H$},
    	xmin=0.5, xmax=4,
    	xtick={0.5, 1, 2, 3, 4},
    	ylabel={$\tau_w$},
    	ymin=0,ymax=1,
        clip mode=individual,
    	legend style={at={(0.03, 1.0)},anchor=north west,font=\small},
    	legend cell align={left},
    	style={font=\normalsize}]
    	        
        \addplot[color={black}, style={very thick, dashed}] table[x expr = \thisrow{x} + 0.5, y = tw, col sep=comma, unbounded coords=jump]{./figs/data/tjNS_Re10_wall2.csv} ;
        % \addlegendentry{Navier--Stokes};
        
        \addplot[color={red!80!black}, style={very thick}] table[x expr = \thisrow{x} + 0.5, y = tw, col sep=comma, unbounded coords=jump]{./figs/data/tjBGK_Re10_wall2.csv};
        % \addlegendentry{Boltzmann--BGK};

	\end{axis}
\end{tikzpicture}}}
        
        \caption{\label{fig:tjunction_rarefied_topwall} Profiles of normalized wall slip velocity (left) and normalized wall shear stress (right) along the upper wall ($y = h/2, z = 0$) for the rarefied three-dimensional T-junction problem computed using a $\mathbb P_3$ approximation and $N_v = 8^3$ with the Navier--Stokes equations (black, dashed) and the Boltzmann--BGK equation (red, solid).}
    \end{figure}
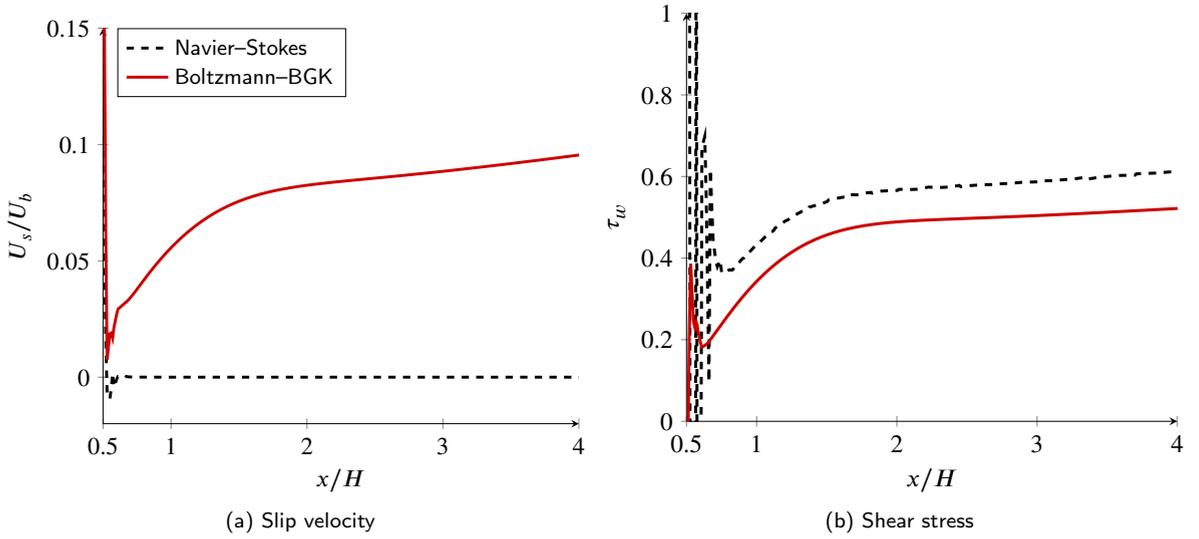

To observe the non-equilibrium effects of the flow in the near-wall region, the wall slip velocity and shear stress were analyzed for the bottom and top walls along the horizontal section of the channel, shown in \cref{fig:tjunction_rarefied_botwall} and \cref{fig:tjunction_rarefied_topwall}, respectively, for the cross-section $z=0$. Along the bottom wall, the wall slip velocity was well-pronounced in the Boltzmann--BGK results, showing slip on the order of $10\%$ of the bulk velocity which could not be predicted via the standard Navier--Stokes approximation. This resulted in a noticeable decrease in the wall shear stress, where the profiles were qualitatively similar but the Navier--Stokes results showed up to a $20\%$ overprediction in the stress compared to the Boltzmann--BGK results. The differences between the two approaches were even more pronounced along the top wall due to the geometric singularity at the convex corner. Similar levels of slip velocity and wall shear stress were predicted by the Boltzmann--BGK approach along the top wall as with the bottom wall, which were not accurately captured via the standard Navier--Stokes approach. Furthermore, due to the singularity at the convex corner, the Navier--Stokes results showed numerical instabilities in the corner region, with highly oscillatory wall shear stress profiles and non-zero wall velocity, the latter of which is possible due to weakly enforced boundary conditions. In comparison, the Boltzmann--BGK predictions did not show any indication of numerical instabilities, resulting in well-behaved wall slip and shear stress profiles even in the corner region. These results highlight the differences between the prediction of rarefied gas dynamics obtained via the standard Navier--Stokes equations and the Boltzmann--BGK equation as well as showcase the ability of the numerical approach in simulating three-dimensional rarefied flows.

\subsection{Laminar boundary layer}
With the previous numerical experiments showing a validation of the numerical approach for flows in the rarefied regime where molecular gas dynamics methods are commonly used, the remaining numerical experiments will focus on more complex flows in the continuum regime for which the Boltzmann equation has not yet been validated for. As an initial validation for continuum flows, we consider the canonical case of a laminar boundary layer for which there exists an approximate analytic solution. To yield flow conditions well into the continuum regime, the Mach number was set as $M = 0.2$ and the Reynolds number was set as $Re = 10^4$ based on the boundary layer reference length $L$.

The domain was set as $\Omega^{\mathbf{x}} = [-2L, L] \times [0, 2L]$. Due to the symmetry of the problem, only the half domain ($y > 0$) was solved. Along the bottom of the domain ($y = 0$), specular wall boundary conditions were used for $x < 0$ while diffuse wall boundary conditions were used for $x \geq 0$, such that the origin of the boundary layer coincided with the origin of the domain. For the remaining boundaries, Dirichlet boundary conditions corresponding to the freestream flow were used. The problem was solved using a $\mathbb P_3$ approximation with $N_v = 16^2$. A quadrilateral mesh with $N_e = 64 \times 32$ elements was generated, with the boundary layer region resolved by $32$ elements in the streamwise direction. At the origin, the mesh spacing was set as $\Delta x = L/200$ and $\Delta y = L/500$, with uniform progression towards the outer boundaries.

   \begin{figure}[tbhp]
        \centering
        \subfloat[Streamwise velocity]{\adjustbox{width=0.45\linewidth, valign=b}{\begin{tikzpicture}[spy using outlines={rectangle, height=3cm,width=2.3cm, magnification=3, connect spies}]
	\begin{axis} [name=plot1,
		axis line style={latex-latex},
	    axis x line=left,
        axis y line=left,
    	xmin=0, xmax=6,
    	xlabel={$\eta$},
		ylabel={$U/U_{\infty}$},
    	ymin=0, ymax=1.05,
    	ytick={0, 0.2, 0.4, 0.6, 0.8, 1.0},
        clip mode=individual,
    	legend style={at={(0.97, 0.03)}, anchor=south east},
    	legend cell align={left},
    	style={font=\normalsize},
        % width=3*\axisdefaultheight,
        % height=\axisdefaultheight
        ]

        \addplot[color=black, style={thick} ]
        table[x=eta, y=u, col sep=comma]{./figs/data/bl_blasius.csv};
        \addlegendentry{Blasius};
                
        \addplot[color=black, style={very thin}, only marks, mark=o, mark options={scale=0.7}, mark repeat = 10, mark phase = 2]
        table[x=eta, y=u, col sep=comma]{./figs/data/blNS_x1.csv}; 
        \addlegendentry{$Re_x = 1{\cdot}10^3$};
        
        \addplot[color=black, style={very thin}, only marks, mark=triangle, mark options={scale=0.8}, mark repeat = 10, mark phase = 4]
        table[x=eta, y=u, col sep=comma]{./figs/data/blNS_x2.csv};
        \addlegendentry{$Re_x = 2{\cdot}10^3$};
        
        \addplot[color=black, style={very thin}, only marks, mark=square, mark options={scale=0.6}, mark repeat = 10, mark phase = 6]
        table[x=eta, y=u, col sep=comma]{./figs/data/blNS_x3.csv};
        \addlegendentry{$Re_x = 3{\cdot}10^3$};
                      
        \addplot[color=black, style={very thin}, only marks, mark=diamond, mark options={scale=0.8}, mark repeat = 10, mark phase = 8]
        table[x=eta, y=u, col sep=comma]{./figs/data/blNS_x4.csv};
        \addlegendentry{$Re_x = 4{\cdot}10^3$};
                      
        \addplot[color=black, style={very thin}, only marks, mark=x, mark options={scale=0.8}, mark repeat = 10, mark phase = 10]
        table[x=eta, y=u, col sep=comma]{./figs/data/blNS_x5.csv};
        \addlegendentry{$Re_x = 5{\cdot}10^3$};

        \addplot[color=red, style={very thin}, only marks, mark=o, mark options={scale=0.7}, mark repeat = 10, mark phase = 2]
        table[x=eta, y=u, col sep=comma]{./figs/data/blBGK_x1.csv}; 
        
        \addplot[color=red, style={very thin}, only marks, mark=triangle, mark options={scale=0.8}, mark repeat = 10, mark phase = 4]
        table[x=eta, y=u, col sep=comma]{./figs/data/blBGK_x2.csv};
        
        \addplot[color=red, style={very thin}, only marks, mark=square, mark options={scale=0.6}, mark repeat = 10, mark phase = 6]
        table[x=eta, y=u, col sep=comma]{./figs/data/blBGK_x3.csv};
                      
        \addplot[color=red, style={very thin}, only marks, mark=diamond, mark options={scale=0.8}, mark repeat = 10, mark phase = 8]
        table[x=eta, y=u, col sep=comma]{./figs/data/blBGK_x4.csv};
                      
        \addplot[color=red, style={very thin}, only marks, mark=x, mark options={scale=0.8}, mark repeat = 10, mark phase = 10]
        table[x=eta, y=u, col sep=comma]{./figs/data/blBGK_x5.csv};

	\end{axis}
\end{tikzpicture}}}
        \subfloat[Normal velocity]{\adjustbox{width=0.45\linewidth, valign=b}{\begin{tikzpicture}[spy using outlines={rectangle, height=3cm,width=2.3cm, magnification=3, connect spies}]
	\begin{axis} [name=plot1,
		axis line style={latex-latex},
	    axis x line=left,
        axis y line=left,
    	xmin=0, xmax=6,
    	xlabel={$\eta$},
		ylabel={$V/\sqrt{\nu U_{\infty} / x}$},
    	ymin=0, ymax=1.05,
    	ytick={0, 0.2, 0.4, 0.6, 0.8, 1.0},
        clip mode=individual,
    	legend style={at={(0.97, 0.03)}, anchor=south east},
    	legend cell align={left},
    	style={font=\normalsize},
        % width=3*\axisdefaultheight,
        % height=\axisdefaultheight
        ]
    
        \addplot[color=black, style={thick} ]
        table[x=eta, y=v, col sep=comma]{./figs/data/bl_blasius.csv};
        % \addlegendentry{Blasius};
                
        \addplot[color=black, style={very thin}, only marks, mark=o, mark options={scale=0.7}, mark repeat = 10, mark phase = 2]
        table[x=eta, y=v, col sep=comma]{./figs/data/blNS_x1.csv}; 
        % \addlegendentry{$Re_x = 1{\cdot}10^3$};
        
        \addplot[color=black, style={very thin}, only marks, mark=triangle, mark options={scale=0.8}, mark repeat = 10, mark phase = 4]
        table[x=eta, y=v, col sep=comma]{./figs/data/blNS_x2.csv};
        % \addlegendentry{$Re_x = 2{\cdot}10^3$};
        
        \addplot[color=black, style={very thin}, only marks, mark=square, mark options={scale=0.6}, mark repeat = 10, mark phase = 6]
        table[x=eta, y=v, col sep=comma]{./figs/data/blNS_x3.csv};
        % \addlegendentry{$Re_x = 3{\cdot}10^3$};
                      
        \addplot[color=black, style={very thin}, only marks, mark=diamond, mark options={scale=0.8}, mark repeat = 10, mark phase = 8]
        table[x=eta, y=v, col sep=comma]{./figs/data/blNS_x4.csv};
        % \addlegendentry{$Re_x = 4{\cdot}10^3$};
                      
        \addplot[color=black, style={very thin}, only marks, mark=x, mark options={scale=0.8}, mark repeat = 10, mark phase = 10]
        table[x=eta, y=v, col sep=comma]{./figs/data/blNS_x5.csv};
        % \addlegendentry{$Re_x = 5{\cdot}10^3$};

        \addplot[color=red, style={very thin}, only marks, mark=o, mark options={scale=0.7}, mark repeat = 10, mark phase = 2]
        table[x=eta, y=v, col sep=comma]{./figs/data/blBGK_x1.csv}; 
        
        \addplot[color=red, style={very thin}, only marks, mark=triangle, mark options={scale=0.8}, mark repeat = 10, mark phase = 4]
        table[x=eta, y=v, col sep=comma]{./figs/data/blBGK_x2.csv};
        
        \addplot[color=red, style={very thin}, only marks, mark=square, mark options={scale=0.6}, mark repeat = 10, mark phase = 6]
        table[x=eta, y=v, col sep=comma]{./figs/data/blBGK_x3.csv};
                      
        \addplot[color=red, style={very thin}, only marks, mark=diamond, mark options={scale=0.8}, mark repeat = 10, mark phase = 8]
        table[x=eta, y=v, col sep=comma]{./figs/data/blBGK_x4.csv};
                      
        \addplot[color=red, style={very thin}, only marks, mark=x, mark options={scale=0.8}, mark repeat = 10, mark phase = 10]
        table[x=eta, y=v, col sep=comma]{./figs/data/blBGK_x5.csv};

	\end{axis}
\end{tikzpicture}}}
        
        \caption{\label{fig:boundarylayer_profiles}Normalized streamwise (left) and vertical (right) velocity profiles for the laminar boundary layer problem at varying streamwise locations computed with the Navier--Stokes equations (black markers) and the Boltzmann--BGK equation (red markers) using a $\mathbb P_3$ approximation with $N_v = 16^2$. Results shown in comparison to the Blasius prediction. }
    \end{figure}
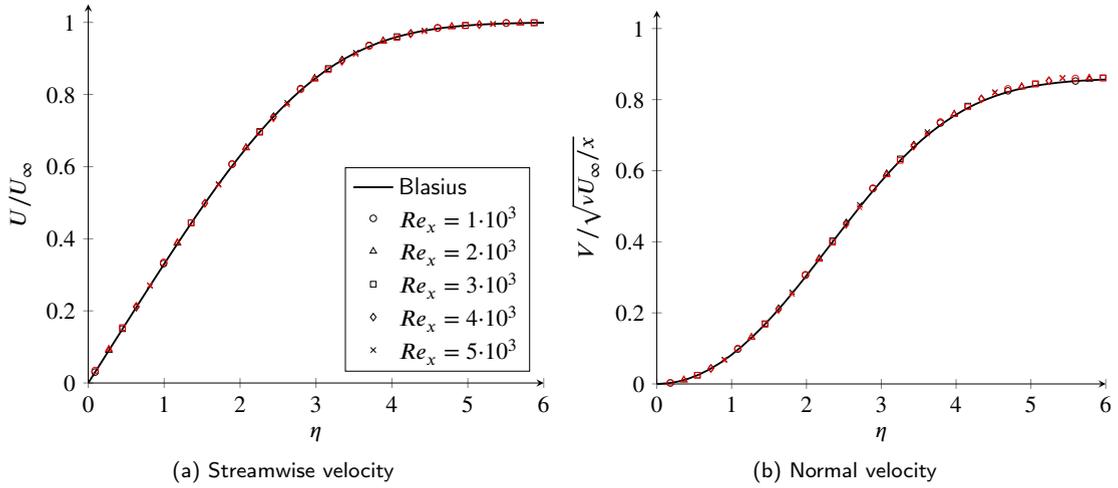

The flow was advanced to a steady state after which the flow at varying streamwise locations in the boundary layer was analyzed, represented through the local Reynolds number $Re_x$. The normalized streamwise and normal velocities as computed by the Boltzmann--BGK approach are shown in \cref{fig:boundarylayer_profiles} in comparison to the Navier--Stokes approach on the same mesh. These profiles are shown with respect to the self-similar Blasius boundary layer solution, given in terms of the self-similarity variable
\begin{equation*}
    \eta = y \sqrt{\frac{U_{\infty}}{\nu x}},
\end{equation*}
where $\nu = \mu/\rho$ is the kinematic viscosity. It can be seen that the streamwise velocity profiles were predicted with very good agreement with the Blasius prediction at all streamwise locations, and the Navier--Stokes and Boltzmann--BGK results were virtually indistinguishable. Furthermore, similar behavior was observed for the normal velocity profiles which are typically much more difficult to accurately resolve. 

   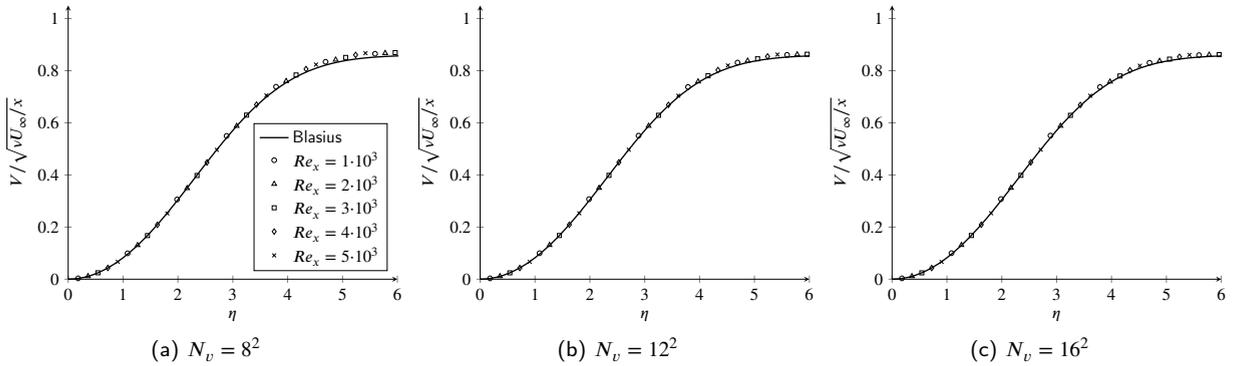
\begin{figure}[tbhp]
        \centering
        \subfloat[$N_v = 8^2$]{\adjustbox{width=0.33\linewidth, valign=b}{\begin{tikzpicture}[spy using outlines={rectangle, height=3cm,width=2.3cm, magnification=3, connect spies}]
	\begin{axis} [name=plot1,
		axis line style={latex-latex},
	    axis x line=left,
        axis y line=left,
    	xmin=0, xmax=6,
    	xlabel={$\eta$},
		ylabel={$V/\sqrt{\nu U_{\infty} / x}$},
    	ymin=0, ymax=1.05,
    	ytick={0, 0.2, 0.4, 0.6, 0.8, 1.0},
        clip mode=individual,
    	legend style={at={(0.97, 0.03)}, anchor=south east},
    	legend cell align={left},
    	style={font=\normalsize},
        % width=3*\axisdefaultheight,
        % height=\axisdefaultheight
        ]
    
        \addplot[color=black, style={thick} ]
        table[x=eta, y=v, col sep=comma]{./figs/data/bl_blasius.csv};
        \addlegendentry{Blasius};
    
        \addplot[color=black, style={very thin}, only marks, mark=o, mark options={scale=0.7}, mark repeat = 10, mark phase = 2]
        table[x=eta, y=v, col sep=comma]{./figs/data/blBGK8x8_x1.csv}; 
        \addlegendentry{$Re_x = 1{\cdot}10^3$};
        
        \addplot[color=black, style={very thin}, only marks, mark=triangle, mark options={scale=0.8}, mark repeat = 10, mark phase = 4]
        table[x=eta, y=v, col sep=comma]{./figs/data/blBGK8x8_x2.csv};
        \addlegendentry{$Re_x = 2{\cdot}10^3$};
        
        \addplot[color=black, style={very thin}, only marks, mark=square, mark options={scale=0.6}, mark repeat = 10, mark phase = 6]
        table[x=eta, y=v, col sep=comma]{./figs/data/blBGK8x8_x3.csv};
        \addlegendentry{$Re_x = 3{\cdot}10^3$};
                      
        \addplot[color=black, style={very thin}, only marks, mark=diamond, mark options={scale=0.8}, mark repeat = 10, mark phase = 8]
        table[x=eta, y=v, col sep=comma]{./figs/data/blBGK8x8_x4.csv};
        \addlegendentry{$Re_x = 4{\cdot}10^3$};
                      
        \addplot[color=black, style={very thin}, only marks, mark=x, mark options={scale=0.8}, mark repeat = 10, mark phase = 10]
        table[x=eta, y=v, col sep=comma]{./figs/data/blBGK8x8_x5.csv};
        \addlegendentry{$Re_x = 5{\cdot}10^3$};

	\end{axis}
\end{tikzpicture}}}
        \subfloat[$N_v = 12^2$]{\adjustbox{width=0.33\linewidth, valign=b}{\begin{tikzpicture}[spy using outlines={rectangle, height=3cm,width=2.3cm, magnification=3, connect spies}]
	\begin{axis} [name=plot1,
		axis line style={latex-latex},
	    axis x line=left,
        axis y line=left,
    	xmin=0, xmax=6,
    	xlabel={$\eta$},
		ylabel={$V/\sqrt{\nu U_{\infty} / x}$},
    	ymin=0, ymax=1.05,
    	ytick={0, 0.2, 0.4, 0.6, 0.8, 1.0},
        clip mode=individual,
    	legend style={at={(0.97, 0.03)}, anchor=south east},
    	legend cell align={left},
    	style={font=\normalsize},
        % width=3*\axisdefaultheight,
        % height=\axisdefaultheight
        ]
    
        \addplot[color=black, style={thick} ]
        table[x=eta, y=v, col sep=comma]{./figs/data/bl_blasius.csv};
        % \addlegendentry{Blasius};
    
        \addplot[color=black, style={very thin}, only marks, mark=o, mark options={scale=0.7}, mark repeat = 10, mark phase = 2]
        table[x=eta, y=v, col sep=comma]{./figs/data/blBGK12x12_x1.csv}; 
        % \addlegendentry{$Re_x = 1{\cdot}10^3$};
        
        \addplot[color=black, style={very thin}, only marks, mark=triangle, mark options={scale=0.8}, mark repeat = 10, mark phase = 4]
        table[x=eta, y=v, col sep=comma]{./figs/data/blBGK12x12_x2.csv};
        % \addlegendentry{$Re_x = 2{\cdot}10^3$};
        
        \addplot[color=black, style={very thin}, only marks, mark=square, mark options={scale=0.6}, mark repeat = 10, mark phase = 6]
        table[x=eta, y=v, col sep=comma]{./figs/data/blBGK12x12_x3.csv};
        % \addlegendentry{$Re_x = 3{\cdot}10^3$};
                      
        \addplot[color=black, style={very thin}, only marks, mark=diamond, mark options={scale=0.8}, mark repeat = 10, mark phase = 8]
        table[x=eta, y=v, col sep=comma]{./figs/data/blBGK12x12_x4.csv};
        % \addlegendentry{$Re_x = 4{\cdot}10^3$};
                      
        \addplot[color=black, style={very thin}, only marks, mark=x, mark options={scale=0.8}, mark repeat = 10, mark phase = 10]
        table[x=eta, y=v, col sep=comma]{./figs/data/blBGK12x12_x5.csv};
        % \addlegendentry{$Re_x = 5{\cdot}10^3$};

	\end{axis}
\end{tikzpicture}}}
        \subfloat[$N_v = 16^2$]{\adjustbox{width=0.33\linewidth, valign=b}{\begin{tikzpicture}[spy using outlines={rectangle, height=3cm,width=2.3cm, magnification=3, connect spies}]
	\begin{axis} [name=plot1,
		axis line style={latex-latex},
	    axis x line=left,
        axis y line=left,
    	xmin=0, xmax=6,
    	xlabel={$\eta$},
		ylabel={$V/\sqrt{\nu U_{\infty} / x}$},
    	ymin=0, ymax=1.05,
    	ytick={0, 0.2, 0.4, 0.6, 0.8, 1.0},
        clip mode=individual,
    	legend style={at={(0.97, 0.03)}, anchor=south east},
    	legend cell align={left},
    	style={font=\normalsize},
        % width=3*\axisdefaultheight,
        % height=\axisdefaultheight
        ]
    
        \addplot[color=black, style={thick} ]
        table[x=eta, y=v, col sep=comma]{./figs/data/bl_blasius.csv};
        % \addlegendentry{Blasius};
    
        \addplot[color=black, style={very thin}, only marks, mark=o, mark options={scale=0.7}, mark repeat = 10, mark phase = 2]
        table[x=eta, y=v, col sep=comma]{./figs/data/blBGK_x1.csv}; 
        % \addlegendentry{$Re_x = 1{\cdot}10^3$};
        
        \addplot[color=black, style={very thin}, only marks, mark=triangle, mark options={scale=0.8}, mark repeat = 10, mark phase = 4]
        table[x=eta, y=v, col sep=comma]{./figs/data/blBGK_x2.csv};
        % \addlegendentry{$Re_x = 2{\cdot}10^3$};
        
        \addplot[color=black, style={very thin}, only marks, mark=square, mark options={scale=0.6}, mark repeat = 10, mark phase = 6]
        table[x=eta, y=v, col sep=comma]{./figs/data/blBGK_x3.csv};
        % \addlegendentry{$Re_x = 3{\cdot}10^3$};
                      
        \addplot[color=black, style={very thin}, only marks, mark=diamond, mark options={scale=0.8}, mark repeat = 10, mark phase = 8]
        table[x=eta, y=v, col sep=comma]{./figs/data/blBGK_x4.csv};
        % \addlegendentry{$Re_x = 4{\cdot}10^3$};
                      
        \addplot[color=black, style={very thin}, only marks, mark=x, mark options={scale=0.8}, mark repeat = 10, mark phase = 10]
        table[x=eta, y=v, col sep=comma]{./figs/data/blBGK_x5.csv};
        % \addlegendentry{$Re_x = 5{\cdot}10^3$};

	\end{axis}
\end{tikzpicture}}}
        
        \caption{\label{fig:boundarylayer_res} Comparison of normalized vertical velocity profiles for the laminar boundary layer problem at varying streamwise locations computed using the Boltzmann--BGK approach and a $\mathbb P_3$ approximation with varying velocity space resolution. }
    \end{figure}
    
   \begin{figure}[tbhp]
        \centering
        \subfloat[$N_v = 8^2$]{\adjustbox{width=0.33\linewidth, valign=b}{\begin{tikzpicture}[spy using outlines={rectangle, height=3cm,width=2.3cm, magnification=3, connect spies}]
	\begin{axis} [name=plot1,
		axis line style={latex-latex},
	    axis x line=left,
        axis y line=left,
    	xmin=0, xmax=1,
    	xlabel={$x/L$},
		ylabel={$\tau_w$},
    	ymin=0, ymax=0.04,
        clip mode=individual,
    	legend style={at={(0.97, 0.97)}, anchor=north east},
    	legend cell align={left},
    	style={font=\normalsize},
        % width=3*\axisdefaultheight,
        % height=\axisdefaultheight
        ]

        \addplot[color=black, style={thick} ]
        table[x=x, y=twb, col sep=comma]{./figs/data/bl_tw.csv};
        \addlegendentry{Blasius};
                
        \addplot[color=black, style={very thin}, only marks, mark=o, mark options={scale=0.7}, mark repeat = 2, mark phase = 0]
        table[x=x, y=tw8, col sep=comma]{./figs/data/bl_tw.csv}; 
        \addlegendentry{Boltzmann--BGK};
                
	\end{axis}
\end{tikzpicture}}}
        \subfloat[$N_v = 12^2$]{\adjustbox{width=0.33\linewidth, valign=b}{\begin{tikzpicture}[spy using outlines={rectangle, height=3cm,width=2.3cm, magnification=3, connect spies}]
	\begin{axis} [name=plot1,
		axis line style={latex-latex},
	    axis x line=left,
        axis y line=left,
    	xmin=0, xmax=1,
    	xlabel={$x/L$},
		ylabel={$\tau_w$},
    	ymin=0, ymax=0.04,
        clip mode=individual,
    	legend style={at={(0.97, 0.97)}, anchor=north east},
    	legend cell align={left},
    	style={font=\normalsize},
        % width=3*\axisdefaultheight,
        % height=\axisdefaultheight
        ]

        \addplot[color=black, style={thick} ]
        table[x=x, y=twb, col sep=comma]{./figs/data/bl_tw.csv};
        % \addlegendentry{Blasius};
                
        \addplot[color=black, style={very thin}, only marks, mark=o, mark options={scale=0.7}, mark repeat = 2, mark phase = 0]
        table[x=x, y=tw12, col sep=comma]{./figs/data/bl_tw.csv}; 
        % \addlegendentry{Boltzmann--BGK};
                
	\end{axis}
\end{tikzpicture}}}
        \subfloat[$N_v = 16^2$]{\adjustbox{width=0.33\linewidth, valign=b}{\begin{tikzpicture}[spy using outlines={rectangle, height=3cm,width=2.3cm, magnification=3, connect spies}]
	\begin{axis} [name=plot1,
		axis line style={latex-latex},
	    axis x line=left,
        axis y line=left,
    	xmin=0, xmax=1,
    	xlabel={$x/L$},
		ylabel={$\tau_w$},
    	ymin=0, ymax=0.04,
        clip mode=individual,
    	legend style={at={(0.97, 0.97)}, anchor=north east},
    	legend cell align={left},
    	style={font=\normalsize},
        % width=3*\axisdefaultheight,
        % height=\axisdefaultheight
        ]

        \addplot[color=black, style={thick} ]
        table[x=x, y=twb, col sep=comma]{./figs/data/bl_tw.csv};
        % \addlegendentry{Blasius};
                
        \addplot[color=black, style={very thin}, only marks, mark=o, mark options={scale=0.7}, mark repeat = 2, mark phase = 0]
        table[x=x, y=tw16, col sep=comma]{./figs/data/bl_tw.csv}; 
        % \addlegendentry{Boltzmann--BGK};
                
	\end{axis}
\end{tikzpicture}}}
        
        \caption{\label{fig:boundarylayer_tw} Comparison of wall shear stress profiles for the laminar boundary layer problem computed using the Boltzmann--BGK approach and a $\mathbb P_3$ approximation with varying velocity space resolution. }
    \end{figure}
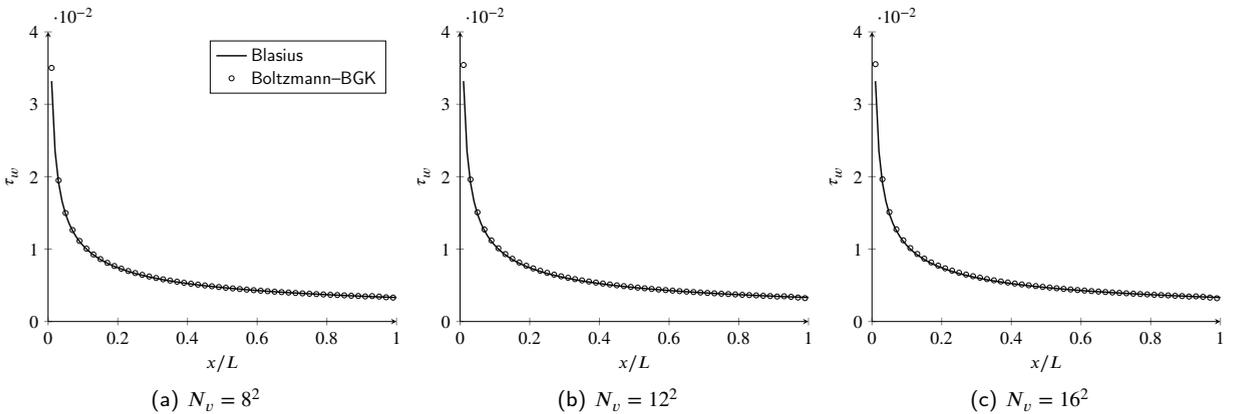

To explore the effects of the velocity space resolution on the ability of the Boltzmann--BGK approach in predicting momentum transfer effects in the continuum regime, a qualitative convergence study was performed for the laminar boundary layer. Three levels of resolution were chosen, corresponding to $N_v = 8^2$, $12^2$, and $16^2$ velocity nodes, the latter of which was the level of resolution used for the previous numerical experiment. A comparison of the normal velocity profiles at these velocity space resolution levels is shown in \cref{fig:boundarylayer_res}. It can be seen that the profiles were essentially identical down to as few as eight velocity nodes per dimension, which is consistent with the observations of the velocity space convergence study performed for the bent microchannel. Furthermore, a comparison of the streamwise distribution of the wall shear stress was performed, shown in \cref{fig:boundarylayer_tw}. As expected from the results of the normal velocity profiles, the predicted wall shear stress profiles showed excellent agreement with the Blasius prediction across all levels of velocity space resolution. These results indicate that, like in the rarefied flow case, accurate prediction of momentum transfer effects for flows in the continuum regime may be achieved with as few as eight velocity nodes per dimension, which opens up the possibility of simulating complex continuum flows via the Boltzmann--BGK at a reasonable computational cost.

\subsection{Transitional Taylor--Couette flow}
With the validation of the Boltzmann--BGK approach for basic wall-bounded continuum flows, the approach was then extended to wall-bounded transitional flows with three-dimensional hydrodynamic instabilities. The flow between concentric rotating cylinders, or Taylor--Couette flow \citep{Taylor1923}, presents an ideal test case due to its simple numerical setup yet complex flow physics, ranging from steady laminar flow to fully-developed turbulent flow. In particular, the numerical setup of \citet{Wang2021}, consisting of a perturbed laminar solution which transitions to turbulence, was chosen as it was proposed as a benchmark for unsteady scale-resolving simulations of wall-bounded transitional and turbulent flows. Furthermore, this case has the added benefit of presenting a proper validation of shear-induced transition to turbulence on curved walls.

The problem was solved on an annular domain $r \in [1,2]$, $\theta \in [0, 2\pi)$, $z \in [0, 2\pi]$, with the inner cylinder wall rotating at a fixed angular velocity $U_{\theta} = 1$ and the outer cylinder wall fixed. Along the $z$ direction, periodic boundary conditions were enforced, whereas for the inner and outer cylinder walls, diffuse wall boundary conditions were used. Per \citet{Wang2021}, the Reynolds number, based on the annulus width $d=1$, was set as $Re = 4000$ as these conditions yield rich flow physics with an initially laminar solution transitioning to fully-developed turbulent flow. The Mach number was set as $0.3$.

The initial conditions were set based on a perturbation of the steady incompressible laminar solution to the Taylor--Couette flow. In terms of the cylindrical coordinate system, these flow conditions are given as
    \begin{subequations}
    \begin{align}
    \rho &= \rho^*, \\
    u_{r} &= u_{r}^* + u_{r}', \\
    u_{\theta} &= u_{\theta}^* + u_{\theta}', \\
    u_{z} &= u_{z}^*,\\
    P &= P^* + P',
    \end{align}
    \end{subequations}
where the superscript ${\cdot}^*$ denotes the laminar solution and ${\cdot}'$ denotes the perturbation. The laminar solution can be given as
    \begin{subequations}
    \begin{align}
    \rho^* &= 1, \\
    u_{r}^* &= 0, \\
    u_{\theta}^* &= Ar + \frac{B}{r}, \\
    u_{z}^* &= 0,\\
    P^* &= \rho\left[\frac{1}{2}A^2 r^2 + 2 A B \log (r) - \frac{1}{2}\frac{B^2}{r^2}  \right] + C,
    \end{align}
    \end{subequations}
where $A = -1/3$ and $B = 4/3$ for the given geometry. The integration constant $C$ can be calculated by setting the reference pressure $P_0 = 1/(\gamma M^2)$ at the inner wall, i.e.,
\begin{equation}
    C = -\frac{1}{2}\rho \left(A^2 - B^2\right) + P_0.
\end{equation}
The boundary conditions at the wall (i.e., the velocity and temperature) are taken from the laminar solution. To drive the initially laminar flow to transition, a perturbation is superimposed upon the laminar solution profile. From \citet{Wang2021}, the perturbation takes on the form
    \begin{subequations}
    \begin{align}
    u_{r}' &= \epsilon \sin (\theta) \sin \left(\pi(r - 1) \right) \sin (z), \\
    u_{\theta}' &= \epsilon \cos (\theta) \sin \left(\pi(r - 1) \right) \sin (z), \\
    P' &= \frac{1}{2} \epsilon^2 \cos (2\theta)\sin \left(2\pi(r - 1) \right) \sin (2z),
    \end{align}
    \end{subequations}
where $\epsilon = 0.1$.

\begin{figure}[htbp!]
    \centering
    \subfloat[Baseline mesh] {
    \adjustbox{width=0.4\linewidth, valign=b}{\includegraphics[width=\textwidth]{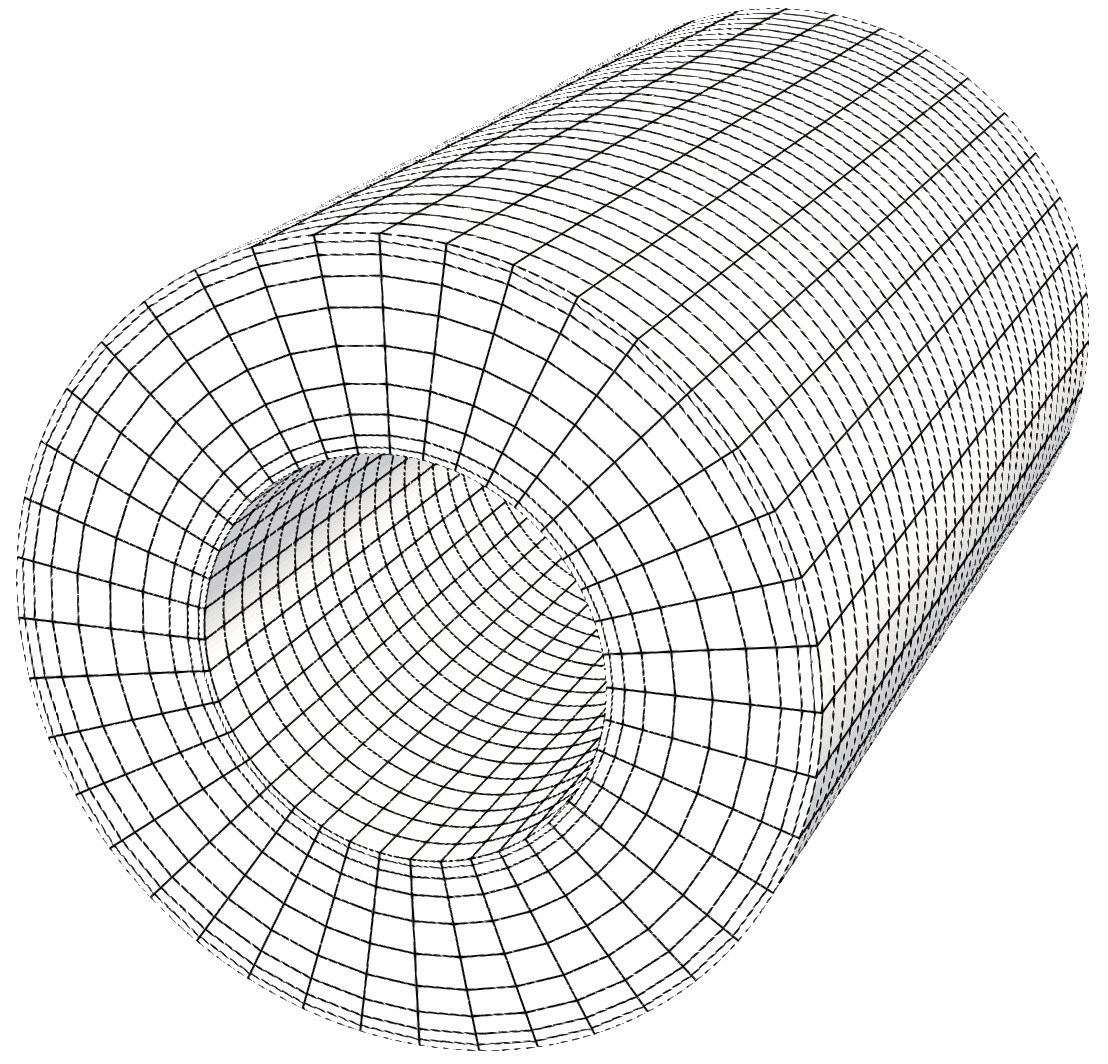}}}
    \hspace{0.1\textwidth}
    \subfloat[Reference mesh] {
    \adjustbox{width=0.4\linewidth, valign=b}{\includegraphics[width=\textwidth]{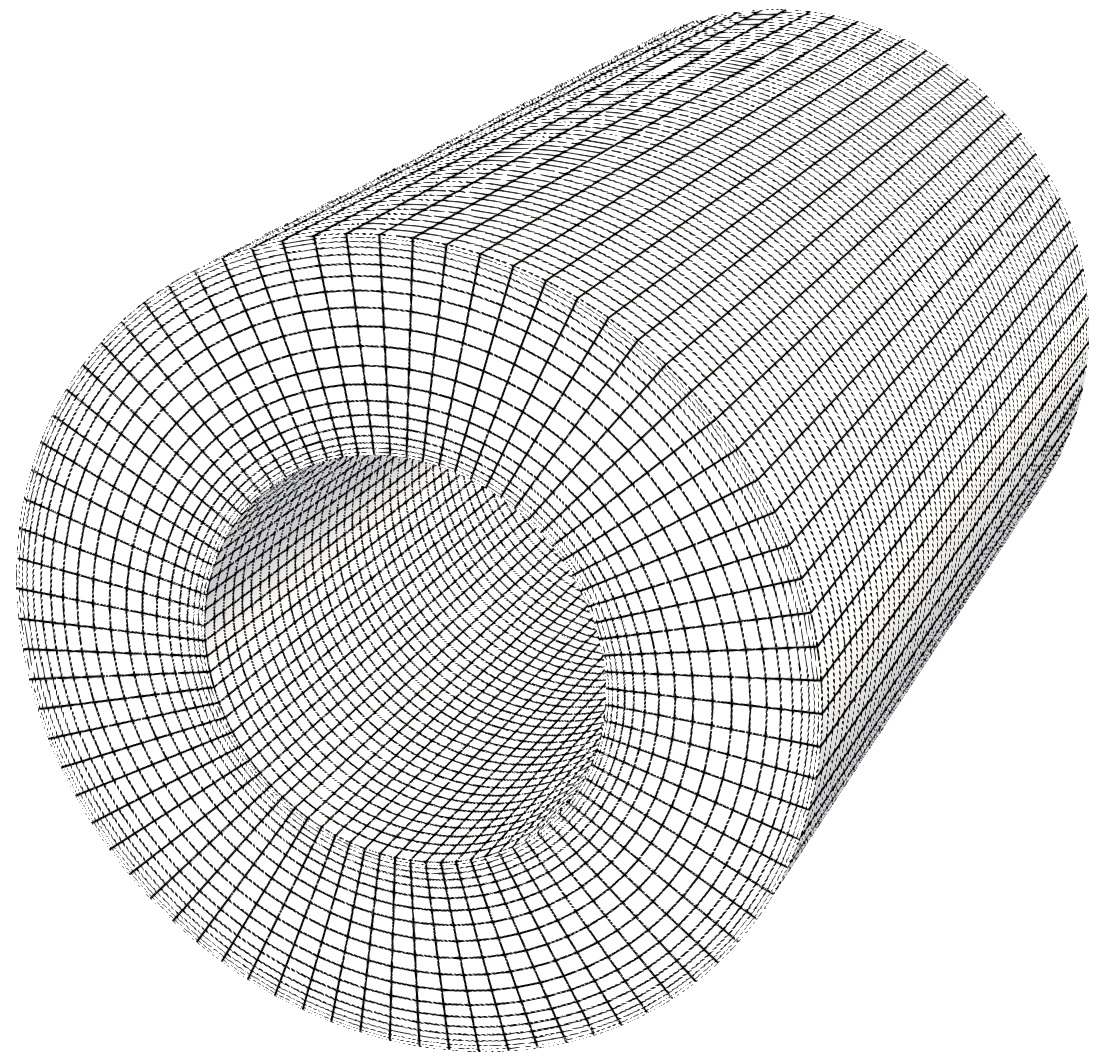}}}
    \caption{\label{fig:tc_mesh} Baseline (left) and reference (right) mesh for the Taylor--Couette flow case. }
\end{figure}

A baseline hexahedral mesh was generated using $N_e = 10 \times 40 \times 32$ elements along the radial, polar, and longitudinal directions, respectively, shown in \cref{fig:tc_mesh}. The wall normal spacing at the inner and outer walls was set as $\Delta r = 0.007$, and the curvature of the wall surfaces was approximated by cubic polynomials. A reference mesh was generated by uniformly subdividing the baseline mesh along each direction, resulting in $N_e = 20 \times 80 \times 64$ elements, also shown in \cref{fig:tc_mesh}. The problem was solved using a $\mathbb P_5$ approximation on the baseline mesh with $N_v = 16^3$ using both the Boltzmann--BGK approach and a standard Navier--Stokes approach, yielding approximately 11.3 billion degrees of freedom for the former. This level of spatial resolution was deemed by \citet{Wang2021} to be sufficient to adequately resolve the transition mechanisms in the flow. Furthermore, a reference solution was computed on the reference mesh using the Navier--Stokes equations with a $\mathbb P_5$ approximation. We remark here that for the baseline mesh, the Navier--Stokes approach diverged without anti-aliasing. The approximate computational cost per characteristic time was 560 GPU hours for the Boltzmann--BGK approach and 1.2 GPU hours for the Navier--Stokes approach, although the computational cost of the former was reduced to 44 GPU hours when the velocity space resolution was decreased to $N_v = 8^3$.

   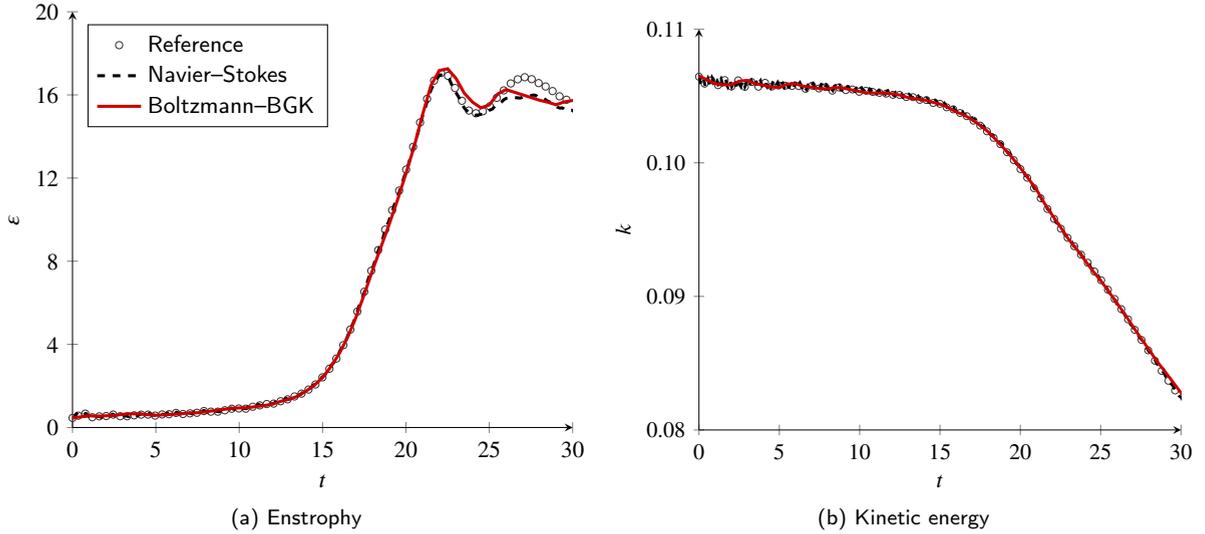
\begin{figure}[tbhp]
        \centering
        \subfloat[Enstrophy] {
        \adjustbox{width=0.48\linewidth, valign=b}{\begin{tikzpicture}[spy using outlines={rectangle, height=3cm,width=2.5cm, magnification=3, connect spies}]
    \begin{axis}
    [
        axis line style={latex-latex},
        axis y line=left,
        axis x line=left,
        clip mode=individual,
        xmode=linear, 
        ymode=linear,
        xlabel = {$t$},
        ylabel = {$\varepsilon$},
        xmin = 0, xmax = 30,
        ymin = 0, ymax = 20,
        legend cell align={left},
        legend style={at={(0.03, 0.97)}, anchor=north west},
        xtick = {0,5,10,15,20,25,30},
        ytick = {0,4,8,12,16,20},
        x tick label style={/pgf/number format/.cd, fixed, fixed zerofill, precision=0, /tikz/.cd},
        % y tick label style={/pgf/number format/.cd, fixed, fixed zerofill, precision=1, /tikz/.cd}
        % scale = 0.9
    ]
        
        \addplot[color=black!80, style={thin}, mark options={scale=0.7}, mark=o, only marks, mark repeat = 20, mark phase = 0] table[x=t, y=enst, col sep=comma]{./figs/data/taylorcouette_ref_enstrophy.csv};
        \addlegendentry{Reference};
        
        \addplot[color=black, style={very thick, dashed}] table[x=t, y=enst, col sep=comma]{./figs/data/taylorcouette_NS_p5_enstrophy.csv};
        \addlegendentry{Navier--Stokes};
        
        \addplot[ color=red!80!black, style={very thick},  mark options={scale=0.7}] table[x=t, y=enst, col sep=comma, mark=*]{./figs/data/taylorcouette_BGK8x8_p5_enstrophy.csv};
        \addlegendentry{Boltzmann--BGK};
        
    \end{axis}

\end{tikzpicture}}}
        \subfloat[Kinetic energy] {
        \adjustbox{width=0.48\linewidth, valign=b}{\begin{tikzpicture}[spy using outlines={rectangle, height=3cm,width=2.5cm, magnification=3, connect spies}]
    \begin{axis}
    [
        axis line style={latex-latex},
        axis y line=left,
        axis x line=left,
        clip mode=individual,
        xmode=linear, 
        ymode=linear,
        xlabel = {$t$},
        ylabel = {$k$},
        xmin = 0, xmax = 30,
        ymin = .08, ymax = .11,
        legend cell align={left},
        legend style={at={(0.03, 0.97)}, anchor=north west},
        xtick = {0,5,10,15,20,25,30},
        ytick = {.08,.09, .1,.11},
        x tick label style={/pgf/number format/.cd, fixed, fixed zerofill, precision=0, /tikz/.cd},
        y tick label style={/pgf/number format/.cd, fixed, fixed zerofill, precision=2, /tikz/.cd}
        % scale = 0.9
    ]
        
        \addplot[color=black!80, style={thin}, mark options={scale=0.7}, mark=o, only marks, mark repeat = 20, mark phase = 0] table[x=t, y=ke, col sep=comma]{./figs/data/taylorcouette_ref_ke.csv};
        % \addlegendentry{Reference};
        
        \addplot[color=black, style={very thick, dashed}] table[x=t, y=ke, col sep=comma]{./figs/data/taylorcouette_NS_p5_ke.csv};
        % % \addlegendentry{Navier--Stokes};
        
        \addplot[ color=red!80!black, style={very thick},  mark options={scale=0.7}] table[x=t, y=ke, col sep=comma, mark=*]{./figs/data/taylorcouette_BGK8x8_p5_ke.csv};
        % \addlegendentry{Boltzmann--BGK};
        
    \end{axis}

\end{tikzpicture}}}
        
        \caption{\label{fig:taylorcouette} Profiles of volume-averaged enstrophy (left) and kinetic energy (right) over time for the three-dimensional transitional Taylor--Couette flow problem computed with the Navier--Stokes equations (black, dashed) and the Boltzmann--BGK equation (red, solid) on the baseline mesh using a $\mathbb P_5$ approximation with $N_v = 16^3$. Reference solution shown with black markers.}
    \end{figure}

The quantities of interest in the flow were the volume-averaged enstrophy and kinetic energy, computed as
\begin{equation}
    \varepsilon = \frac{1}{|\Omega^{\mathbf{x}}|} \int_{\Omega^{\mathbf{x}}} \rho \boldsymbol{\omega}{\cdot}\boldsymbol{\omega}\ \mathrm{d}V,
\end{equation}
and
\begin{equation}
    k = \frac{1}{|\Omega^{\mathbf{x}}|} \int_{\Omega^{\mathbf{x}}} \frac{1}{2} \rho \mathbf{U}{\cdot}\mathbf{U}\ \mathrm{d}V,
\end{equation}
respectively, where $\boldsymbol{\omega} = \boldsymbol{\nabla} \times \mathbf{U}$ is the vorticity and $|\Omega^{\mathbf{x}}| = 6 \pi^2$ is the domain volume. The enstrophy and kinetic energy over time as computed by the Boltzmann--BGK approach and Navier--Stokes approach on the baseline mesh are shown in \cref{fig:taylorcouette} in comparison to the reference results. For the given resolution, the Boltzmann--BGK approach showed excellent agreement with the Navier--Stokes approach, with both the rapid rise in enstrophy and associated decay in kinetic energy dissipation rate well-predicted up to and around the enstrophy peak at $t = 22$. Near the peak, the Boltzmann--BGK approach showed marginally higher enstrophy than the Navier--Stokes approach and reference results. After the peak, both methods showed marginally less enstrophy than the reference results which can most likely be attributed to the underresolution of the now turbulent flow.

      \begin{figure}[htbp!]
        \centering
        \subfloat[Reference] {
        \adjustbox{width=0.357\linewidth,valign=b}{\includegraphics[width=\textwidth]{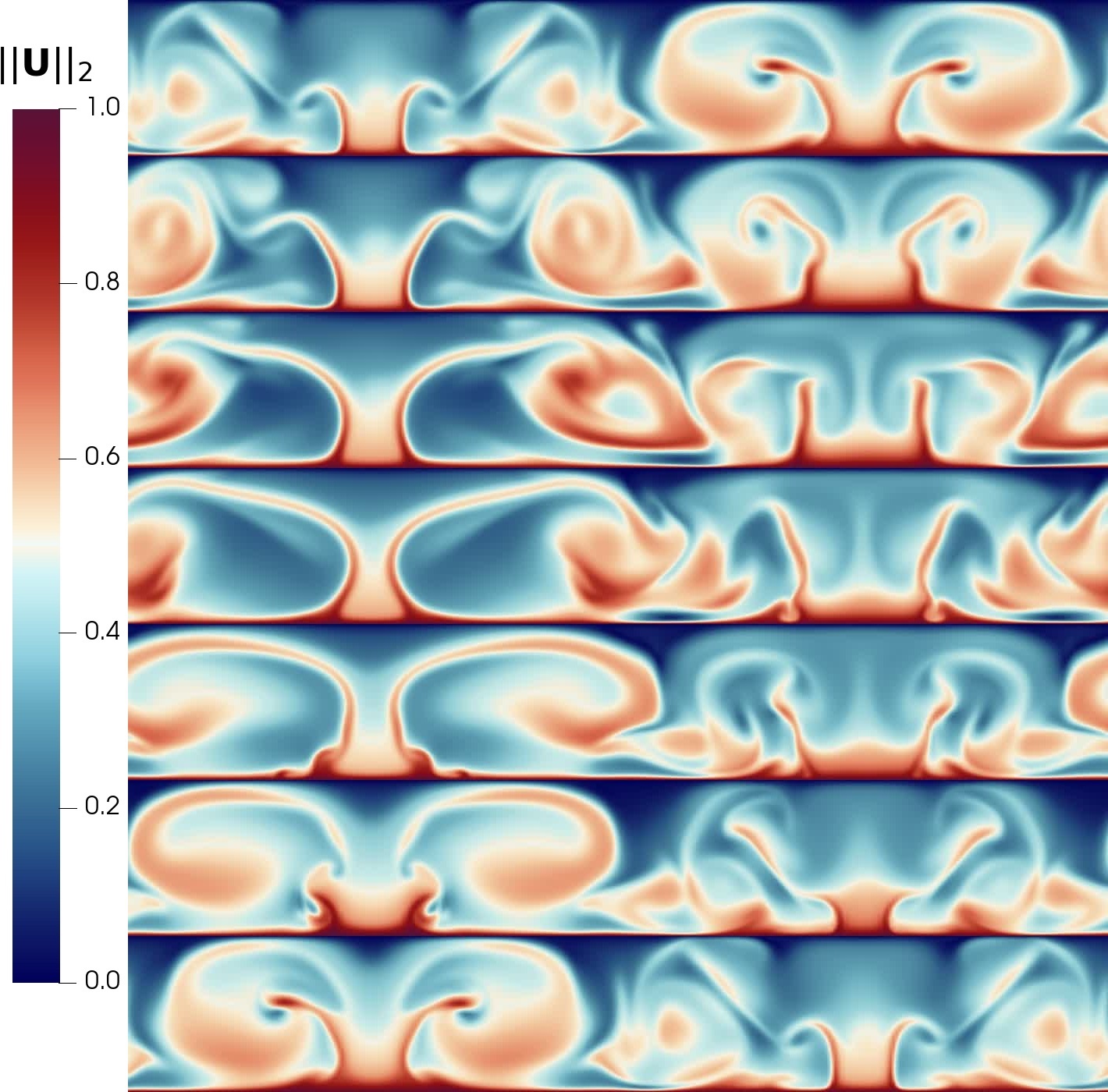}}}
        \subfloat[Navier--Stokes] {
        \adjustbox{width=0.315\linewidth,valign=b}{\includegraphics[width=\textwidth]{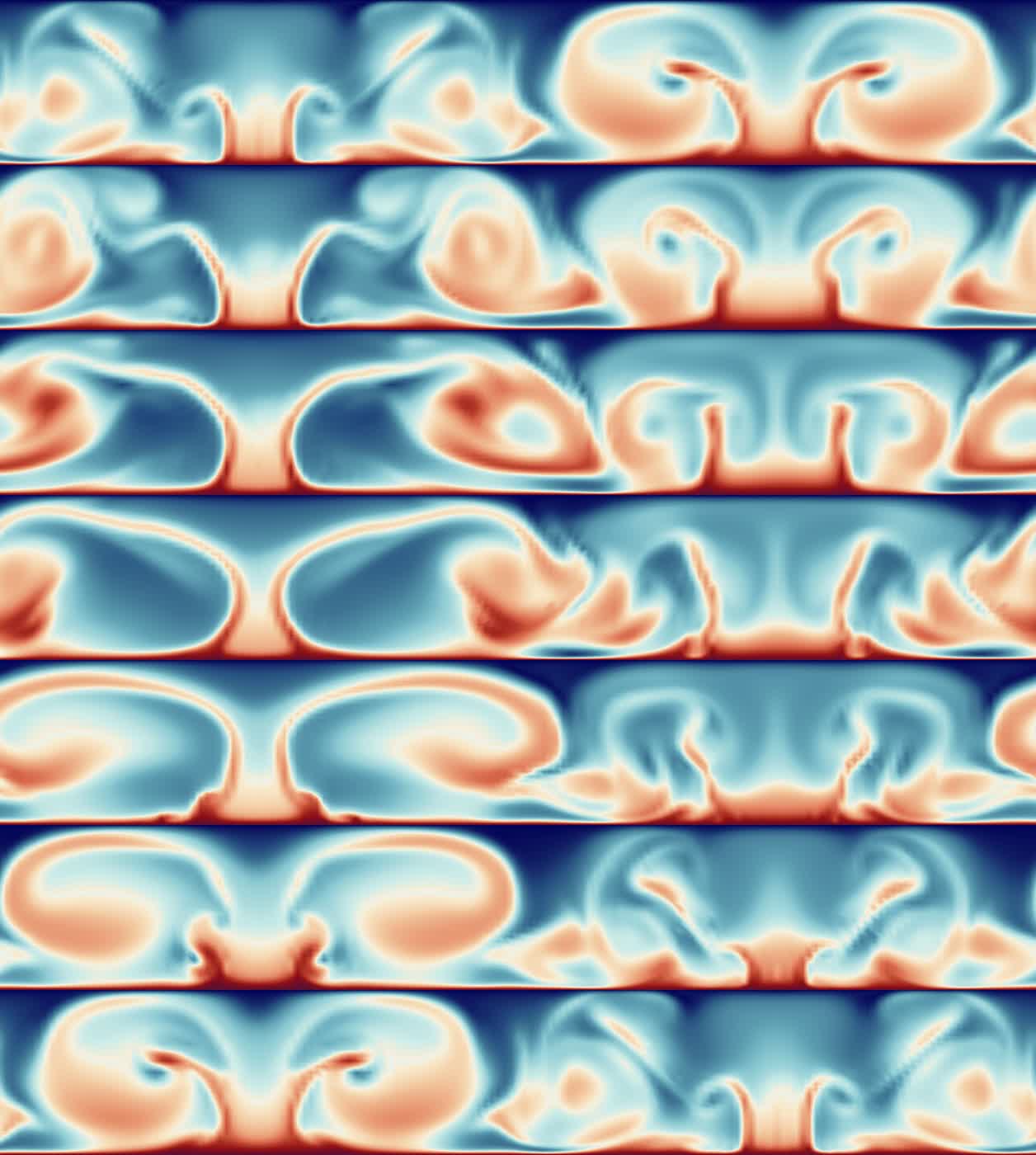}}}
        \subfloat[Boltzmann--BGK ($N_v = 16^3$)] {
        \adjustbox{width=0.315\linewidth,valign=b}{\includegraphics[width=\textwidth]{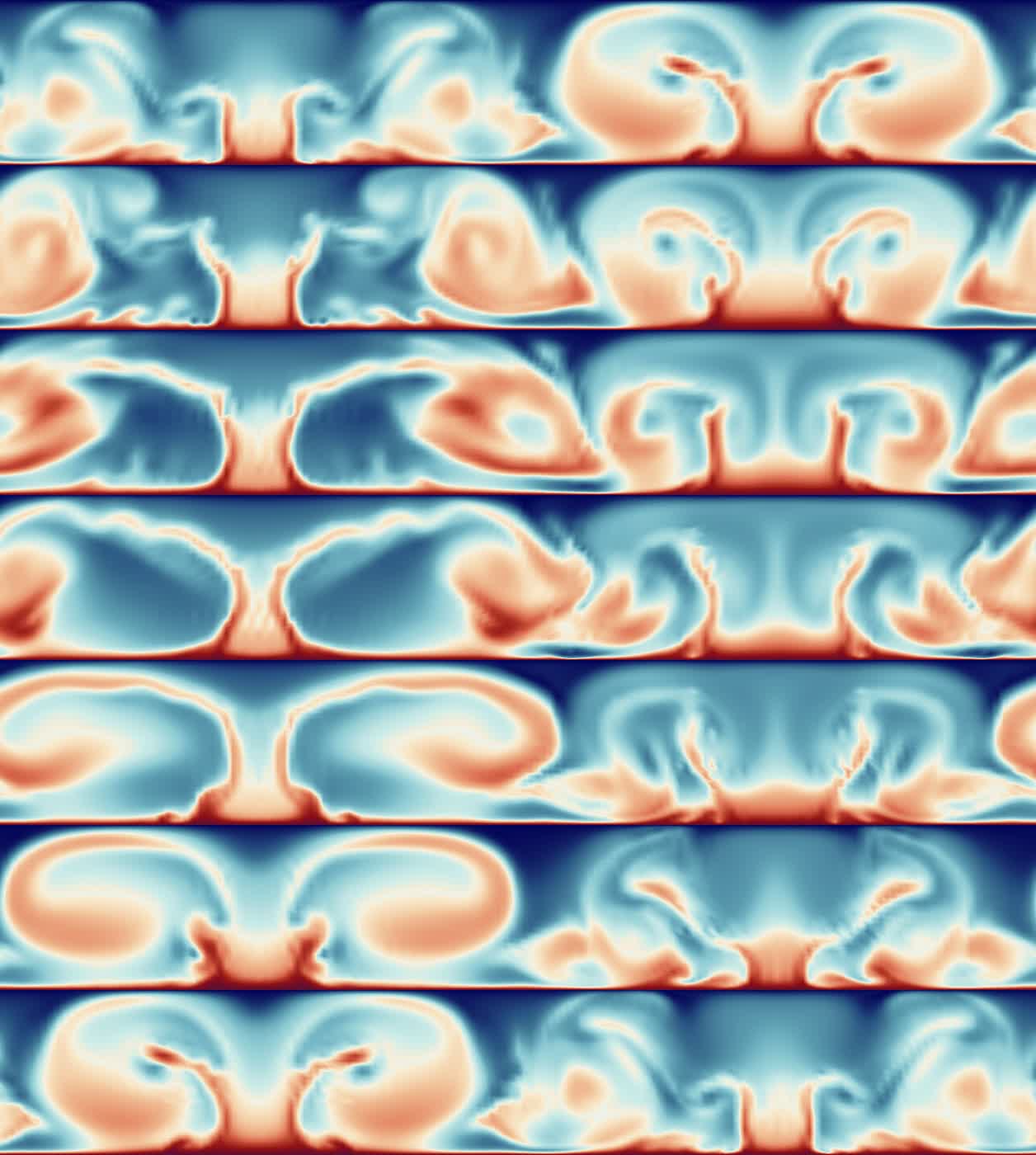}}}
        \caption{\label{fig:taylorcouette_velocity} Contours of velocity magnitude for the three-dimensional transitional Taylor--Couette flow problem at $t = 22$
        on angularly equispaced cross-sections from $\theta = 0$ (top) to $\theta = \pi$ (bottom). Results computed with the Navier--Stokes equations (middle) and the Boltzmann--BGK equation (right) on the baseline mesh using a $\mathbb P_5$ approximation with $N_v = 16^3$. Reference solution shown on left.}
    \end{figure}
      \begin{figure}[htbp!]
        \centering
        \subfloat[Reference] {
        \adjustbox{width=0.357\linewidth,valign=b}{\includegraphics[width=\textwidth]{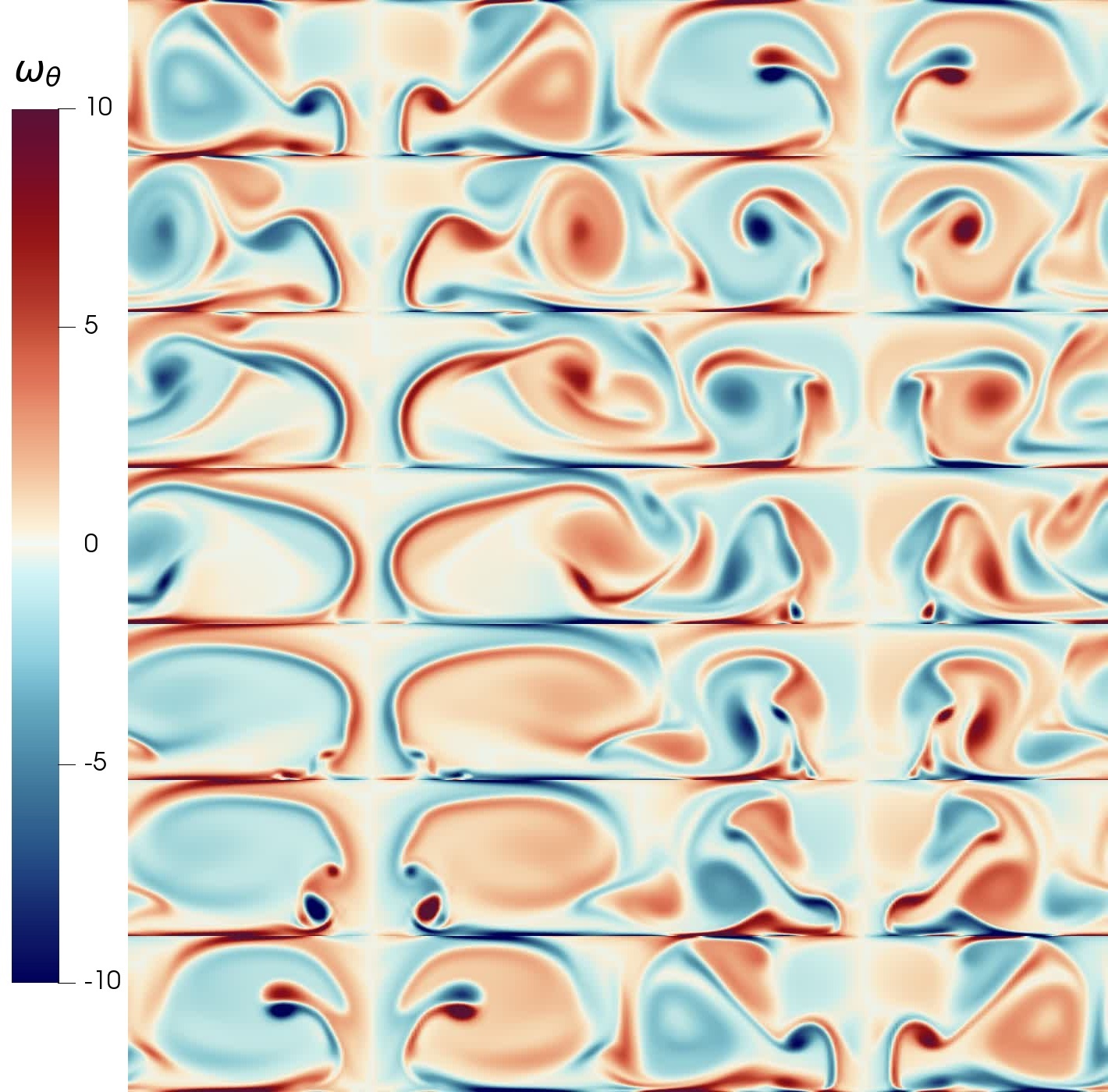}}}
        \subfloat[Navier--Stokes] {
        \adjustbox{width=0.315\linewidth,valign=b}{\includegraphics[width=\textwidth]{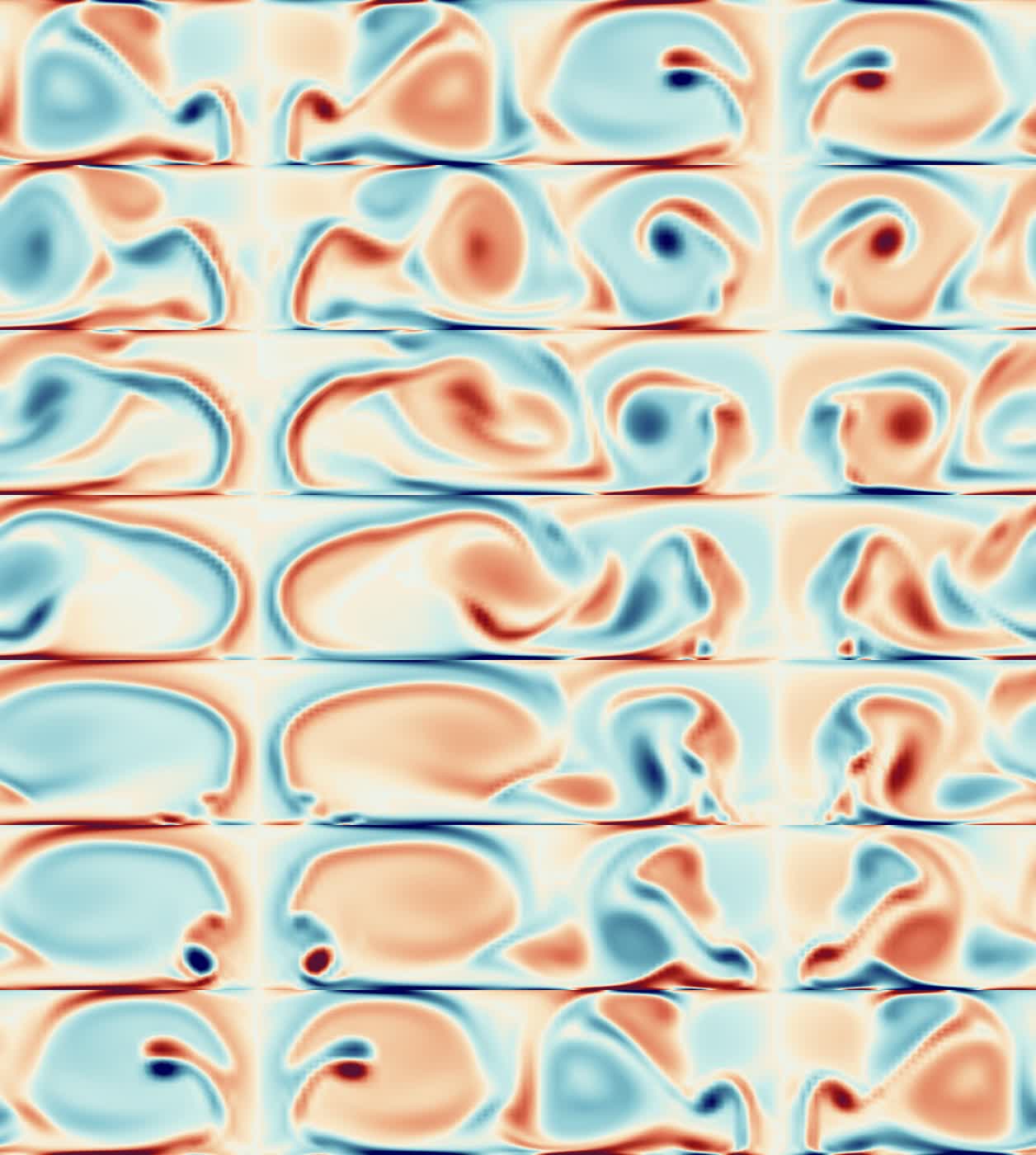}}}
        \subfloat[Boltzmann--BGK ($N_v = 16^3$)] {
        \adjustbox{width=0.315\linewidth,valign=b}{\includegraphics[width=\textwidth]{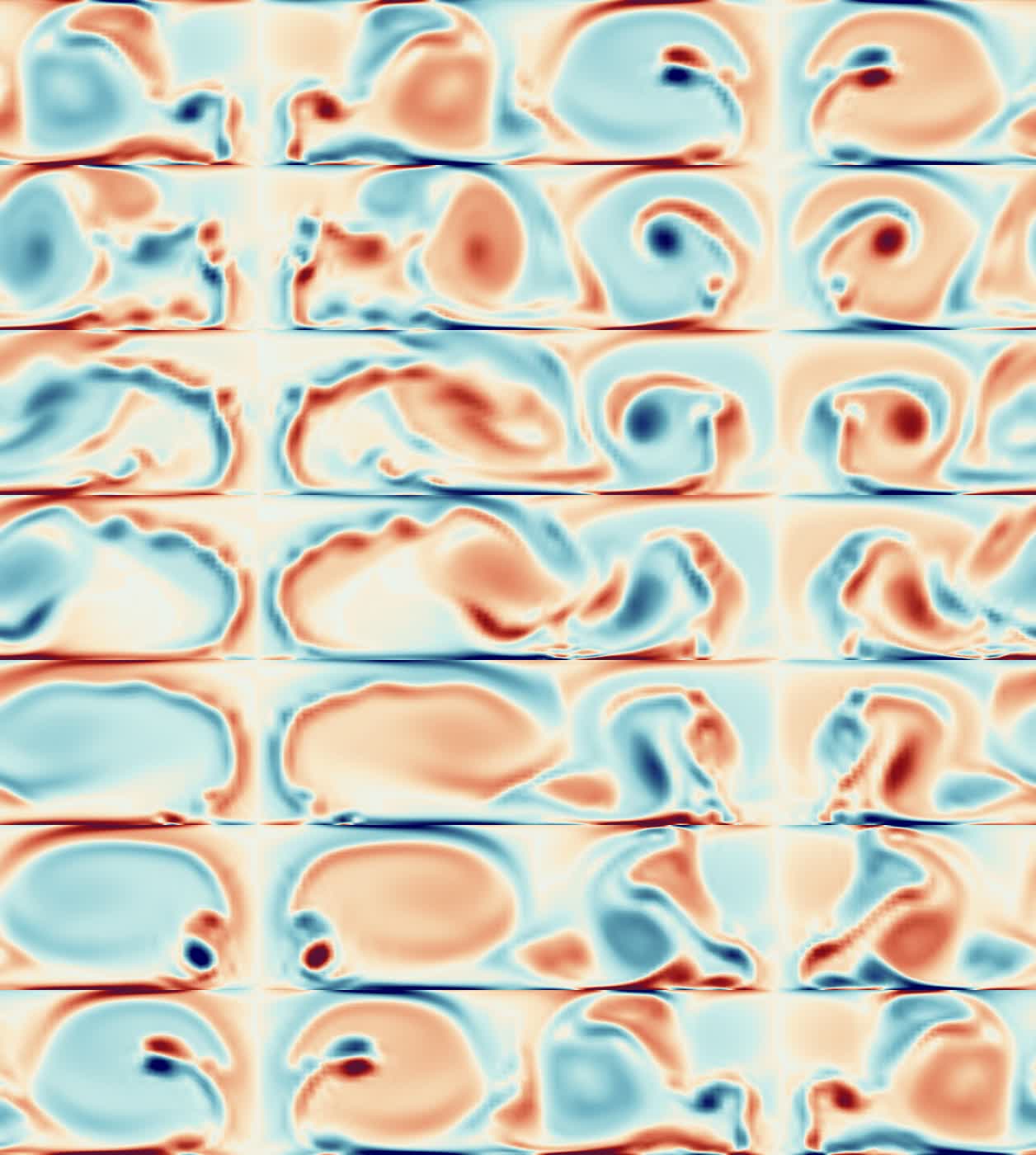}}}
        \caption{\label{fig:taylorcouette_vorticity} Contours of vorticity along the cross-section normal direction for the three-dimensional transitional Taylor--Couette flow problem at $t = 22$ on angularly equispaced cross-sections from $\theta = 0$ (top) to $\theta = \pi$ (bottom). Results computed with the Navier--Stokes equations (middle) and the Boltzmann--BGK equation (right) on the baseline mesh using a $\mathbb P_5$ approximation with $N_v = 16^3$. Reference solution shown on left.}
    \end{figure}

The better analyze the ability of the Boltzmann--BGK approach in predicting the dominant structures in the flow, the velocity and vorticity profiles were extracted at various angularly equispaced cross-sections. The contours of velocity magnitude and vorticity (aligned with the cross-section normal direction) computed by the Boltzmann--BGK and Navier--Stokes approaches on the baseline mesh are shown in \cref{fig:taylorcouette_velocity} and \cref{fig:taylorcouette_vorticity}, respectively, in comparison to the reference results. It can be seen that the dominant flow structures are well predicted by the Boltzmann--BGK approach, with the rollup of the vortices in the flow showing good agreement with the Navier--Stokes results on the same mesh and the reference results. However, whereas the Navier--Stokes results were visually very similar to the reference results for the given resolution, the Boltzmann--BGK results showed some notable differences. Particularly, the rollup of the shear layer showed signs of instabilities that were not present in the Navier--Stokes results or the reference results. If these results were shown to be converged with respect to both the spatial and velocity domains, this would indicate that the Boltzmann--BGK approach does not accurately predict hydrodynamic instabilities, which could be attributed to the exponential nature of the BGK approximation in comparison to the quadratic nature of the true collision operator. We therefore attempt to deduce the underlying numerical mechanisms behind these discrepancies. 

      \begin{figure}[htbp!]
        \centering
        \subfloat[Reference] {
        \adjustbox{width=0.357\linewidth,valign=b}{\includegraphics[width=\textwidth]{figs/taylorcouette_ref_velocity.jpg}}}
        \subfloat[Boltzmann--BGK ($N_v = 8^3$)] {
        \adjustbox{width=0.315\linewidth,valign=b}{\includegraphics[width=\textwidth]{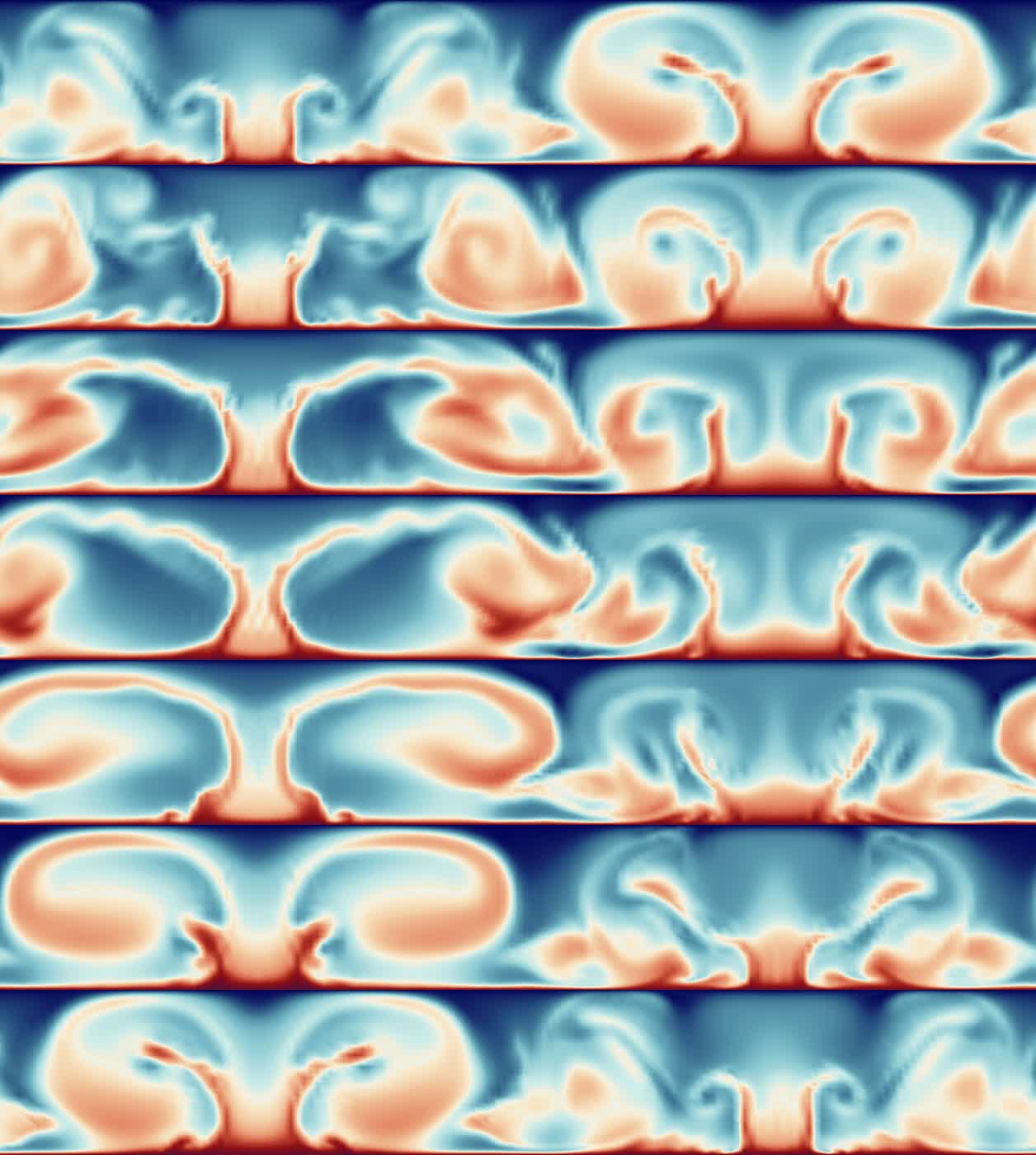}}}
        \subfloat[Boltzmann--BGK ($N_v = 16^3$)] {
        \adjustbox{width=0.315\linewidth,valign=b}{\includegraphics[width=\textwidth]{figs/taylorcouette_BGK16_velocity.jpg}}}
        \caption{\label{fig:taylorcouette_velocity_nv_comp} Contours of velocity magnitude for the three-dimensional transitional Taylor--Couette flow problem at $t = 22$ on angularly equispaced cross-sections from $\theta = 0$ (top) to $\theta = \pi$ (bottom). Results computed with Boltzmann--BGK equation on the baseline mesh using a $\mathbb P_5$ approximation with $N_v = 8^3$ (middle) and $N_v = 16^3$ (right). Reference solution shown on left.}
    \end{figure}
      \begin{figure}[htbp!]
        \centering
        \subfloat[Reference] {
        \adjustbox{width=0.357\linewidth,valign=b}{\includegraphics[width=\textwidth]{figs/taylorcouette_ref_vorticity.jpg}}}
        \subfloat[Boltzmann--BGK ($N_v = 8^3$)] {
        \adjustbox{width=0.315\linewidth,valign=b}{\includegraphics[width=\textwidth]{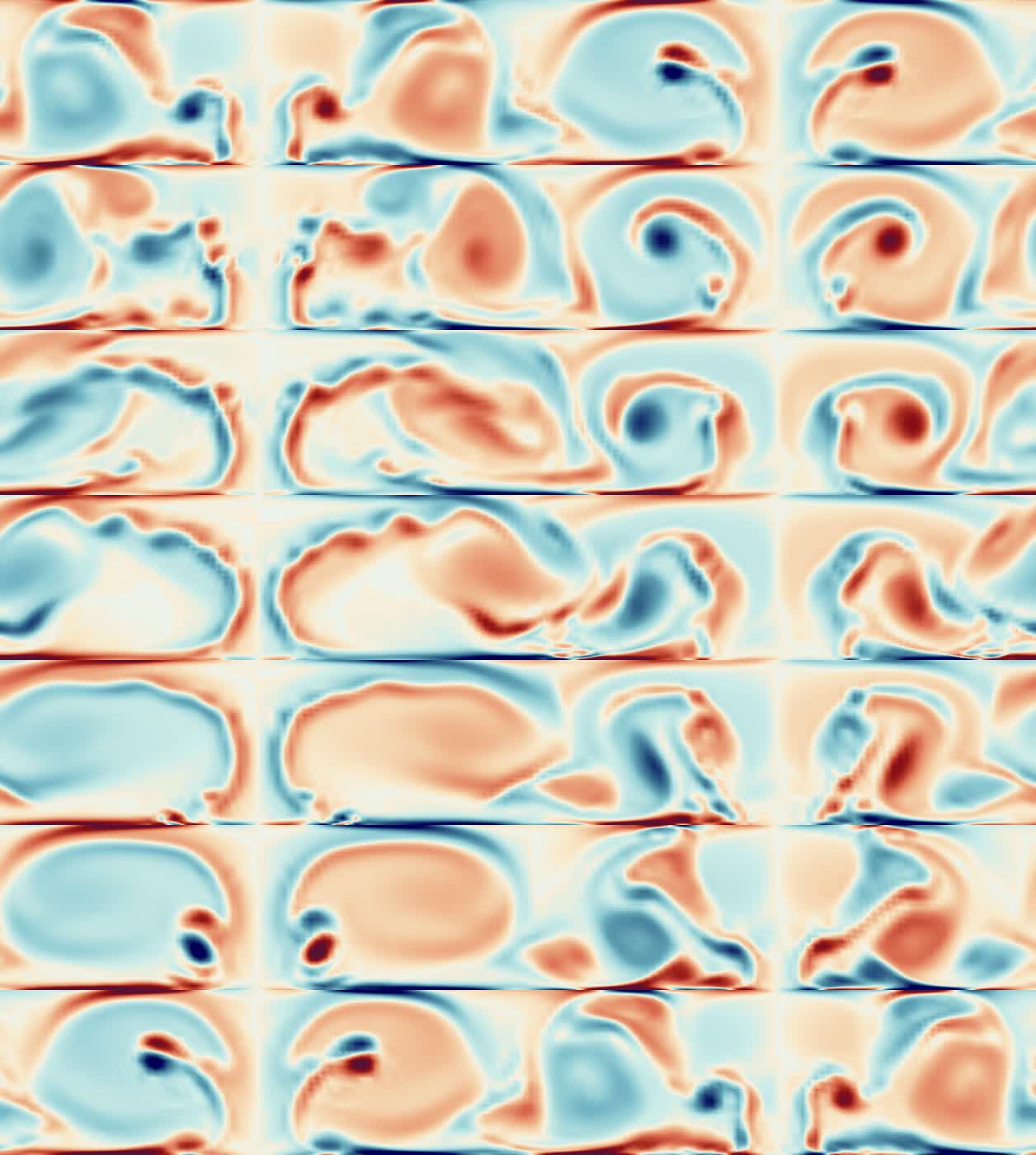}}}
        \subfloat[Boltzmann--BGK ($N_v = 16^3$)] {
        \adjustbox{width=0.315\linewidth,valign=b}{\includegraphics[width=\textwidth]{figs/taylorcouette_BGK16_vorticity.jpg}}}
        \caption{\label{fig:taylorcouette_vorticity_nv_comp}
        Contours of vorticity along the cross-section normal direction for the three-dimensional transitional Taylor--Couette flow problem at $t = 22$ on angularly equispaced cross-sections from $\theta = 0$ (top) to $\theta = \pi$ (bottom). Results computed with Boltzmann--BGK equation on the baseline mesh using a $\mathbb P_5$ approximation with $N_v = 8^3$ (middle) and $N_v = 16^3$ (right). Reference solution shown on left. }
    \end{figure}

The most intuitive explanation for why the Boltzmann--BGK approach does not predict the flow physics in a consistent manner with the Navier--Stokes approach \textit{on the same spatial mesh} would be that the velocity space is not sufficiently resolved. Given the velocity space resolution of the numerical experiments, $N_v = 16^3$, this would be in sharp contrast to the observations for the required level of resolution for steady rarefied and continuum flows, indicating that much higher resolution is required to simulate the more complex unsteady flow physics. To assess this claim, the numerical experiment was repeated using a different level of velocity space resolution, $N_v = 8^3$, which corresponds to the level of resolution deemed to be sufficient for the previously simulated steady flows. A comparison of the predicted velocity magnitude contours and vorticity contours between the two levels of velocity space resolution is shown in \cref{fig:taylorcouette_velocity_nv_comp} and \cref{fig:taylorcouette_vorticity_nv_comp}, respectively. It can be seen that the two results were visually indistinguishable, both showing the same exact instabilities across the shear layer that are not present in the reference results. Therefore, it is sufficient to say that the presented results are converged in velocity space, and that the discrepancies between the Navier--Stokes approximation and the Boltzmann--BGK approximation are not as a result of a lack of resolution in the velocity domain. Furthermore, these results also indicate that the level of velocity space resolution needed to be converged in velocity space does not differ much between steady rarefied/continuum flows and unsteady transitional/turbulent flows.

      \begin{figure}[htbp!]
        \centering
        \subfloat[Reference] {
        \adjustbox{width=0.357\linewidth,valign=b}{\includegraphics[width=\textwidth]{figs/taylorcouette_ref_velocity.jpg}}}
        \subfloat[Boltzmann--BGK ($N_e = 10 \times 40 \times 32$)] {
        \adjustbox{width=0.315\linewidth,valign=b}{\includegraphics[width=\textwidth]{figs/taylorcouette_BGK8_velocity.jpg}}}
        \subfloat[Boltzmann--BGK ($N_e = 15 \times 60 \times 48$)] {
        \adjustbox{width=0.315\linewidth,valign=b}{\includegraphics[width=\textwidth]{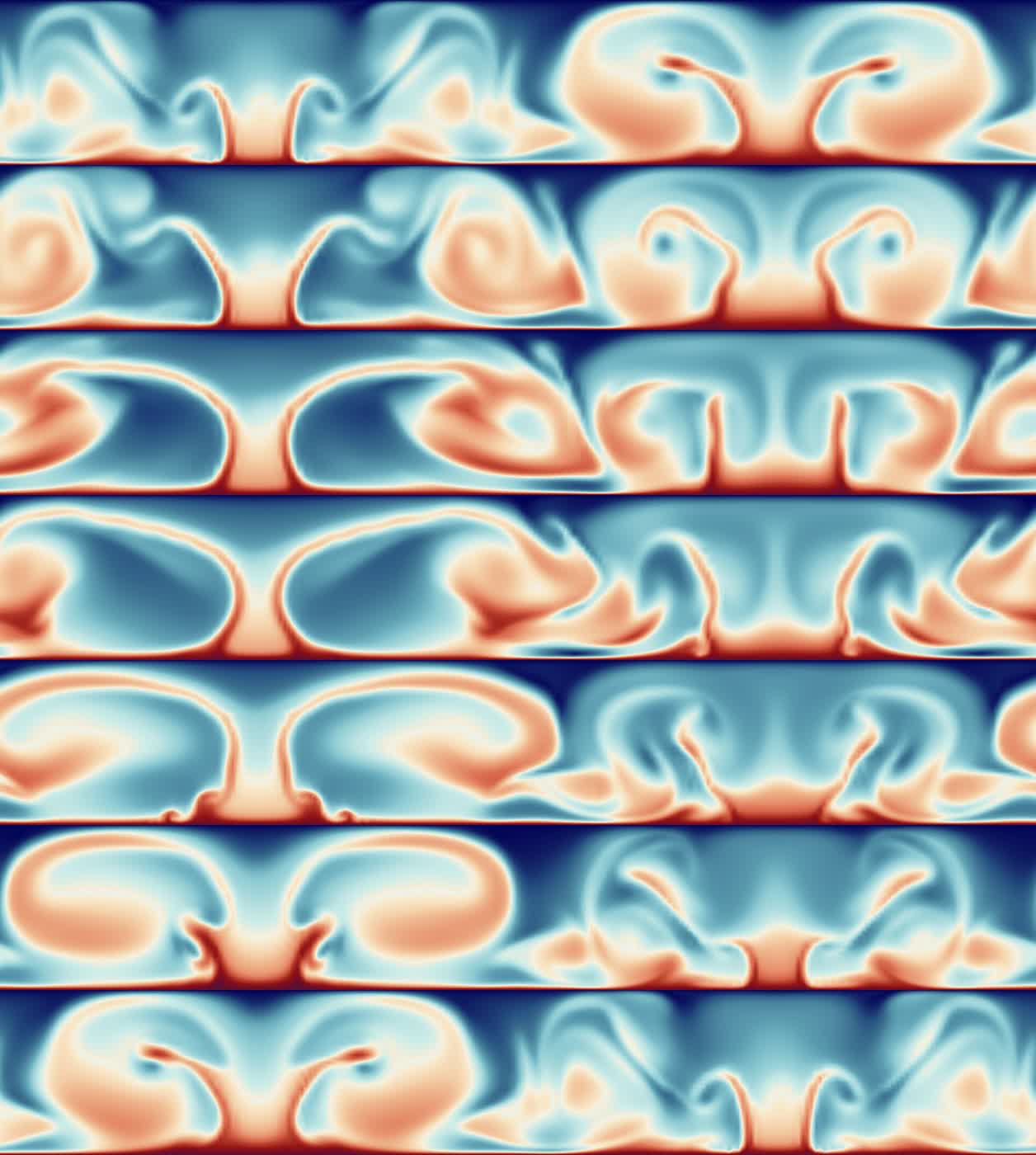}}}
        \caption{\label{fig:taylorcouette_velocity_nx_comp} Contours of velocity magnitude for the three-dimensional transitional Taylor--Couette flow problem at $t = 22$ on angularly equispaced cross-sections from $\theta = 0$ (top) to $\theta = \pi$ (bottom). Results computed with Boltzmann--BGK equation on the baseline mesh (middle) and refined mesh (right) using a $\mathbb P_5$ approximation with $N_v = 8^3$. Reference solution shown on left.}
    \end{figure}
      \begin{figure}[htbp!]
        \centering
        \subfloat[Reference] {
        \adjustbox{width=0.357\linewidth,valign=b}{\includegraphics[width=\textwidth]{figs/taylorcouette_ref_vorticity.jpg}}}
        \subfloat[Boltzmann--BGK ($N_e = 10 \times 40 \times 32$)] {
        \adjustbox{width=0.315\linewidth,valign=b}{\includegraphics[width=\textwidth]{figs/taylorcouette_BGK8_vorticity.jpg}}}
        \subfloat[Boltzmann--BGK ($N_e = 15 \times 60 \times 48$)] {
        \adjustbox{width=0.315\linewidth,valign=b}{\includegraphics[width=\textwidth]{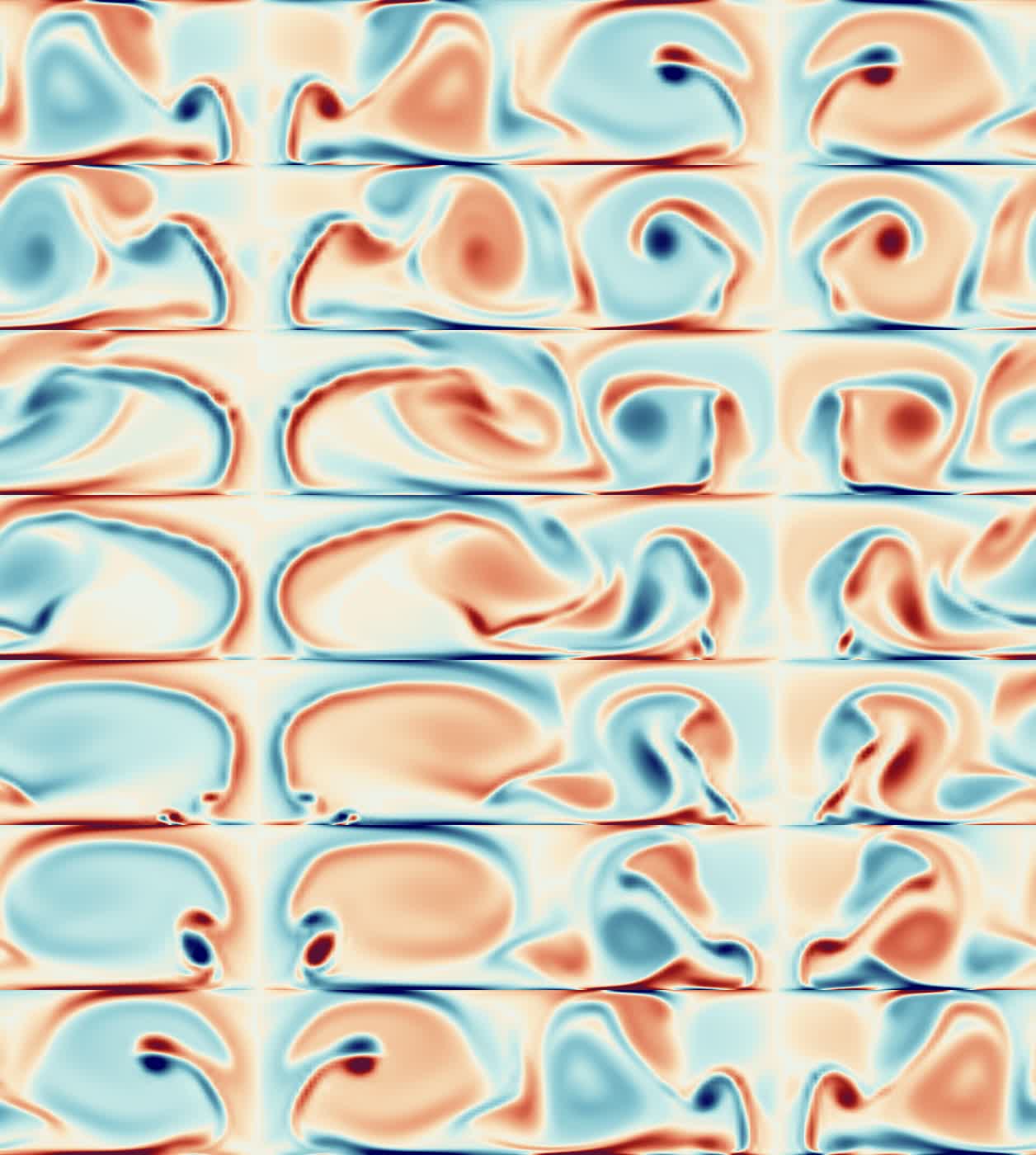}}}
        \caption{\label{fig:taylorcouette_vorticity_nx_comp} Contours of vorticity along the cross-section normal direction for the three-dimensional transitional Taylor--Couette flow problem at $t = 22$ on angularly equispaced cross-sections from $\theta = 0$ (top) to $\theta = \pi$ (bottom). Results computed with Boltzmann--BGK equation on the baseline mesh (middle) and refined mesh (right) using a $\mathbb P_5$ approximation with $N_v = 8^3$. Reference solution shown on left.}
    \end{figure}

Another possible explanation for the discrepancies could be attributed to underresolution in the spatial domain. We remark again that the Navier--Stokes simulations on the baseline mesh were unstable without anti-aliasing, diverging shortly after the enstrophy peak at $t=22$, indicating that the flow is necessarily underresolved on the baseline mesh. It was observed by the authors that when anti-aliasing was disabled for the Navier--Stokes approach on the baseline mesh, the results also showed instabilities in the flow prior to the simulation diverging. To verify whether the discrepancies between the Boltzmann--BGK approach and Navier--Stokes approach could be attributed to inadequate modeling of the transport term in the Boltzmann equation (i.e., underresolved spatial domain), a refined mesh was generated with $N_e = 15 \times 60 \times 48$ elements. A comparison of the predicted velocity magnitude contours and vorticity contours between the baseline mesh and refined mesh is shown in \cref{fig:taylorcouette_velocity_nx_comp} and \cref{fig:taylorcouette_vorticity_nx_comp}, respectively. It can be seen that the increased spatial resolution significantly reduces the discrepancies between the Boltzmann--BGK results and the reference results, with very good agreement in the prediction of the structures including the shear layers. These observations indicate a very interesting characteristic of numerical approximations of the Boltzmann--BGK equation: \textit{to accurately predict hydrodynamic instabilities consistently with the Navier--Stokes equations, it is much more important to accurately resolve particle transport than particle collision,} i.e., it is more important to have a highly-resolved spatial domain than a highly-resolved velocity domain. These findings suggest that highly-accurate spatial schemes such as high-order methods could be very beneficial for performing scale-resolving simulations via the Boltzmann--BGK approach. 

\subsection{SD7003 at Re = 60,000}
As a final test case for the Boltzmann--BGK approach for complex fluid flows, the flow around an SD7003 airfoil at a Reynolds number of $Re = 60,000$ and angle of attack of $\alpha = 8^{\circ}$ was simulated. At these conditions, a variety of complex flow phenomena can be observed, including the laminar separation of the flow around the curved leading edge, the subsequent transition of the laminar shear layer to turbulence, and the reattachment of the turbulent shear layer which develops into a turbulent boundary layer and wake \citep{Garmann2012}. It can be notably difficult to accurately predict these flow features as they can be highly sensitive to the flow conditions and numerical resolution, and as such, this case has been used as a benchmark for scale-resolving simulations in many works \citep{Garmann2012,Cox2021,Beck2014,Visbal2009, Galbraith2010, Vermeire2017}.

\begin{figure}[htbp!]
        \centering
        \adjustbox{width=0.7\linewidth, valign=b}{\includegraphics[width=\textwidth]{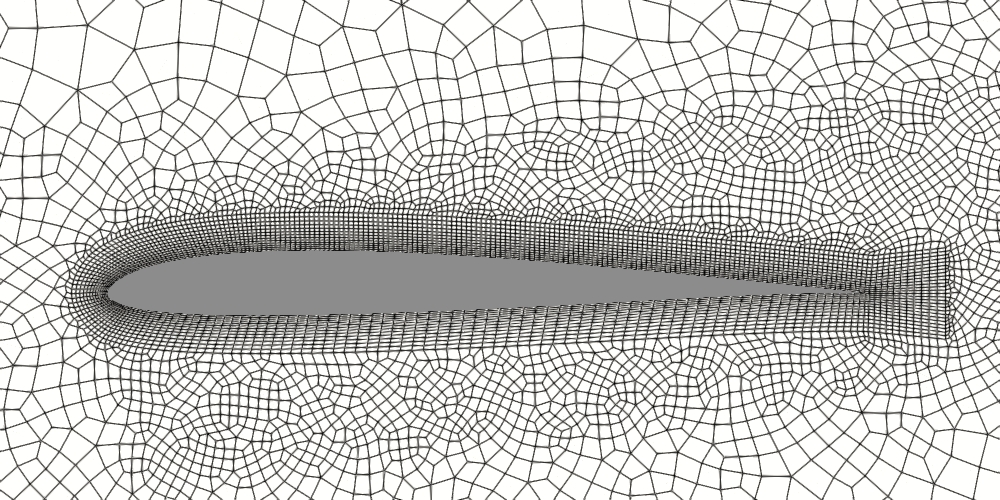}}
        \caption{\label{fig:sd7003_mesh} Cross-section of the near-airfoil region of the mesh used for the $Re = 60,000$ SD7003 airfoil case. }
    \end{figure}

The problem was solved using an identical domain and mesh as the work of \citet{Vermeire2017}. The domain extent was set as $\Omega^{\mathbf{x}} = [-10c, 20c] \times [-10c, 10c] \times [0, 0.2c]$, where $c$ is the chord length. An unstructured hexahedral mesh was used, shown in \cref{fig:sd7003_mesh}, with approximately $1.4{\cdot}10^5$ elements and a near-wall normal mesh spacing of $\Delta y_n \approx 2{\cdot}10^{-3}$. The curvature of the airfoil surface was represented with quadratic polynomials. At the farfield, Dirichlet boundary conditions corresponding to freestream conditions of $Re = 60,000$ and $M = 0.3$ were used. At the airfoil surface, diffuse wall boundary conditions were used with the temperature set identically to the freestream, and along the spanwise direction, periodicity was enforced.

A comparison was made between the Boltzmann--BGK approach and a standard Navier--Stokes approach using $\mathbb P_3$ and $\mathbb P_4$ approximations. We remark here that since this case is quite sensitive, particularly so at this Mach number \citep{Yuan2005}, and there is a fair amount of variation in the reported results in the literature, we use the results of the Navier--Stokes approach only for comparison and not necessarily as a reference result. Given the observations drawn from the simulation of the Taylor--Couette flow, a velocity space resolution of $N_v = 8^3$ was deemed sufficient for the Boltzmann--BGK approach. The resulting total number of degrees of freedom for the $\mathbb P_3$ and $\mathbb P_4$ approximations was $4.5$ billion and $8.8$ billion, respectively. For the $\mathbb P_4$ case, the approximate computational cost of the Boltzmann--BGK approach was 500 GPU hours per flow over chord (computed across 32 NVIDIA V100 GPUs) in comparison to 8 GPU hours for the Navier--Stokes approach. To initially develop the flow, the simulations were impulsively started at $\mathbb P_1$ and advanced until a characteristic time of $t = 10$, nondimensionalized by the freestream velocity and chord length, after which the approximation order was set to the operating conditions. At $t = 20$, the flow was assumed to be well-developed and the gathering of the flow statistics was performed over the range $t \in [20, 30]$. For the averaged quantities to be presented, both temporal and spanwise averaging was performed. 

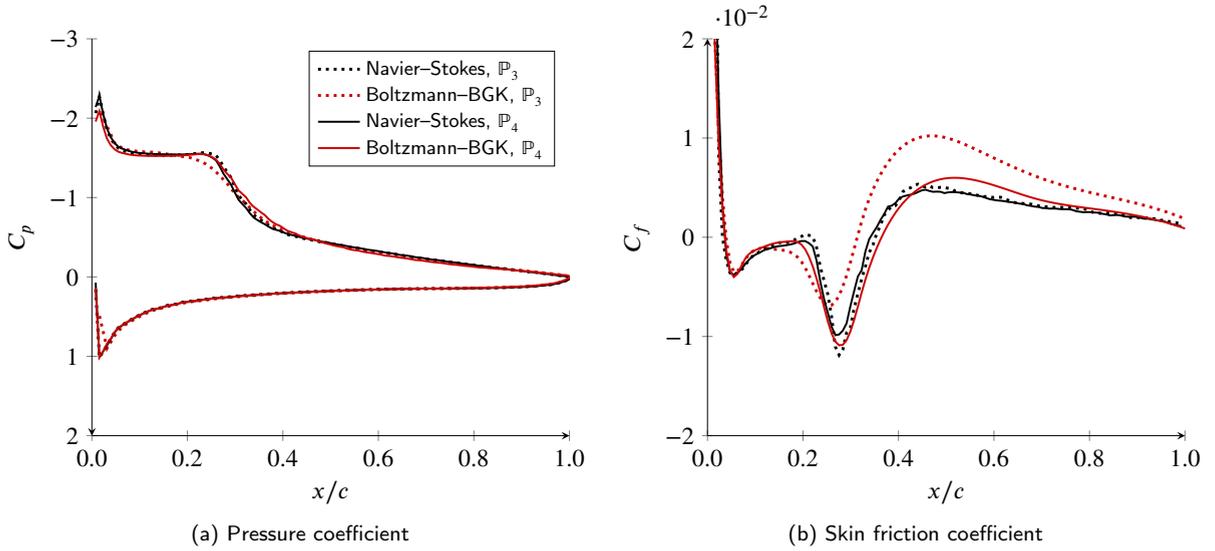
\begin{figure}[htbp!]
        \centering
        \subfloat[Pressure coefficient]{\adjustbox{width=0.48\linewidth, valign=b}{    \begin{tikzpicture}[spy using outlines={rectangle, height=3cm,width=2.3cm, magnification=3, connect spies}]
		\begin{axis}[name=plot1,
		    axis line style={latex-latex},
		    axis x line=left,
            axis y line=left,
            clip mode=individual,
		    xlabel={$x/c$},
		    xtick={0,0.2,0.4,0.6,0.8,1},
    		xmin=0,
    		xmax=1,
    	    x tick label style={
        		/pgf/number format/.cd,
            	fixed,
            	fixed zerofill,
            	precision=1,
        	    /tikz/.cd},
    		ylabel={$C_p$},
    		ytick={-3, -2, -1, 0, 1, 2},
    		ymin=-3,
    		ymax=2,
            % reverse legend,
    		y dir=reverse,
    		y tick label style={
        		/pgf/number format/.cd,
            	% fixed,
            	% fixed zerofill,
            	precision=1,
        	    /tikz/.cd},
    		legend style={at={(0.97, 0.97)},anchor=north east,font=\small,nodes={scale=0.9, transform shape}},
    		legend cell align={left},
    		style={font=\normalsize}]

    %   		\addplot[color=black!75, style={ultra thin}, only marks, mark=+, mark options={scale=0.7}]
				% table[x=x,y expr={\thisrow{cp}*-1},col sep=comma,unbounded coords=jump]{./figs/data/sd7003_beckp7_cp.csv};
    % 		\addlegendentry{\citet{Beck2014}}

    %   		\addplot[color=black!75, style={ultra thin}, only marks, mark=diamond, mark options={scale=0.7}]
				% table[x=x,y expr={\thisrow{cp}*-1},col sep=comma,unbounded coords=jump]{./figs/data/sd7003_beckp3_cp.csv};
    % 		\addlegendentry{\citet{Beck2014}}
      
    %   		\addplot[color=black!75, style={ultra thin}, only marks, mark=o, mark options={scale=0.7}]
				% table[x=x,y expr={\thisrow{cp}*-1},col sep=comma,unbounded coords=jump]{./figs/data/sd7003_cox_cp.csv};
    % 		\addlegendentry{\citet{Cox2021}}

			\addplot[color=black, style={dotted, very thick}]
				table[x=x,y=cp,col sep=comma,unbounded coords=jump]{./figs/data/sd7003_ns_p3_cp.csv};
    		\addlegendentry{Navier--Stokes, $\mathbb P_3$}

			\addplot[color=red!80!black, style={dotted, very thick}]
				table[x=x,y=cp,col sep=comma,unbounded coords=jump]{./figs/data/sd7003_bgk_p3_cp.csv};
    		\addlegendentry{Boltzmann--BGK, $\mathbb P_3$}

			\addplot[color=black, style={thick}]
				table[x=x,y=cp,col sep=comma,unbounded coords=jump]{./figs/data/sd7003_ns_p4_cp.csv};
    		\addlegendentry{Navier--Stokes, $\mathbb P_4$}
      
			\addplot[color=red!80!black, style={thick}]
				table[x=x,y=cp,col sep=comma,unbounded coords=jump]{./figs/data/sd7003_bgk_p4_cp.csv};
    		\addlegendentry{Boltzmann--BGK, $\mathbb P_4$}

		\end{axis} 		
	\end{tikzpicture}}}
        ~
        \subfloat[Skin friction coefficient]{\adjustbox{width=0.48\linewidth, valign=b}{    \begin{tikzpicture}[spy using outlines={rectangle, height=3cm,width=2.3cm, magnification=3, connect spies}]
		\begin{axis}[name=plot1,
		    axis line style={latex-latex},
		    axis x line=left,
            axis y line=left,
            clip mode=individual,
		    xlabel={$x/c$},
		    xtick={0,0.2,0.4,0.6,0.8,1},
    		xmin=0,
    		xmax=1,
    	    x tick label style={
        		/pgf/number format/.cd,
            	fixed,
            	fixed zerofill,
            	precision=1,
        	    /tikz/.cd},
    		ylabel={$C_f$},
    		ytick={-0.02, -0.01, 0, 0.01, 0.02},
    		ymin=-.02,
    		ymax=.02,
    		y tick label style={
        		/pgf/number format/.cd,
            	% fixed,
            	% fixed zerofill,
            	precision=1,
        	    /tikz/.cd},
    		legend style={at={(0.97, 0.97)},anchor=north east,font=\small},
    		legend cell align={left},
    		style={font=\normalsize}]
      
    %   		\addplot[color=black!75, style={ultra thin}, only marks, mark=o, mark options={scale=0.7}]
				% table[x=x,y=cf,col sep=comma,unbounded coords=jump]{./figs/data/sd7003_cox_cf.csv};
    
    %   		\addplot[color=black!75, style={ultra thin}, only marks, mark=diamond, mark options={scale=0.7}]
				% table[x=x,y=cf,col sep=comma,unbounded coords=jump]{./figs/data/sd7003_beckp3_cf.csv};

    %   		\addplot[color=black!75, style={ultra thin}, only marks, mark=+, mark options={scale=0.7}]
				% table[x=x,y=cf,col sep=comma,unbounded coords=jump]{./figs/data/sd7003_beckp7_cf.csv};

			\addplot[color=black, style={dotted, very thick}]
				table[x=x,y=cf,col sep=comma,unbounded coords=jump]{./figs/data/sd7003_ns_p3_cf.csv};
   	
			\addplot[color=red!80!black, style={dotted, very thick}]
				table[x=x,y=cf,col sep=comma,unbounded coords=jump]{./figs/data/sd7003_bgk_p3_cf.csv};
   	
			\addplot[color=black, style={thick}]
				table[x=x,y=cf,col sep=comma,unbounded coords=jump]{./figs/data/sd7003_ns_p4_cf.csv}; 
   	
			\addplot[color=red!80!black, style={thick}]
				table[x=x,y=cf,col sep=comma,unbounded coords=jump]{./figs/data/sd7003_bgk_p4_cf.csv} ;

		\end{axis} 		
	\end{tikzpicture}}}
        \caption{\label{fig:sd7003} Surface pressure coefficient (left) and suction-side skin friction coefficient (right) for the $Re = 60,000$ SD7003 airfoil case computed with the Navier--Stokes equations (black) and the Boltzmann--BGK equation (red) using a $\mathbb P_3$ (dotted) and $\mathbb P_4$ (solid) approximation. }
    \end{figure}

The profiles of the average surface pressure coefficient and skin friction coefficient for the Boltzmann--BGK and Navier--Stokes approaches computed with a $\mathbb P_3$ and $\mathbb P_4$ approximation are shown in \cref{fig:sd7003}. It can be seen in the surface pressure coefficient profiles that all approaches predict the pressure plateau on the suction side associated with the laminar separation bubble and the sharp adverse pressure gradient near the reattachment point. For the Navier--Stokes results, the pressure profiles were well-converged, with nearly identical surface pressure coefficient distributions computed by the $\mathbb P_3$ and $\mathbb P_4$ approximations. However, the Boltzmann--BGK results showed larger discrepancies with respect to the resolution, with the $\mathbb P_3$ results showing some notable differences compared to the $\mathbb P_4$ results which were in good agreement with the Navier--Stokes results. Similar observations could be drawn with the skin friction coefficient distributions which highlighted the differences more drastically. While the Navier--Stokes approach showed similar results with the $\mathbb P_3$ and $\mathbb P_4$ approximation, the Boltzmann--BGK approach showed significant differences between the two approximation orders. At $\mathbb P_3$, the reattachment point was predicted far forward of the Navier--Stokes results, with a strong overprediction of the skin friction aft of the reattachment point. However, when the resolution was increased with the $\mathbb P_4$ approximation, the Boltzmann--BGK results showed good agreement with the Navier--Stokes results, with only a minor overprediction of the skin friction aft of the reattachment point. These observations are consistent with the case of the Taylor--Couette flow where it was also seen that the accuracy of the Boltzmann--BGK approach was more affected by spatially underresolved flows. 

    \begin{figure}[htbp!]
        \centering
        \subfloat[Navier--Stokes] {
        \adjustbox{width=0.48\linewidth,valign=b}{\includegraphics[width=\textwidth]{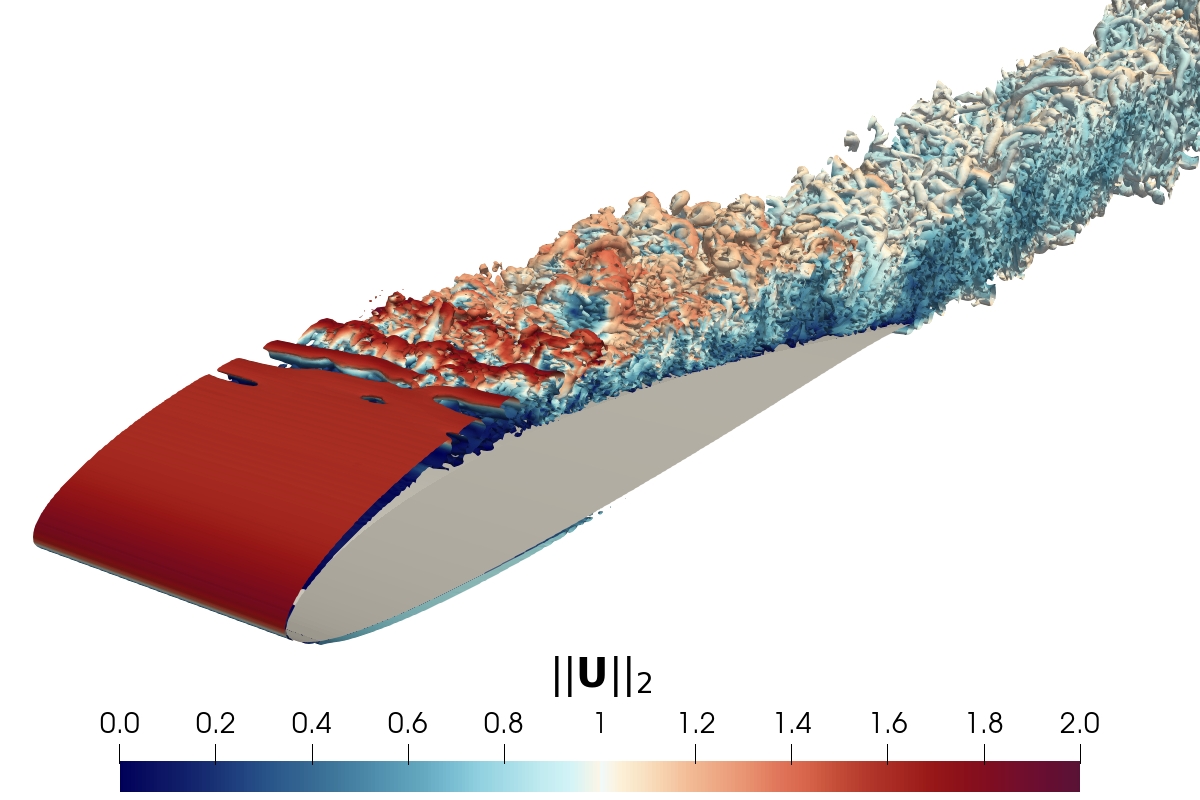}}}
        \subfloat[Boltzmann--BGK] {
        \adjustbox{width=0.48\linewidth,valign=b}{\includegraphics[width=\textwidth]{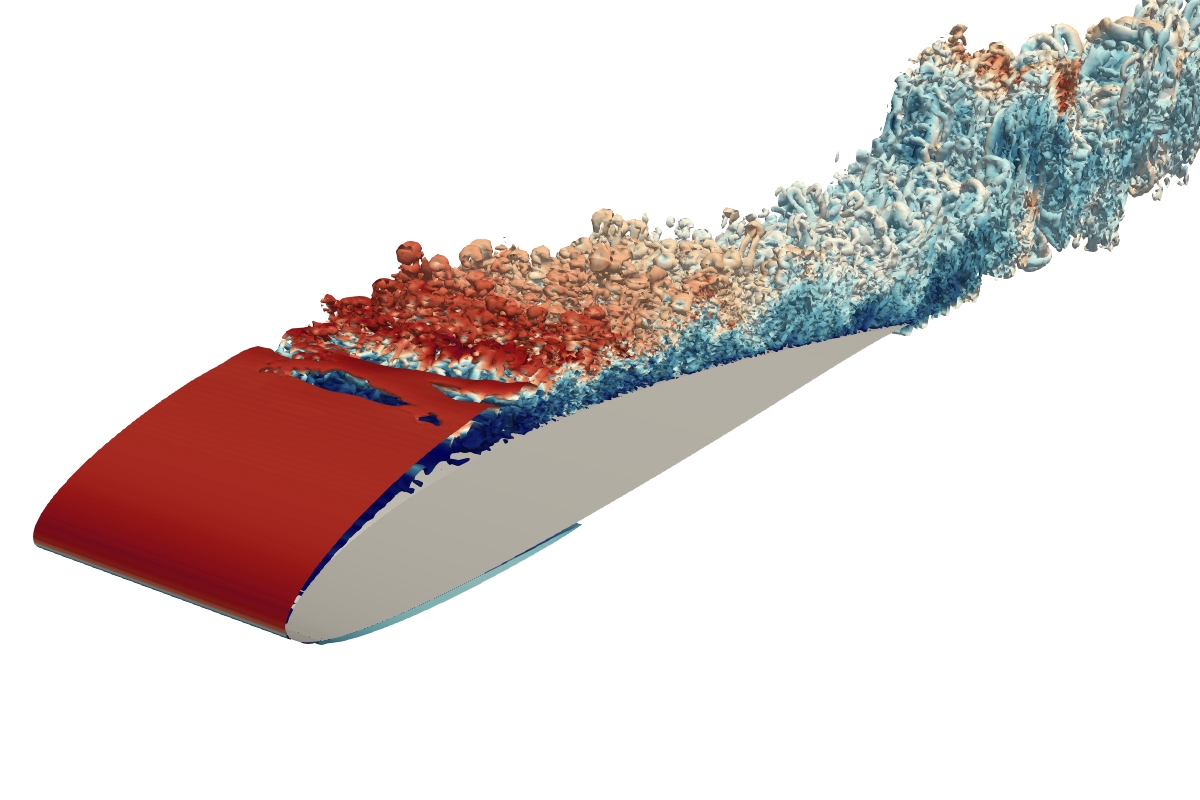}}}
        \caption{\label{fig:sd7003_qcrit} Isocontour of Q-criterion colored by velocity magnitude for the $Re = 60,000$ SD7003 airfoil case computed with the Navier--Stokes equations (left) and the Boltzmann--BGK equation (right) using a $\mathbb P_4$ approximation.  }
    \end{figure}
    
    \begin{figure}[htbp!]
        \centering
        \subfloat[Navier--Stokes] {
        \adjustbox{width=0.48\linewidth,valign=b}{\includegraphics[width=\textwidth]{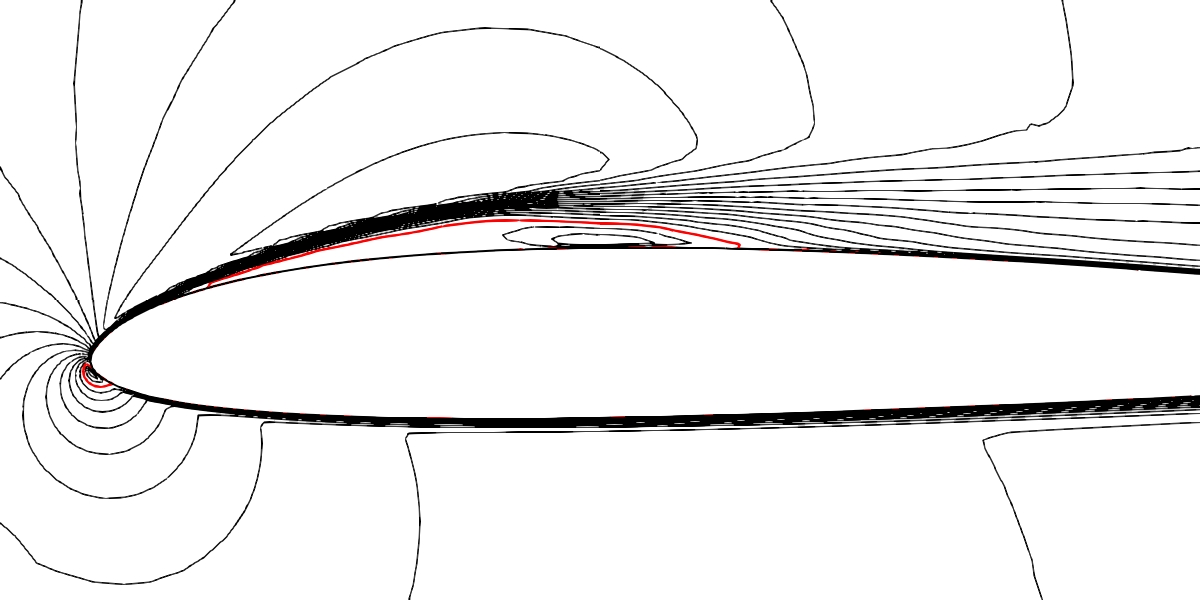}}}
        \subfloat[Boltzmann--BGK] {
        \adjustbox{width=0.48\linewidth,valign=b}{\includegraphics[width=\textwidth]{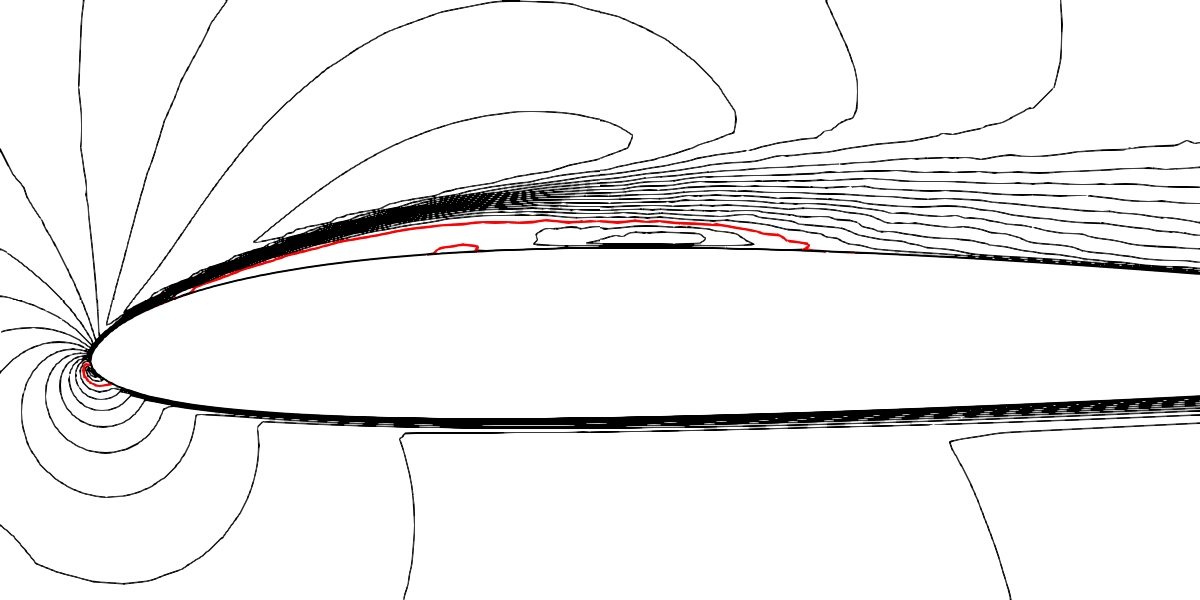}}}
        \caption{\label{fig:sd7003_contours} Isocontours of average streamwise velocity (equispaced on the range [-0.2, 1.5]) for the $Re = 60,000$ SD7003 airfoil case computed with the Navier--Stokes equations (left) and the Boltzmann--BGK equation (right) using a $\mathbb P_4$ approximation. Red isocontour represents zero average streamwise velocity. }
    \end{figure}

To evaluate the ability of the Boltzmann--BGK approach in resolving the dominant structures in the flow, the instantaneous flow was visualized using isocontours of Q-criterion, shown in \cref{fig:sd7003_qcrit} for both the Navier--Stokes approach and the Boltzmann--BGK approach at $\mathbb P_4$. It can be seen that the typical flow phenomena associated with the test case could be observed in both approaches, with the laminar separation bubble transitioning to a turbulent boundary layer and wake. While the laminar separation region was predicted by the Boltzmann--BGK approach quite consistently with the Navier--Stokes approach, it can be seen that the shear layer and subsequent turbulent boundary layer was more energized in the Boltzmann--BGK results. These observations can also be seen in \cref{fig:sd7003_contours} which shows the average isocontours of streamwise velocity. In comparison to the Navier--Stokes results, the Boltzmann--BGK results showed more spreading of the contours across the shear layer, indicating a higher degree of momentum transport. As similar observations were drawn in the Taylor--Couette case, this indicates that for spatially underresolved flows, the Boltzmann--BGK approach tends to predict more energized flow with a higher degree of momentum transport. However, it is expected that these results would converge to the Navier--Stokes results with increasing resolution based on the behavior of the skin friction coefficient profiles at the two levels of resolution.

    \begin{figure}[htbp!]
        \centering
        \adjustbox{width=0.98\linewidth, valign=b}{     \begin{tikzpicture}[spy using outlines={rectangle, height=3cm,width=2.3cm, magnification=3, connect spies}]
		\begin{axis}[name=plot1,
		    axis line style={latex-latex},
		    axis x line=left,
            axis y line=left,        
            width=\axisdefaultwidth,
            height=0.4*\axisdefaultwidth,
            clip mode=individual,
		    xlabel={$U_t/U_\infty$},
		    xtick={0, 2, 4, 6, 8, 10, 12, 14},
    		xmin=-0.25,
    		xmax=16,
    		x tick label style={
        		/pgf/number format/.cd,
            	fixed,
            	precision=1,
        	    /tikz/.cd},
    		ylabel={$y_n/c$},
    		ytick={0, .01, .02, .03, .04, .05},
    		ymin=0,
    		ymax=0.05,
    		y tick label style={
        		/pgf/number format/.cd,
            	% fixed,
            	% precision=1,
        	    /tikz/.cd},
    		legend style={at={(1.0,0.5)},anchor=west},
    		legend cell align={left},
    		style={font=\small},
    		scale = 2]

            \foreach \x in {1,...,8}{    				
    			\addplot[color={black}, style={very thick, dotted}]
    				table[x expr={\thisrow{u\x}+ 2*\x - 2}, y = y, col sep=comma, unbounded coords=jump]{./figs/data/sd7003_ns_p3_u2.csv};
        
    			\addplot[color={black}, style={thick}]
    				table[x expr={\thisrow{u\x}+ 2*\x - 2}, y = y, col sep=comma, unbounded coords=jump]{./figs/data/sd7003_ns_p4_u2.csv};
        
    			\addplot[color={red!80!black}, style={very thick, dotted}]
    				table[x expr={\thisrow{u\x}+ 2*\x - 2}, y = y, col sep=comma, unbounded coords=jump]{./figs/data/sd7003_bgk_p3_u2.csv};
        
    			\addplot[color={red!80!black}, style={thick}]
    				table[x expr={\thisrow{u\x}+ 2*\x - 2}, y = y, col sep=comma, unbounded coords=jump]{./figs/data/sd7003_bgk_p4_u2.csv};
            }
            
            \node at (0.8, .048) {\tiny$x/c = 0.1$};
            \node at (2.85, .048) {\tiny$x/c = 0.15$};
            \node at (4.95, .048) {\tiny$x/c = 0.2$};
            \node at (6.8, .048) {\tiny$x/c = 0.25$};
            \node at (8.75, .048) {\tiny$x/c = 0.3$};
            \node at (10.6, .048) {\tiny$x/c = 0.35$};
            \node at (12.55, .048) {\tiny$x/c = 0.4$};
            \node at (14.4, .048) {\tiny$x/c = 0.45$};
    		
    		% \addlegendentry{Navier--Stokes, $\mathbb P_3$}
    		% \addlegendentry{Boltzmann--BGK, $\mathbb P_3$}

		\end{axis}

	\end{tikzpicture}}
        \caption{\label{fig:sd7003_vel} Wall-tangential average velocity profiles for the $Re = 60,000$ SD7003 airfoil case at varying chord-wise locations on the suction side computed with the Navier--Stokes equations (black) and the Boltzmann--BGK equation (red) using a $\mathbb P_3$ (dotted) and $\mathbb P_4$ (solid) approximation. Legend identical to \cref{fig:sd7003}. Profiles are shifted $+0$, $+2$, ..., $+14$ along the abscissa, respectively. }
    \end{figure}
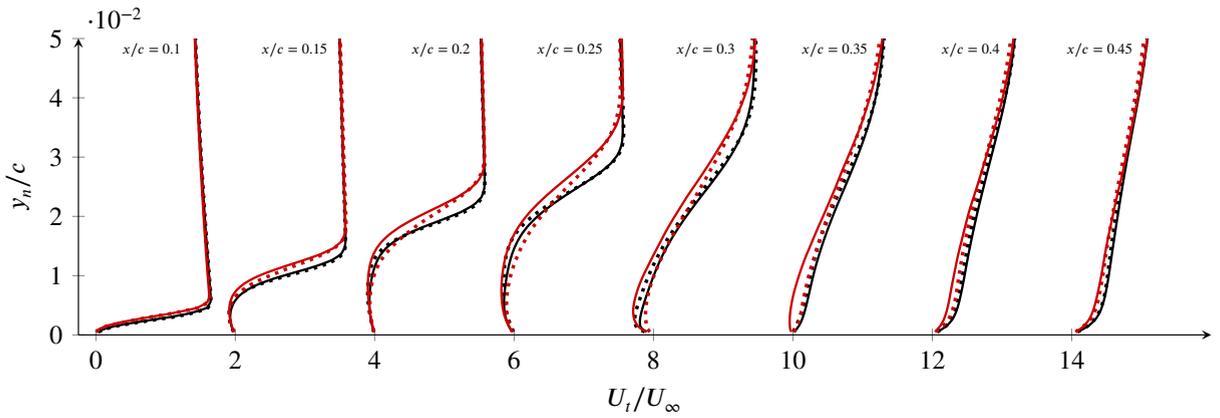
    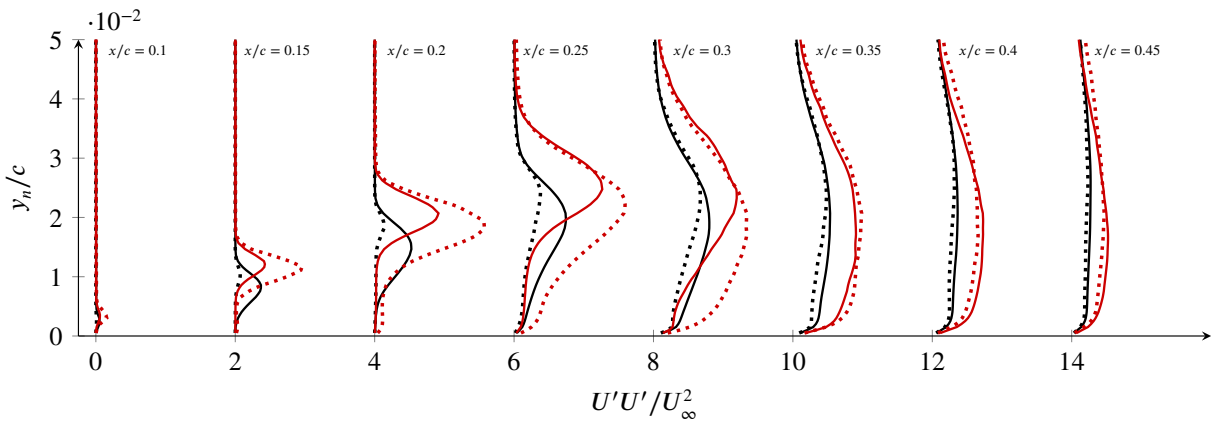
\begin{figure}[htbp!]
        \centering
        \adjustbox{width=0.98\linewidth, valign=b}{     \begin{tikzpicture}[spy using outlines={rectangle, height=3cm,width=2.3cm, magnification=3, connect spies}]
		\begin{axis}[name=plot1,
		    axis line style={latex-latex},
		    axis x line=left,
            axis y line=left,        
            width=\axisdefaultwidth,
            height=0.4*\axisdefaultwidth,
            clip mode=individual,
		    xlabel={$U'U'/U_\infty^2$},
		    xtick={0, 2, 4, 6, 8, 10, 12, 14},
    		xmin=-0.25,
    		xmax=16,
    		x tick label style={
        		/pgf/number format/.cd,
            	fixed,
            	precision=1,
        	    /tikz/.cd},
    		ylabel={$y_n/c$},
    		ytick={0, .01, .02, .03, .04, .05},
    		ymin=0,
    		ymax=0.05,
    		y tick label style={
        		/pgf/number format/.cd,
            	% fixed,
            	% precision=1,
        	    /tikz/.cd},
    		legend style={at={(1.0,0.5)},anchor=west},
    		legend cell align={left},
    		style={font=\small},
    		scale = 2]

            \foreach \x in {1,...,8}{    				
    			\addplot[color={black}, style={very thick, dotted}]
    				table[x expr={5*\thisrow{uu\x}+ 2*\x - 2}, y = y, col sep=comma, unbounded coords=jump]{./figs/data/sd7003_ns_p3_uu2.csv};
        
    			\addplot[color={black}, style={thick}]
    				table[x expr={5*\thisrow{uu\x}+ 2*\x - 2}, y = y, col sep=comma, unbounded coords=jump]{./figs/data/sd7003_ns_p4_uu2.csv};
        
    			\addplot[color={red!80!black}, style={very thick, dotted}]
    				table[x expr={5*\thisrow{uu\x}+ 2*\x - 2}, y = y, col sep=comma, unbounded coords=jump]{./figs/data/sd7003_bgk_p3_uu2.csv};
        
    			\addplot[color={red!80!black}, style={thick}]
    				table[x expr={5.0*\thisrow{uu\x}+ 2*\x - 2}, y = y, col sep=comma, unbounded coords=jump]{./figs/data/sd7003_bgk_p4_uu2.csv};
            }
            
            \node at (0.6, .048) {\tiny$x/c = 0.1$};
            \node at (2.6, .048) {\tiny$x/c = 0.15$};
            \node at (4.6, .048) {\tiny$x/c = 0.2$};
            \node at (6.6, .048) {\tiny$x/c = 0.25$};
            \node at (8.7, .048) {\tiny$x/c = 0.3$};
            \node at (10.8, .048) {\tiny$x/c = 0.35$};
            \node at (12.8, .048) {\tiny$x/c = 0.4$};
            \node at (14.8, .048) {\tiny$x/c = 0.45$};
    		
    		% \addlegendentry{Navier--Stokes, $\mathbb P_3$}
    		% \addlegendentry{Boltzmann--BGK, $\mathbb P_3$}

		\end{axis}

	\end{tikzpicture}}
        \caption{\label{fig:sd7003_uu} Streamwise velocity variance profiles for the $Re = 60,000$ SD7003 airfoil case at varying chord-wise locations on the suction side computed with the Navier--Stokes equations (black) and the Boltzmann--BGK equation (red) using a $\mathbb P_3$ (dotted) and $\mathbb P_4$ (solid) approximation. Legend identical to \cref{fig:sd7003}. Profiles are shifted $+0$, $+2$, ..., $+14$ along the abscissa, respectively, with a scaling factor of 5. }
    \end{figure}

For a more quantitative comparison of the flow, the average flow profiles were extracted at various chord-wise locations on the suction side of the airfoil. The wall-tangential velocity profiles, computed with respect to the surface normals at the given locations, are shown in \cref{fig:sd7003_vel}. It can be seen that across the suction side of the airfoil, both approaches and approximation orders show notably similar velocity profiles with respect to how much variation is shown in the reported results in the literature. In fact, even though the Boltzmann--BGK results show an overprediction of the height of the separation bubble, which is consistent with the velocity contours in \cref{fig:sd7003_contours}, the velocity profiles suggest a scaling for both the Navier--Stokes and Boltzmann--BGK results with respect to the separation bubble height. 

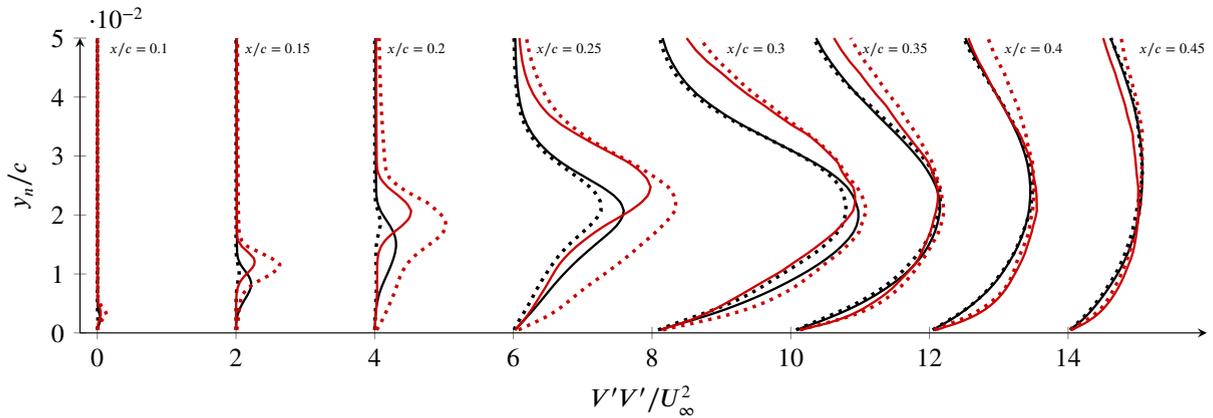
\begin{figure}[htbp!]
        \centering
        \adjustbox{width=0.98\linewidth, valign=b}{     \begin{tikzpicture}[spy using outlines={rectangle, height=3cm,width=2.3cm, magnification=3, connect spies}]
		\begin{axis}[name=plot1,
		    axis line style={latex-latex},
		    axis x line=left,
            axis y line=left,        
            width=\axisdefaultwidth,
            height=0.4*\axisdefaultwidth,
            clip mode=individual,
		    xlabel={$V'V'/U_\infty^2$},
		    xtick={0, 2, 4, 6, 8, 10, 12, 14},
    		xmin=-0.25,
    		xmax=16,
    		x tick label style={
        		/pgf/number format/.cd,
            	fixed,
            	precision=1,
        	    /tikz/.cd},
    		ylabel={$y_n/c$},
    		ytick={0, .01, .02, .03, .04, .05},
    		ymin=0,
    		ymax=0.05,
    		y tick label style={
        		/pgf/number format/.cd,
            	% fixed,
            	% precision=1,
        	    /tikz/.cd},
    		legend style={at={(1.0,0.5)},anchor=west},
    		legend cell align={left},
    		style={font=\small},
    		scale = 2]

            \foreach \x in {1,...,8}{    				
    			\addplot[color={black}, style={very thick, dotted}]
    				table[x expr={20*\thisrow{vv\x}+ 2*\x - 2}, y = y, col sep=comma, unbounded coords=jump]{./figs/data/sd7003_ns_p3_vv2.csv};
        
    			\addplot[color={black}, style={thick}]
    				table[x expr={20*\thisrow{vv\x}+ 2*\x - 2}, y = y, col sep=comma, unbounded coords=jump]{./figs/data/sd7003_ns_p4_vv2.csv};
        
    			\addplot[color={red!80!black}, style={very thick, dotted}]
    				table[x expr={20*\thisrow{vv\x}+ 2*\x - 2}, y = y, col sep=comma, unbounded coords=jump]{./figs/data/sd7003_bgk_p3_vv2.csv};
        
    			\addplot[color={red!80!black}, style={thick}]
    				table[x expr={20.0*\thisrow{vv\x}+ 2*\x - 2}, y = y, col sep=comma, unbounded coords=jump]{./figs/data/sd7003_bgk_p4_vv2.csv};
            }
            
            \node at (0.6, .048) {\tiny$x/c = 0.1$};
            \node at (2.6, .048) {\tiny$x/c = 0.15$};
            \node at (4.6, .048) {\tiny$x/c = 0.2$};
            \node at (6.8, .048) {\tiny$x/c = 0.25$};
            \node at (9.5, .048) {\tiny$x/c = 0.3$};
            \node at (11.5, .048) {\tiny$x/c = 0.35$};
            \node at (13.5, .048) {\tiny$x/c = 0.4$};
            \node at (15.5, .048) {\tiny$x/c = 0.45$};
    		
    		% \addlegendentry{Navier--Stokes, $\mathbb P_3$}
    		% \addlegendentry{Boltzmann--BGK, $\mathbb P_3$}

		\end{axis}

	\end{tikzpicture}}
        \caption{\label{fig:sd7003_vv} Normal velocity variance profiles for the $Re = 60,000$ SD7003 airfoil case at varying chord-wise locations on the suction side computed with the Navier--Stokes equations (black) and the Boltzmann--BGK equation (red) using a $\mathbb P_3$ (dotted) and $\mathbb P_4$ (solid) approximation. Legend identical to \cref{fig:sd7003}. Profiles are shifted $+0$, $+2$, ..., $+14$ along the abscissa, respectively, with a scaling factor of 20.  }
    \end{figure}

A similar comparison was performed for the streamwise and normal velocity variance profiles, shown in \cref{fig:sd7003_uu} and \cref{fig:sd7003_vv}, respectively. The overprediction of the height of the separation bubble by the Boltzmann--BGK approach is also evident in the variance profiles, with the peaks of the profiles shown farther from the airfoil surface in the region $0.1 \leq x/c \leq 0.25$. However, it must be noted that these discrepancies in separation bubble height were on the order of $1\%$ of the chord. A notably interesting observation from these variance profiles is the effect of underresolution on the flow behavior. It can be seen that for the Navier--Stokes results, the less resolved case ($\mathbb P_3$) shows lower variance in the flow, i.e., the flow is less energized. When the resolution was increased with $\mathbb P_4$, higher variance was observed, indicating that underresolution for the Navier--Stokes approach tends to have a dissipative effect on the flow, which is expected. In contrast, the Boltzmann--BGK results showed the complete opposite behavior. Noticeably more variance was observed in the less resolved Boltzmann--BGK results than the more resolved results, indicating that the flow is actually more energized when underresolved, i.e., underresolved flow simulations via the Boltzmann--BGK approach tend to be much less dissipative than their Navier--Stokes counterparts. It can be seen that with increasing resolution, the magnitude of the peaks of the variance profiles seems to be qualitatively converging between the Boltzmann--BGK results and the Navier--Stokes results. 

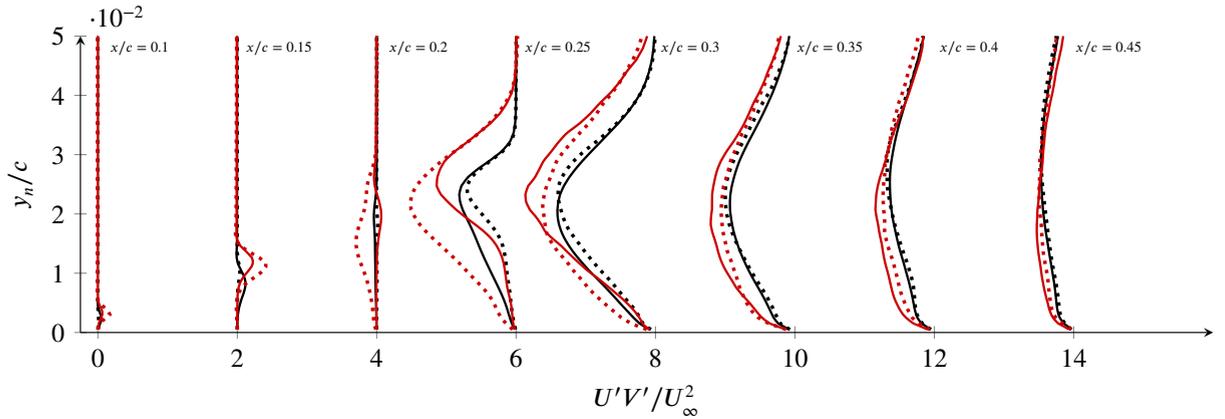
\begin{figure}[htbp!]
        \centering
        \adjustbox{width=0.98\linewidth, valign=b}{     \begin{tikzpicture}[spy using outlines={rectangle, height=3cm,width=2.3cm, magnification=3, connect spies}]
		\begin{axis}[name=plot1,
		    axis line style={latex-latex},
		    axis x line=left,
            axis y line=left,        
            width=\axisdefaultwidth,
            height=0.4*\axisdefaultwidth,
            clip mode=individual,
		    xlabel={$U'V'/U_\infty^2$},
		    xtick={0, 2, 4, 6, 8, 10, 12, 14},
    		xmin=-0.25,
    		xmax=16,
    		x tick label style={
        		/pgf/number format/.cd,
            	fixed,
            	precision=1,
        	    /tikz/.cd},
    		ylabel={$y_n/c$},
    		ytick={0, .01, .02, .03, .04, .05},
    		ymin=0,
    		ymax=0.05,
    		y tick label style={
        		/pgf/number format/.cd,
            	% fixed,
            	% precision=1,
        	    /tikz/.cd},
    		legend style={at={(1.0,0.5)},anchor=west},
    		legend cell align={left},
    		style={font=\small},
    		scale = 2]

            \foreach \x in {1,...,8}{    				
    			\addplot[color={black}, style={very thick, dotted}]
    				table[x expr={20*\thisrow{uv\x}+ 2*\x - 2}, y = y, col sep=comma, unbounded coords=jump]{./figs/data/sd7003_ns_p3_uv2.csv};
        
    			\addplot[color={black}, style={thick}]
    				table[x expr={20*\thisrow{uv\x}+ 2*\x - 2}, y = y, col sep=comma, unbounded coords=jump]{./figs/data/sd7003_ns_p4_uv2.csv};
        
    			\addplot[color={red!80!black}, style={very thick, dotted}]
    				table[x expr={20*\thisrow{uv\x}+ 2*\x - 2}, y = y, col sep=comma, unbounded coords=jump]{./figs/data/sd7003_bgk_p3_uv2.csv};
        
    			\addplot[color={red!80!black}, style={thick}]
    				table[x expr={20.0*\thisrow{uv\x}+ 2*\x - 2}, y = y, col sep=comma, unbounded coords=jump]{./figs/data/sd7003_bgk_p4_uv2.csv};
            }
            
            \node at (0.6, .048) {\tiny$x/c = 0.1$};
            \node at (2.6, .048) {\tiny$x/c = 0.15$};
            \node at (4.6, .048) {\tiny$x/c = 0.2$};
            \node at (6.6, .048) {\tiny$x/c = 0.25$};
            \node at (8.5, .048) {\tiny$x/c = 0.3$};
            \node at (10.5, .048) {\tiny$x/c = 0.35$};
            \node at (12.5, .048) {\tiny$x/c = 0.4$};
            \node at (14.5, .048) {\tiny$x/c = 0.45$};
    		
    		% \addlegendentry{Navier--Stokes, $\mathbb P_3$}
    		% \addlegendentry{Boltzmann--BGK, $\mathbb P_3$}

		\end{axis}

	\end{tikzpicture}}
        \caption{\label{fig:sd7003_uv} Streamwise-normal velocity covariance profiles for the $Re = 60,000$ SD7003 airfoil case at varying chord-wise locations on the suction side computed with the Navier--Stokes equations (black) and the Boltzmann--BGK equation (red) using a $\mathbb P_3$ (dotted) and $\mathbb P_4$ (solid) approximation. Legend identical to \cref{fig:sd7003}. Profiles are shifted $+0$, $+2$, ..., $+14$ along the abscissa, respectively, with a scaling factor of 20.  }
    \end{figure}

To quantify the differences in the momentum transport effects between the two approaches, a comparison of the streamwise-normal velocity covariance was also performed, shown in \cref{fig:sd7003_uv}. The overprediction in the streamwise-normal velocity covariance indicative of an overprediction in momentum transport was evident with the Boltzmann--BGK results, particularly so with the $\mathbb P_3$ approximation around $x/c = 0.25$. Similar observations could be drawn for the covariance profiles as with the variance profiles, with the less-resolved Boltzmann--BGK results showing the highest degree of (negative) covariance and the less-resolved Navier--Stokes results showing the lowest. These results are consistent with the observations drawn from the visualization of the instantaneous Q-criterion and the average streamwise velocity isocontours.  

\section{Discussion}\label{sec:discussion}
Given the results of the numerical experiments performed, we present a discussion of the overarching observations on the use of the Boltzmann--BGK approach in this section.

\paragraph{Accuracy and consistency.} Throughout our  numerical experiments, it can be seen that if sufficient levels of spatial and velocity domain resolution are used, the Boltzmann--BGK approach accurately predicts complex flow physics and momentum transfer effects for wall-bounded and shear-induced flows. For non-equilibrium flows, the approach was validated against both canonical test cases and more complex microchannel applications, with good results in comparison to reference data. Additionally, we presented novel results for three-dimensional rarefied flows in the form of the flow through a T-junction. For continuum flows, the approach showed good agreement with analytical predictions and Navier--Stokes results for fundamental flows such as laminar boundary layers. Furthermore, the ability of the approach to accurately predict fundamental three-dimensional hydrodynamic instabilities was shown through the simulation of a transitional Taylor--Couette flow. A final test case consisting of the separated flow over an SD7003 airfoil highlighted the capability of the approach for significantly more complex flow physics and discrepancies between the hydrodynamic equations and the Boltzmann--BGK approach.

\paragraph{Velocity space resolution.} The results of the numerical experiments showed that for both rarefied and continuum flows, accurate and converged results could be obtained using very few degrees of freedom in the velocity domain. It was observed that resolution levels of as low as eight velocity space nodes per dimension were sufficient for all numerical experiments, from two-dimensional steady rarefied flows to three-dimensional transitional and turbulent flows. In fact, for the discrete velocity model approach used in this work, this level of velocity space resolution is only marginally higher than the minimum resolution needed to meet the realizability conditions for the model (i.e., the level of resolution needed for there to exist a solution to the nonlinear optimization problem). Furthermore, this highlights the necessity of using discretely-conservative velocity models, such as the one implemented in this work, as the level of resolution needed to ensure conservation to an acceptable tolerance would be significantly higher than the level of resolution needed to accurately predict complex flows. 

\paragraph{Spatial resolution.} The effect of the spatial resolution on the predictions of fluid flows via the Boltzmann--BGK approach is arguably the most principal finding of this work. It was observed that if the flow was spatially well-resolved, the Boltzmann--BGK approach could accurately predict complex hydrodynamic instabilities such as transition to turbulence consistently with the Navier--Stokes approach of the same resolution. However, when the flow was not spatially well-resolved, interesting differences could be seen between the Boltzmann--BGK approach and the Navier--Stokes approach. It was found that for spatially underresolved flows, the Boltzmann--BGK approach would be noticeably less dissipative than its Navier--Stokes counterpart, showing a more energized flow and spurious predictions of hydrodynamic instabilities. This is in complete contrast to approximations of the Navier--Stokes equations, where numerical dissipation from underresolution in the flow tends to act as a somewhat physically consistent subgrid-scale model for the underresolved components of the flow. The fact that these two approaches show completely different behavior under the effect of underresolution in the flow is not unexpected. For the Navier--Stokes equations, numerical dissipation can directly affect the evolution of momentum in the flow, acting as a subgrid-scale model, whereas for the Boltzmann--BGK equation, underresolution in the spatial domain only introduces errors in particle transport without affecting particle collision, the actual mechanism for modeling viscous effects. Therefore, it is expected that the arguments for intrinsic subgrid-scale modeling via numerical dissipation (i.e., implicit large eddy simulation) would not directly extend to the Boltzmann--BGK approach.

However, it must be noted that the spurious prediction of hydrodynamic instabilities by the Boltzmann--BGK approach may not be baseless. As was observed with the Taylor--Couette flow, underresolved flows computed via the Navier--Stokes equations may also show a similar form of instability which stems from aliasing errors. While this numerical instability may not directly affect the Boltzmann--BGK approach due to the linearity of the transport term, it is possible that the predicted flow is somewhat consistent with the flow that would be recovered by the hydrodynamic equations with the given discretization order. 

Regardless, the results indicate that to resolve complex hydrodynamic instabilities, the accurate prediction of particle transport (i.e., high spatial resolution) seems to be of much higher importance than the accurate prediction of particle collision (i.e., high velocity domain resolution). This observation suggests that the Boltzmann--BGK approach is better suited for highly-resolved simulations (e.g., direct numerical simulation) and that numerical methods with higher spatial accuracy (e.g., high-order schemes) are particularly well-suited for simulating complex fluid flows via the Boltzmann--BGK equation.
%Furthermore, this opens up the possibility for radically different approaches for turbulence modeling and subgrid-scale modeling from the perspective of the Boltzmann equation, which is of a completely different mathematical character than the Navier--Stokes equations and seems to respond entirely differently to numerical discretization effects. 

\paragraph{Stability and robustness.} It was observed in the numerical experiments that high-order approximations of the Boltzmann--BGK equation were significantly more robust and less prone to numerical stability issues than the Navier--Stokes approach for a given mesh and problem setup. Even for flows such as the rarefied flat plate which should ostensibly be trivial for the hydrodynamic equations, poor mesh quality due to the presence of large aspect ratio cells caused the Navier--Stokes approach to diverge even at only moderately high approximation orders. In contrast, there were no issues with the Boltzmann--BGK approach on such problems. Furthermore, for underresolved transitional and turbulent flows, the Navier--Stokes equations required overintegration of the flux term to mitigate aliasing errors stemming from the nonlinearity of the equations. No such additional stabilization was required for the Boltzmann--BGK approach. These observations indicate that for high-order simulations of fluid flows, the Boltzmann--BGK approach may be a more robust alternative to the Navier--Stokes approach.

\paragraph{Computational cost.} As mentioned, the approximation of the Boltzmann--BGK equation comes with a significantly higher computational cost than the Navier--Stokes equation. The computational cost differences between the two can be attributed to three main sources.

First, the total degrees of freedom is significantly higher in the Boltzmann--BGK approach due to the higher dimensionality. However, based on the results of the numerical experiments, it appears to be sufficient to utilize a velocity space resolution of approximately $N_v = 8^d$, where $d$ is the dimensionality of the problem. This results in approximately an order of magnitude more degrees of freedom for two-dimensional flows and two orders of magnitude more degrees of freedom for three-dimensional flows. While this difference is substantial, with the emergence of progressively larger high-performance computing clusters, it is not unreasonable to attempt to solve the Boltzmann--BGK equation for complex fluid flows. We remark that based on our experiences with an efficient GPU implementation of the presented numerical approach, this tends to be primarily an issue of having enough memory to account for the increased number of degrees of freedom. When enough memory is available, either through parallelizing across more GPUs or through the use of GPUs with more memory, the actual compute time tends to be very reasonable, typically much less so than the compute time needed for a Navier--Stokes calculation of an equivalent total number of degrees of freedom. Therefore, the computational cost associated with the large increase in total number of degrees of freedom is not as much of a problem as it may seem, e.g., the Taylor--Couette and SD7003 cases with $\mathcal O(10^{10})$ degrees of freedom could be reasonably run on 32 32 GiB NVIDIA V100 GPUs.

Secondly, the time step restrictions between the Boltzmann--BGK approach and the Navier--Stokes approach can drastically differ depending on the flow regime being simulated. In the continuum limit (low Mach number or high Reynolds number), the stiffness of the collision operator in the Boltzmann--BGK approach can detrimentally affect the maximum admissible time step for explicit discretizations. However, for highly-resolved simulations, the time step limits imposed by the stiffness of the source term were not drastically different than the time step limits given by the standard CFL condition for the transport term, typically well within an order of magnitude. For the Navier--Stokes approach, the presence of second-order viscous terms can also impose even stricter time step constraints, especially so for well-resolved flows at higher Reynolds numbers. As such, the use of explicit time stepping for highly-resolved flow simulations via the Boltzmann--BGK approach was not found to be significantly disadvantageous in comparison to the Navier--Stokes approach. Furthermore, in the rarefied regime, the maximum admissible time step for the Boltzmann--BGK approach, limited by the transport term, was typically much higher than the Navier--Stokes approach, limited by the viscous term. In such regimes, the time step restrictions of the Boltzmann--BGK approach were much more favorable. 

Finally, the computational cost of solving one time step of the Boltzmann--BGK equation versus one time step of the Navier--Stokes equations with an equivalent number of degrees of freedom can drastically differ. For the Boltzmann--BGK equation, the temporal evolution can be represented as a simple linear advection step with a nonlinear source term. As the transport term is first-order, there is no need to evaluate gradients of the solution which is a substantial cost in the approximation of the Navier--Stokes equation. Furthermore, the computation of the source term, which requires nonlinear optimization for the discrete velocity model, is a very compute heavy task, such that is can be implemented very efficiently on GPU computing architectures with minimal additional cost. As such, the evaluation of a single step of the Boltzmann--BGK equation is typically much more efficient than the Navier--Stokes equation, where second-order viscous terms require the evaluation of solution gradients. Additionally, the nonlinearity of the Navier--Stokes equation can introduce further computational costs. For underresolved flows, it is typically necessary to apply some sort of anti-aliasing to mitigate aliasing errors, and due to the nonlinearity of the flux term, it is required in many cases to use an adaptive time stepping scheme to account for the changing time step restrictions in the flow. These numerical difficulties can further decrease the gap between the computational cost of solving the Boltzmann--BGK equation and the Navier--Stokes equations. 
\section{Conclusions}\label{sec:conclusion}
In this work, we explore the capability of simulating complex fluid flows via the Boltzmann--BGK equation and present a comprehensive validation for the effects of wall boundary conditions on momentum transfer in the flow in the low Mach limit. The numerical approach was first validated against canonical wall-bounded flows in both the rarefied and continuum regimes. The approach was then applied to more complex problems including three-dimensional rarefied flows and transitional/turbulent flows, the latter of which are, to the authors' knowledge, the first instances of such flows computed by directly solving the Boltzmann equation. The results of the numerical experiments indicate that the Boltzmann--BGK approach can accurately predict momentum transfer and non-equilibrium effects in the rarefied regime as well as complex hydrodynamic instabilities in the continuum regime. It was found that to predict these flow instabilities and momentum transfer effects in a manner consistent with the hydrodynamic equations, a highly-resolved spatial domain was of much higher importance than a highly-resolved velocity domain, indicating that the accurate prediction of particle transport is more important than the accurate prediction of particle collision and that high-order spatial schemes may be a promising approach for effectively solving the Boltzmann equation. Furthermore, it was found that given a discretely-conservative velocity model, converged results could be obtained with very few degrees of freedom in the velocity domain, typically as few as eight nodes per dimension. However, the behavior of the Boltzmann--BGK approach for spatially underresolved flows showed marked differences in comparison to the Navier--Stokes approach, with notably less dissipation in the flow and spurious onset of hydrodynamic instabilities, which suggests that the use of numerical dissipation as an intrinsic subgrid-scale model for fluid flows does not readily extend to the Boltzmann--BGK approach.

The results of this work present a validation of the Boltzmann--BGK approach for wall-bounded flows, showcasing the capability of the approach for simulating complex non-equilibrium and hydrodynamic effects. Further work is necessary to comprehensively evaluate the capabilities of molecular gas dynamics approaches such as the Boltzmann equation for fluid dynamics applications, including a validation of the approach for predicting energy transfer effects and surface heat fluxes in the high Mach regime, shock-driven flow instabilities and interactions, and multi-scale and high-temperature behavior in hypersonic and high-enthalpy fluid dynamics applications. Furthermore, as seen from the observations drawn from the numerical experiments, there may also be a need for alternate approaches for subgrid-scale modeling for underresolved flows. Ultimately, the results of this work present opportunities for novel approaches to analyzing fundamental flow phenomena such as transition to turbulence as well as developing alternate turbulence modeling approaches through the perspective of the evolution of a particle distribution function.

\section*{Acknowledgements}
\label{sec:ack}
 TD and LM would like to acknowledge the computational resources provided by the Princeton Institute for Computational Science and Engineering. FW would like to acknowledge funding provided by the  U.S. Air Force Office of Scientific Research Young Investigator Program via grant FA9550-23-1-0232 under the direction of Dr. Fariba Fahroo.

\bibliographystyle{unsrtnat}
\bibliography{reference}


% Show the list of todo's in the document.  Needed to avoid stupid warnings/errors when using the todo package
%\todos

\end{document}